\documentclass{article}

\usepackage{microtype}
\usepackage{graphicx}
\usepackage{tabularx}
\usepackage{array}
\usepackage{subfigure}
\usepackage{booktabs} % for professional tables
\usepackage{caption}
\usepackage{url}

\usepackage{hyperref}

\usepackage[accepted]{icml2025}

\usepackage{amsmath}
\usepackage{amssymb}
\usepackage{mathtools}
\usepackage{amsthm}

\usepackage[capitalize,noabbrev]{cleveref}

\theoremstyle{plain}

\theoremstyle{definition}

\theoremstyle{remark}

\usepackage[textsize=tiny]{todonotes}

\newcommand{\bench}{DecisionUnlearn-Bench}

\title{DecisionUnlearn-Bench: Unlearning Probabilistic Decisions under Changing Schemas}

\begin{document}

\twocolumn[
\icmltitle{DecisionUnlearn-Bench: Unlearning Probabilistic Decisions under Changing Schemas}




\begin{icmlauthorlist}
\icmlauthor{Tiantong Wu}{yyy}
\icmlauthor{Wei Yang Bryan Lim}{yyy}
% \icmlauthor{Firstname3 Lastname3}{comp}
\end{icmlauthorlist}

\icmlaffiliation{yyy}{Nanyang Technological University, Singapore}

\icmlcorrespondingauthor{Tiantong Wu}{tiantong.wu@ntu.edu.sg}

\icmlkeywords{Machine Learning, ICML}

\vskip 0.3in
]



\printAffiliationsAndNotice{}  % leave blank if no need to mention equal contribution

\begin{abstract}
Machine unlearning aims to reproduce the behavior of a model trained without designated data. For probabilistic decision models, this target also depends on the candidate options supplied at inference. We introduce \bench{}, an empirical framework that compares full, unlearned, and retain-only models across deletion granularities, label renderings, and candidate sets. Across 20 eligible model configurations on the BANKING77 dataset, full-to-oracle displacement increases consistently from example to association to class deletion. Replacing semantic option labels with opaque codes nearly eliminates the retrained oracle's recognition of deleted classes, even with successful acquisition of the full model, whereas changing the candidate set alters both the oracle and the measured unlearning progress. SimNPO satisfies the screening criterion in 19 of 20 configurations, but its gains vary substantially across seeds and deletion targets. Development tuning on the CLINC150 dataset does not establish a positive pooled effect across 48 intents. At approximately matched suppression, methods can differ markedly in oracle fidelity. Recovery experiments further distinguish initial forgetting from subsequent adaptation and show conclusions that depend on both architecture and metric. These results show that evaluating decision-model unlearning relative to a schema-specific retraining target, with separate reporting of effect size, utility, robustness, and recoverability.
\end{abstract}

\section{Introduction}
\label{sec:introduction}
% CARS Move 1: establish the research territory and its importance.
Machine unlearning asks how to remove the influence of designated training data in a trained model without repeating the entire training process~\citep{ginart2019making,bourtoule2021machine}. A central reference is the model obtained by training on the remaining data. This reference makes the objective more precise than simply reducing accuracy on examples associated with a deletion request. A relevant question is whether the unlearned model would have behaved the same way without those examples.

This distinction matters for models that return probabilities over a specified set of decisions. For example, in intent classification, a request is evaluated against a set of named candidate intents. The model can use both its learned mapping from utterances to intents and the meanings of the supplied labels. Removing every training sample of an intent therefore need not eliminate recognition of that intent. The remaining data, pretrained knowledge, and candidate descriptions may still support the correct decision.

% CARS Move 2: identify the unresolved problem, without claiming that
% retraining-based evaluation or relearning tests are themselves new.
Existing language-model unlearning benchmarks examine forgotten knowledge, retained utility, privacy, and robustness through complementary evaluations~\citep{maini2024tofu,shi2024muse,dorna2025openunlearning}. Preference-based methods also address the instability of directly maximizing the loss on data that needs to be forgotten~\citep{zhang2024npo,fan2025simnpo}. These developments leave an additional evaluation question for decision models:

\begin{center}
    \textit{How should fidelity to retraining be interpreted when the information supplied by the decision schema changes?}
\end{center}

A low forgotten-class accuracy can agree with one retraining target and disagree with another. Similarly, a normalized improvement can change because the unlearned model changes, the oracle changes, or the original deletion effect becomes small.

Three distinctions are consequently necessary. First, the counterfactual effect of deletion must be measured before interpreting an intervention. Second, the amount of probability removed from forgotten options must be separated from where that probability is redistributed. Third, high performance after further adaptation must be separated from the behavior already present before adaptation. Without these distinctions, a method can appear to forget successfully, transfer reliably, or relearn rapidly for reasons that the reported endpoint does not isolate.

% CARS Move 3: occupy the niche with the study, design, and findings.
We address these questions with \bench{}, a study of probabilistic decision unlearning on the BANKING77 and CLINC150 datasets~\citep{casanueva2020efficient,larson2019evaluation}. The benchmark compares a full model, an unlearned model, and a retain-only retrained oracle using their complete option distributions. It separates: (i) example, association, and class deletion, (ii) development selection and subsequent evaluation, and (iii) broad model screening and targeted replication. Controlled changes to label rendering and candidate composition test how the decision interface changes the target itself.

The results identify both useful approximation and clear limits. Class deletion creates the largest distributional shift across all models in the 20-model granularity panel. SimNPO provides reliable gains across the reference BANKING77 seed-bundle grid, yet its performance under the frozen setting is less stable across other seeds and across CLINC150 targets. Opaque labels almost eliminate oracle recognition, but do not consistently increase normalized unlearning progress. Reduced candidate sets can also expose poor fidelity when labels are opaque, showing that schema sensitivity extends beyond residual semantic recognition. Recovery analyses produce different conclusions for Apertus-8B and Qwen3.5-9B models once starting behavior and remaining headroom are separated.

Our contributions are as follows:
\begin{itemize}
 \item An evaluation design that measures displacement and approximation induced by deletion to a retraining target specific to the schema on both deleted training rows and unseen related rows.
 \item An empirical account of how label semantics and candidate composition alter oracle behavior and the interpretation of unlearning fidelity.
 \item A combined analysis of method transfer, probability redistribution, and recovery that separates statistical acceptance from effect size and identifies the limits of conclusions drawn from a fixed decision interface.
\end{itemize}

\section{Related Work}
\label{sec:related}
\paragraph{Unlearning as approximation to retraining.}
Data-deletion formulations and efficient retraining schemes establish the importance of specifying the learning procedure whose output is to be reproduced~\citep{ginart2019making,bourtoule2021machine}. More recent oracle-matching approaches explicitly target output distributions rather than only the correctness of forgotten predictions~\citep{georgiev2025oracle}. We adopt retraining as an operational reference for supervised task adaptation. Our experiments do not remove the dataset from the pre-training or establish equality between the distributions over the trained model parameters.

\paragraph{Language-model unlearning methods and evaluation.}
TOFU~\citep{maini2024tofu} studies synthetic author knowledge with retain-only references, whereas MUSE~\citep{shi2024muse} evaluates several aspects of forgetting and retained utility. The OpenUnlearning~\citep{dorna2025openunlearning} framework consolidates these two methods and evaluations to examine the reliability across different metrics. NPO~\citep{zhang2024npo} and SimNPO~\citep{fan2025simnpo} modify the forgetting objective to improve optimization behavior. RMU~\citep{li2024wmdp} instead intervenes on internal representations. Our contribution is an empirical investigation of these approaches in a decision setting, rather than a new unlearning optimizer.

\paragraph{Decision schemas and recoverability.}
Intent datasets provide explicit candidate meanings and natural distinctions between removing individual records and removing all examples of an intent~\citep{casanueva2020efficient,larson2019evaluation}. We vary those candidate meanings and the supplied option set while retaining the same evaluation inputs. Relearning is already an established unlearning evaluation, including in the SimNPO study~\citep{fan2025simnpo}. Here, it compares adaptation from unlearned and retrained checkpoints while distinguishing starting-point differences, subsequent gains, and transfer through a newly initialized head.

\section{Problem Formulation and Evaluation}
\label{sec:formulation}
\subsection{A schema-specific retraining target}
Let $\mathcal{D}$ be the supervised training set, $\mathcal{D}_\mathcal{F}\subset\mathcal{D}$ a deletion set, and $\mathcal{D}_\mathcal{R}=\mathcal{D}\setminus\mathcal{D}_\mathcal{F}$ the retained set. A decision schema $\mathcal{S}=(\mathcal{C},\ell)$ specifies an ordered candidate set $\mathcal{C}$ and its textual rendering $\ell$, including option names and descriptions. Model $M$ returns a categorical distribution
\begin{equation}
 \mathbf p_M(x;\mathcal{S})=\bigl(p_M(c\mid x;\mathcal{S})\bigr)_{c\in\mathcal{C}},
 \qquad \sum_{c\in\mathcal{C}}p_M(c\mid x;\mathcal{S})=1.
\end{equation}
The term \emph{decision model} refers to this probabilistic interface throughout the paper.

For a training procedure $\mathcal A$, pretrained initialization $M_0$, and training seed $\xi$, we compare
\begin{equation}
\begin{aligned}
 M_{\mathrm{full}}&=\mathcal A(M_0,\mathcal{D},\mathcal{S}_{\rm train};\xi),\\
 M_{\mathcal{O}}&=\mathcal A(M_0,\mathcal{D}_\mathcal{R},\mathcal{S}_{\rm train};\xi),\\
 M_{\mathcal{U}}&=\mathcal U(M_{\mathrm{full}},\mathcal{D}_\mathcal{F},\mathcal{D}_\mathcal{R}),
\end{aligned}
\label{eq:models}
\end{equation}
where $\mathcal U$ denotes an approximate unlearning procedure. The \emph{retrained oracle} $M_\mathcal{O}$ is trained without the deletion set under the task-training recipe. The \emph{no-op} baseline is $M_\mathrm{full}$. Both references are evaluated under the same schema as $M_\mathcal{U}$. In particular, deleted classes remain available as options: deleting their training examples does not remove their outputs from the decision space.

The label intervention changes $\mathcal{S}_{\rm train}$ and the corresponding evaluation rendering, requiring separately trained full and oracle models. The candidate-set intervention changes the evaluation schema of existing checkpoints and performs a new forward pass for each reduced option set. These interventions address different sources of schema dependence.

\subsection{Deletion granularity and evaluation views}
We study three deletion requests. \emph{(i) Example deletion} removes a random 5\% of training rows. \emph{(ii) Association deletion} removes selected sub-intent clusters while retaining other examples of their parent intents. And \emph{(iii) Class deletion} removes all training rows of selected classes. The primary BANKING77 comparison uses ten association clusters or four classes, whereas CLINC150 class bundles contain eight intents.

We distinguish two evaluation views. The \emph{(i) forget-training view} $\mathcal{E}_\mathcal{F}$ contains deleted training rows. The \emph{(ii) forget-generalization view} $\mathcal{E}_\mathcal{G}$ contains unseen evaluation rows associated with the deleted units. For class and association deletion, these are rows of the corresponding class or cluster. For example deletion, these are evaluation rows whose nearest training row belongs to $\mathcal{D}_\mathcal{F}$. Thus, $\mathcal{G}$ does not denote a second deletion set or a generic robustness benchmark. Retained utility is evaluated on $\mathcal{E}_\mathcal{R}$.

\subsection{Distance, progress, and feasibility}
For $v\in\{\mathcal{F},\mathcal{G}\}$, the oracle distance is the mean base-2 Jensen--Shannon divergence:
\begin{equation}
 \operatorname{OD}_v(M;\mathcal{S})=\frac{1}{|\mathcal{E}_v|}
 \sum_{x\in\mathcal{E}_v}\operatorname{JS}_2\!\left(\mathbf p_M\left(x;\mathcal{S}\right),\mathbf p_{M_\mathcal{O}}\left(x;\mathcal{S}\right)\right).
 \label{eq:od}
\end{equation}
We use raw option probabilities, with numerical stabilization at $10^{-12}$. We define
\begin{equation}
\begin{aligned}
 D_{0,v}&=\operatorname{OD}_v(M_\mathrm{full};\mathcal{S}),\\
 \Delta\operatorname{OD}_v&=D_{0,v}-\operatorname{OD}_v(M_\mathcal{U};\mathcal{S}),\\
 P_v&=\frac{\Delta\operatorname{OD}_v}{D_{0,v}}
     =1-\frac{\operatorname{OD}_v(M_\mathcal{U};\mathcal{S})}{D_{0,v}}.
\end{aligned}
\label{eq:progress}
\end{equation}
Here $D_{0,v}$ is \emph{deletion displacement}, $\Delta\operatorname{OD}_v$ is the absolute gain over no-op, and $P_v$ is \emph{normalized oracle progress}. Positive gain means closer agreement with the oracle. Progress equals zero for no-op and one for exact agreement on the evaluated distributions, and negative values indicate movement away from the oracle. Because the denominator depends on the deletion request and schema, we report $D_{0,v}$ and absolute gains alongside progress. We do not interpret ratios with negligible displacement as stable effect sizes.

A configuration is \emph{feasible} when retained accuracy decreases by at most 0.02 and retained expected calibration error increases by at most 0.02 relative to the corresponding full model:
\begin{equation}
\begin{split}
 \operatorname{Acc}_\mathcal{R}(M_\mathcal{U})\geq\operatorname{Acc}_\mathcal{R}(M_\mathrm{full})-0.02,\\
 \mathcal{E}CE_\mathcal{R}(M_\mathcal{U})\leq\mathcal{E}CE_\mathcal{R}(M_\mathrm{full})+0.02.
\end{split}
\end{equation}
Protocol \emph{acceptance} additionally requires positive gains with positive lower 95\% confidence bounds on both views. It specifies no minimum progress fraction and does not imply equivalence to retraining.

\subsection{Selection, replication, and uncertainty}
Development data determine target eligibility, method configurations, and suppression matching. Primary class targets require full-model development accuracy of at least 0.70 and gold probability of at least 0.50. Association targets use an accuracy threshold of 0.65. Targets are selected in hash order among eligible units at the primary seed. Replication preserves target selection rather than silently replacing difficult seed-specific cases. One semantic BANKING77 bundle-02 class falls below the accuracy threshold at seed 42. We disclose this exception in Appendix~\ref{app:design}.

Within a fixed trained configuration, estimates use 5000 paired row-bootstrap draws, with forgotten-class or association-cluster bootstraps as sensitivity analyses. Each progress ratio is recomputed inside the resample. The frozen confirmation family reports Holm-adjusted intersection--union tests based on the larger of the two one-sided view-specific $p$-values. Architecture-screening decisions use the stored per-model rule without correction across models. Reference-model cross-bundle summaries use the stored hierarchical bootstrap over seeds, bundles, and units. Row intervals are conditioned on trained checkpoints and do not quantify uncertainty over future training seeds.

\section{Experimental Design}
\label{sec:design}
BANKING77 provides 77 fine-grained banking intents, while CLINC150 spans 150 in-scope intents across ten domains \citep{casanueva2020efficient,larson2019evaluation}. Our CLINC150 interface contains 151 options, including out-of-scope options. The model panel contains 20 eligible configurations. InternLM3-8B was excluded because checkpoint reloading was not reproducible. Qwen3.5-0.8B is the reference model for method selection and the most extensive replication. We use \emph{model configuration} to denote a named backbone and its size. Multiple configurations can belong to the same model family.

We evaluate gradient ascent (GradAscent), gradient difference with retained-data supervision (GradDiff), its KL-regularized variant (GradDiff-KL), NPO, SimNPO, RMU, and a uniformization control in the experimental blocks, each of which was selected or specified. These are not a complete method-by-model factorial comparison. The main confirmation compares six class-deletion arms, with the broad transfer panel evaluating the frozen reference-model SimNPO configuration. RMU is included in the suppression diagnostic. Temperature scaling provides an additional control for CLINC150.

\begin{table*}[tbp]
\centering
\small
\setlength{\tabcolsep}{5pt}
\renewcommand{\arraystretch}{1.15}
\caption{Observed evaluation design. Repeated evaluations of one checkpoint are not independent training runs.}
\label{tab:design}
\begin{tabularx}{\textwidth}{
    @{}
    >{\raggedright\arraybackslash}p{0.15\textwidth}
    >{\raggedright\arraybackslash}X
    >{\raggedright\arraybackslash}p{0.24\textwidth}
    @{}
}
\toprule
\textbf{Experiment} & \textbf{Observed coverage} & \textbf{Role} \\
\midrule

Granularity &
BANKING77; 20 eligible models; seed 17; three deletion requests &
Descriptive full--oracle comparison \\
\addlinespace[3pt]

Method comparison &
Qwen3.5-0.8B; six class-confirmation arms; four methods across a
$3$-seed $\times$ $3$-bundle grid &
Development selection, confirmation, replication \\
\addlinespace[3pt]

Model transfer &
Frozen SimNPO; 20-model panel including the reference; five-model
seed/bundle comparison &
Single-seed screening and targeted replication \\
\addlinespace[3pt]

Dataset transfer &
CLINC150; frozen transfer, then tuned SimNPO on seven cells across
bundles 02--07 &
Confirmation plus intent-level replication \\
\addlinespace[3pt]

Label rendering &
Qwen3.5-0.8B; seven semantic/opaque pairs: four BANKING77,
three CLINC150 &
Mechanistic contrasts \\
\addlinespace[3pt]

Candidate sets &
Three models; full, neighbor, and equal-size random-distractor
schemas; semantic and selected opaque checkpoints &
Evaluation-only rescoring \\
\addlinespace[3pt]

Suppression &
Reference model; BANKING77 target $s=0.6$, CLINC150 target $s=1$
on development data &
Held-out diagnostic \\
\addlinespace[3pt]

Recovery &
Three models; one trained seed each; three exposure budgets,
two adaptation seeds, three head seeds &
Exploratory; controls for 0.8B and Apertus only \\

\bottomrule
\end{tabularx}
\end{table*}

The main evaluation design separates three questions. First, the granularity screen measures the changes in retain-only retraining across all eligible models. Second, method confirmation and a seed-bundle replication test approximate that target. Third, schema, suppression, and recovery interventions examine why a favorable fixed-interface result may not persist. Table~\ref{tab:design} summarizes observed evaluation coverage. Supplementary tables preserve the full numerical results and the detailed distinction among the pilot, development, confirmation, screening, replication, and exploratory stages.

\section{Results}
\label{sec:results}
\subsection{Deletion granularity determines the size of the counterfactual shift}
\label{sec:granularity}
\begin{figure*}[ht]\centering
  \includegraphics[width=0.95\linewidth,height=0.76\textheight,keepaspectratio]{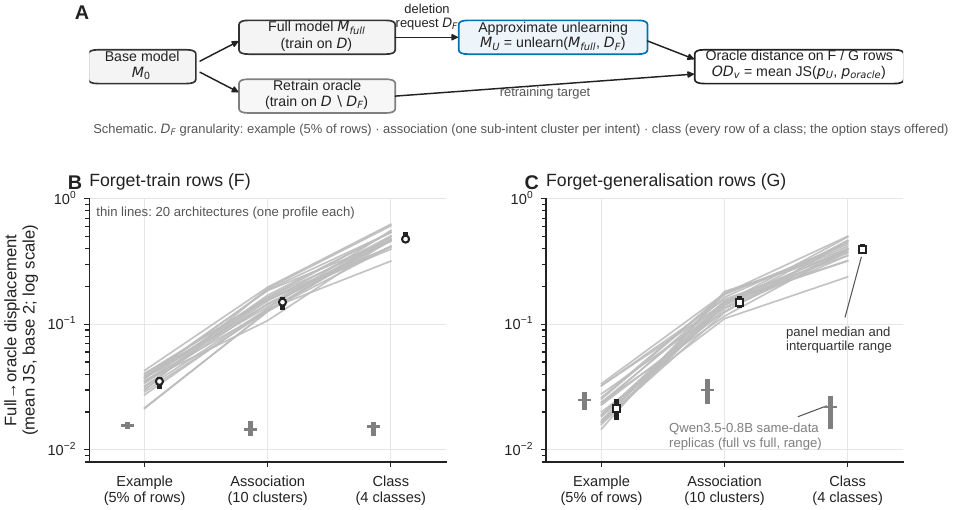}
\caption{Deletion granularity changes the retraining target. (A) Full, unlearned, and retain-only models. (B,C) BANKING77 displacement for 20 eligible configurations, seed 17, on $\mathcal{F}$ and $\mathcal{G}$. Thin lines connect each model's three deletion settings: black marks show the median and interquartile range, not confidence intervals; gray marks show the range of same-data full-model divergence for Qwen3.5-0.8B only. Example, association, and class deletion increase displacement in every model. The schematic's association description refers to selected parent intents, not all 77 intents.}
\label{fig:granularity}
\end{figure*}

Full-to-oracle displacement increases from example to association, then to class deletion in every configuration in the BANKING77 screen (Figure~\ref{fig:granularity}). The panel medians are respectively 0.035, 0.150, and 0.477 on $\mathcal{F}$, and 0.021, 0.149, and 0.396 on $\mathcal{G}$. These are descriptive medians over model configurations, not population estimates.

Same-data replicas are available for Qwen3.5-0.8B at seeds 17, 42, and 99. For this model, example-deletion displacement is approximately within twice the observed replica noise. This supports the protocol's decision not to advance example-deletion methods beyond selection. It does not establish that deleted records were never memorized, or that retaining the full model meets a privacy requirement. The replica comparison is also specific to this reference model.

Association deletion produces a larger target shift, and its final cross-bundle sensitivity analysis finds modest SimNPO gains: $\Delta\operatorname{OD}_\mathcal{F}=0.019$ $[0.007,0.034]$ and $\Delta\operatorname{OD}_\mathcal{G}=0.018$ $[0.005,0.035]$. We retain this result while distinguishing its selection pathway from the primary class-deletion confirmation. The strongest and most extensively replicated intervention evidence concerns class deletion.

\subsection{Label meaning changes the oracle, but does not guarantee easier unlearning}
\label{sec:semantics}
We replace option names and descriptions with fixed opaque codes while preserving utterances, target identities, option order, and the availability of forgotten options. Full models pass the opaque-label acquisition checks. On BANKING77, oracle accuracy on deleted training rows falls from 0.209--0.315 with semantic labels to zero across both bundles and both seeds. The corresponding unseen-row accuracies are also zero under opaque labels. Forgotten-option mass remains small but nonzero, approximately $4.7\times10^{-4}$ to $1.8\times10^{-3}$ on $\mathcal{F}$ (Figure~\ref{fig:schema}A,B).

The CLINC150 contrasts show the same qualitative oracle effect. Semantic-label oracle accuracy on $\mathcal{F}$ is 0.661, 0.505, and 0.781 for bundles 02, 03, and 06. With opaque labels, only one of 791 rows is correct on bundle 02, and none is correct on bundles 03 or 06. Full-to-oracle displacement increases to 0.983, 0.993, and 0.999. Thus, the retrained model's residual recognition depends strongly on the information in the supplied labels in these configurations.

Unlearning progress does not respond uniformly. On BANKING77 bundle 01, SimNPO reaches $P_\mathcal{F}$ of 0.60--0.61 under opaque labels, compared with 0.55--0.59 under semantic labels. On bundle 02, progress decreases from 0.70 to 0.45 at seed 17 and from 0.47 to 0.40 at seed 42. CLINC150 progress changes from 0.22 to 0.28 on bundle 02, from 0.22 to 0.13 on bundle 03, and from -0.24 to 0.02 on bundle 06. The last opaque run is statistically accepted but reproduces only a small fraction of the deletion shift.

The intervention therefore changes the oracle more consistently than it improves approximation to the oracle. Removing semantic labels does not by itself resolve CLINC150's unlearning difficulty. These are mechanistic contrasts, including previously inspected targets, rather than an independent estimate of cross-target success.

\subsection{Candidate composition changes both recognition and fidelity}
\label{sec:candidates}
We next rescore the existing checkpoints under the full option set ($S_0$), the set of forgotten classes with semantic neighbors ($S_1$), and the set of forgotten classes with randomly retained distractors ($S_2$). The two reduced schemas have equal cardinality, with 12 options in BANKING77 and 24 in CLINC150. This comparison controls for subset size between $S_1$ and $S_2$. Comparison with $S_0$ changes both size and composition.

\begin{figure*}[ht]
  \centering
  \includegraphics[width=\linewidth,height=0.76\textheight,keepaspectratio]{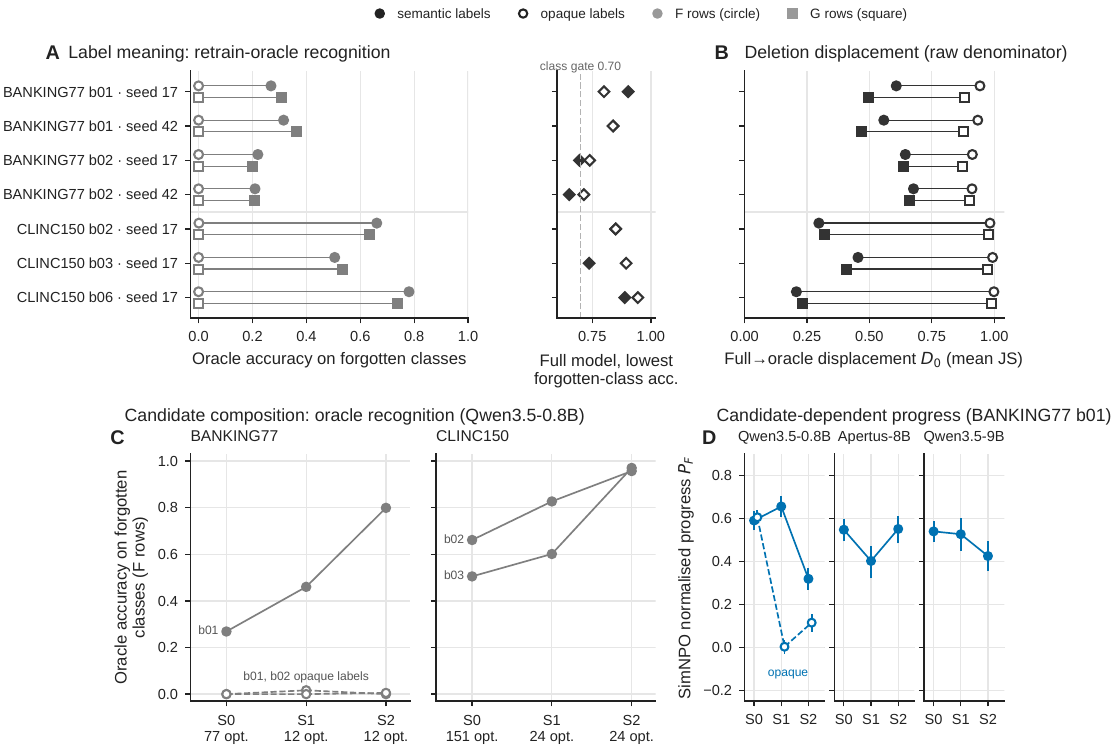}
  \caption{The decision schema changes the oracle and measured progress. (A) Oracle recognition under semantic and opaque labels; the narrow strip reports the lowest full-model development accuracy among forgotten classes. (B) Each condition's displacement denominator. (C) Oracle $\mathcal{F}$ accuracy under full, neighbor, and random-distractor schemas. (D) BANKING77 SimNPO $P_\mathcal{F}$ with row-bootstrap 95\% intervals, dashed open markers denote opaque labels. Reduced schemas are genuinely rescored and retain every forgotten option. Circles and squares in A,B indicate $\mathcal{F}$ and $\mathcal{G}$, option counts appear below C.}
  \label{fig:schema}
  \end{figure*}

For semantic BANKING77 bundle 01, Qwen3.5-0.8B's oracle accuracy on $\mathcal{F}$ increases from 0.269 under $S_0$ to 0.460 under $S_1$ and 0.799 under $S_2$. On $\mathcal{G}$, the corresponding values are 0.306, 0.519, and 0.756. The deletion displacement on $\mathcal{F}$ decreases from 0.608 to 0.438 and 0.148. SimNPO's progress changes from $0.59$ to $0.66$ and then to $0.32$. Other selected methods also lose progress with random distractors. Apertus-8B and Qwen3.5-9B retain positive row-level progress under $S_2$, although class-level uncertainty is wider (Figure~\ref{fig:schema}C,D).

Opaque labels separate this effect from semantic oracle recognition. On BANKING77 bundle 01, oracle recognition remains near zero under reduced schemas, yet SimNPO progress falls from 0.61 under $S_0$ to 0.00 under $S_1$ and 0.12 under $S_2$. Bundle 02 shows the same qualitative failure to preserve full-schema progress (0.45 to 0.04 and 0.12). Reduced-schema sensitivity therefore persists even when the oracle does not substantially re-recognize the forgotten classes.

\subsection{Reliable reference-model gains do not imply uniform transfer}
\label{sec:transfer}
In the reference class-deletion confirmation, SimNPO, GradAscent, GradDiff, and NPO satisfy the feasibility and gain criteria (Table~\ref{tab:methods}). SimNPO has the largest point gains, $\Delta\operatorname{OD}_\mathcal{F}=0.359$ and $\Delta\operatorname{OD}_\mathcal{G}=0.280$, while GradDiff-KL and Uniformization fail the utility/calibration gate despite positive distance gains. These results concern selected configurations, not an exhaustive ranking of optimization algorithms.

Across the reference model's nine seed--bundle cells, SimNPO retains positive hierarchical gains on both views, remains feasible in every cell, and survives the stored leave-one-seed/bundle-out checks. Its pooled gains are 0.337 [0.238,0.443] on $\mathcal{F}$ and 0.302 [0.208,0.412] on $\mathcal{G}$. GradAscent does not yield positive aggregate gains, whereas GradDiff and NPO do, but both fail the full replication rule because feasibility is not maintained in every cell.

\begin{table}[!htb]
\centering
\small
\caption{Reference-model class deletion on BANKING77. Gains are no-op minus unlearned oracle distance. Intervals in the upper block use paired rows; the lower block reports the stored hierarchical seed--bundle--class analysis.}
\label{tab:methods}
\setlength{\tabcolsep}{4pt}
\begin{tabular}{
@{}
>{\raggedright\arraybackslash}p{2.0cm}
>{\raggedright\arraybackslash}p{1cm}
>{\raggedright\arraybackslash}p{1cm}
>{\raggedright\arraybackslash}p{1.2cm}
>{\raggedright\arraybackslash}p{1.4cm}
@{}
}
\toprule
Method & $\Delta\operatorname{OD}_\mathrm{F}$ & $\Delta\operatorname{OD}_\mathrm{G}$ & $P_\mathrm{F}/P_\mathrm{G}$ & Decision \\
\midrule
\multicolumn{5}{@{}l}{\emph{Phase 1: seed 17, bundle 01; $D_{0,\mathrm{F}}=0.608$, $D_{0,\mathrm{G}}=0.495$}}\\
SimNPO & 0.359 & 0.280 & 0.59/0.57 & Accepted \\
GradAscent & 0.332 & 0.241 & 0.55/0.49 & Accepted \\
GradDiff & 0.274 & 0.209 & 0.45/0.42 & Accepted \\
NPO & 0.236 & 0.176 & 0.39/0.36 & Accepted \\
GradDiff-KL & 0.188 & 0.139 & 0.31/0.28 & Infeasible \\
Uniform\-ization & 0.160 & 0.102 & 0.26/0.21 & Infeasible \\
\midrule
\multicolumn{5}{@{}l}{\emph{Phase 2: seeds 17/42/99; bundles 01--03; nine cells per method}}\\
SimNPO & 0.337 & 0.302 & --- & Replicates \\
GradAscent & 0.114 & 0.081 & --- & Does not replicate \\
GradDiff & 0.187 & 0.160 & --- & Infeasible cell(s) \\
NPO & 0.168 & 0.142 & --- & Infeasible cell(s) \\
\bottomrule
\end{tabular}

\par\smallskip
\noindent
\parbox{\linewidth}{\footnotesize
Phase 1 uses 428 $\mathrm{F}$ and 160 $\mathrm{G}$ rows. The full model's retained accuracy is 93.7\%, and ECE is 0.026. Retained accuracy changes for the six methods are -0.1,-0.1,+0.2,+0.2,-2.1,+0.2 percentage points; ECE changes are +0.001,+0.013,+0.011,+0.003,+0.035,+0.046. The supplied confirmation table reports Holm-adjusted $p=0.0005$ for the first five arms; it does not provide an adjusted value for Uniformization. Feasibility is required independently of gain. Phase 2 progress is not pooled; individual results and ranges are in the supplement.
}
\end{table}

\begin{figure*}[ht]\centering
  \includegraphics[width=0.95\linewidth,height=0.76\textheight,keepaspectratio]{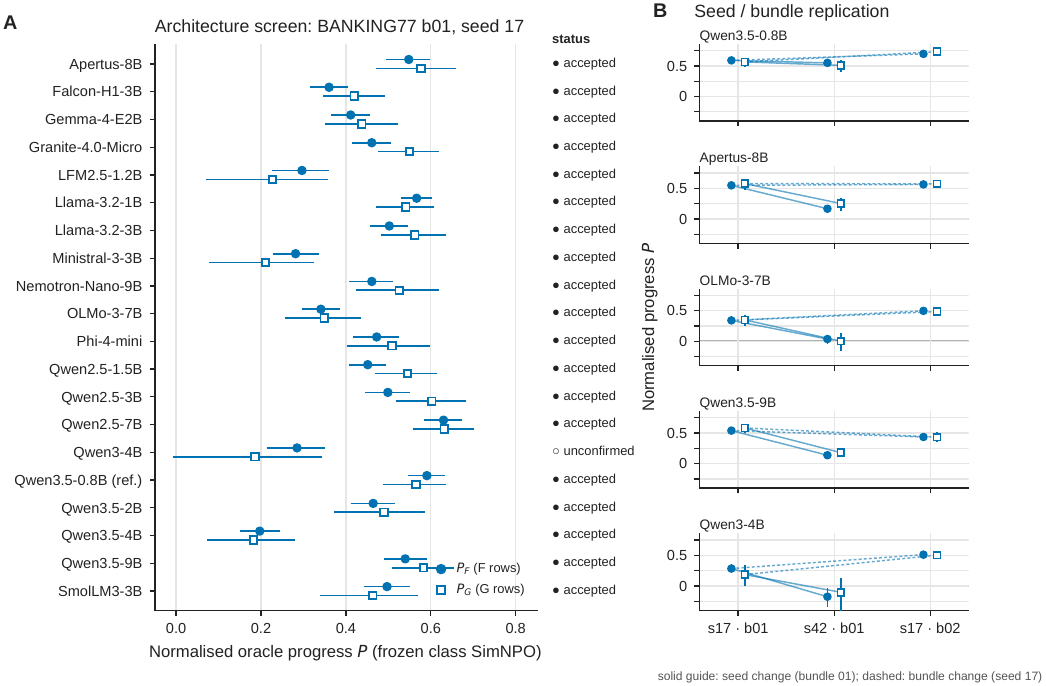}
\caption{Frozen-setting transfer and within-model variation. (A) BANKING77 bundle 01, seed 17, for 20 configurations including the reference model: circles $P_\mathcal{F}$, squares $P_\mathcal{G}$, with conditional row-bootstrap 95\% intervals. Status follows the stored screening rule without correction across models. (B) Five-model replication panel. Solid guides mark seed changes, dashed guides mark bundle changes. Small positive effects can satisfy acceptance, while seed-specific effects can be weak or negative.}
\label{fig:transfer}
  \end{figure*}

The frozen-setting panel contains 20 configurations, including the reference model. 19 satisfy the stored screening criterion, while Qwen3-4B is positive but unconfirmed. Progress on $\mathcal{F}$ ranges from 0.20 to 0.63, with a median of 0.46. However, additional seeds reveal substantial variation (Figure~\ref{fig:transfer}). At seed 42, $P_\mathcal{F}$ is 0.17 for Apertus-8B, 0.14 for Qwen3.5-9B, 0.04 for OLMo-3-7B, and -0.18 for Qwen3-4B. The reference model retains 0.55. Results for seed 17 on bundle 02 are positive across all five replicated models. This purposive, incomplete panel supports within-model contrasts, not a population variance decomposition or a 95\% general success probability.

Dataset transfer is less favorable. On CLINC150 bundle 01, the BANKING77 SimNPO settings increase oracle distance: gains are -0.191 on $\mathcal{F}$ and -0.237 on $\mathcal{G}$. Development tuning produces accepted results on bundles 02 and 03, but later bundles include unconfirmed and negative outcomes. Across 48 intents in six bundles, the pooled gains are 0.022 [-0.045,0.090] on $\mathcal{F}$ and 0.022 [-0.044,0.094] on $\mathcal{G}$ (Table~\ref{tab:transfer}). These intervals do not establish a positive average effect across targets. They also do not establish equivalence to no-op. CLINC150 bundle 08 remains unused.

\begin{table}[!htb]
\centering
\small
\caption{CLINC150 transfer and label-semantic contrasts. Positive row-level gains in selected bundles do not establish a pooled gain across intents. All methods in this table are SimNPO on Qwen3.5-0.8B.}
\label{tab:transfer}
\setlength{\tabcolsep}{4pt}
\begin{tabular}{
@{}
>{\raggedright\arraybackslash}p{3.5cm}
>{\raggedleft\arraybackslash}p{1cm}
>{\raggedleft\arraybackslash}p{1cm}
>{\raggedleft\arraybackslash}p{1.5cm}
@{}
}
\toprule
Configuration & $\Delta\operatorname{OD}_\mathrm{F}$ & $\Delta\operatorname{OD}_\mathrm{G}$ & $P_\mathrm{F}/P_\mathrm{G}$ \\
\midrule
Frozen transfer, b01/s17 & -0.191 & -0.237 & -0.46/-0.67 \\
Tuned, b02/s17 & 0.065 & 0.088 & 0.22/0.28 \\
Tuned, b02/s42 & 0.062 & 0.040 & 0.18/0.11 \\
Tuned, b03/s17 & 0.102 & 0.083 & 0.22/0.20 \\
Tuned, b04/s17 & 0.061 & 0.033 & 0.13/0.08 \\
Tuned, b05/s17 & 0.025 & 0.022 & 0.07/0.07 \\
Tuned, b06/s17 & -0.050 & -0.020 & -0.24/-0.09 \\
Tuned, b07/s17 & -0.071 & -0.051 & -0.24/-0.17 \\
\textbf{Pooled, 48 intents} & \textbf{0.022} & \textbf{0.022} & Not pooled \\
\midrule
\multicolumn{4}{@{}l}{\emph{Opaque-label contrasts: same tuned configuration, seed 17}}\\
Opaque, b02 & 0.275 & 0.287 & 0.28/0.29 \\
Opaque, b03 & 0.132 & 0.111 & 0.13/0.11 \\
Opaque, b06 & 0.023 & 0.024 & 0.02/0.02 \\
\bottomrule
\end{tabular}

\par\smallskip
\noindent
\parbox{\linewidth}{\footnotesize
Bundle-wise intervals resample rows; the pooled row resamples intents after averaging seeds within an intent. Opaque contrasts are exploratory and are excluded from that pooled semantic-label estimate. Their $D_{0,\mathrm{F}}$ values are 0.983, 0.993, and 0.999, versus 0.298, 0.454, and 0.207 with semantic labels. Bundle 08 remains reserved.
}
\end{table}

\subsection{Suppression magnitude is insufficient to characterize fidelity}
\label{sec:geometry}
Let $\mathcal{C}_\mathcal{F}\subset\mathcal{C}$ be the forgotten classes and $m_M(x)=\sum_{c\in\mathcal{C}_\mathcal{F}}p_M(c\mid x;\mathcal{S})$ their total probability mass. The suppression ratio
\begin{equation}
 s=\frac{\sum_{x\in\mathcal{E}_\mathcal{F}\cup\mathcal{E}_\mathcal{G}}[m_{M_\mathrm{full}}(x)-m_{M_\mathcal{U}}(x)]}
         {\sum_{x\in\mathcal{E}_\mathcal{F}\cup\mathcal{E}_\mathcal{G}}[m_{M_\mathrm{full}}(x)-m_{M_\mathcal{O}}(x)]}
\end{equation}
compares the amount removed with the oracle's amount. It is distinct from $P_v$, which uses the complete distribution.

\begin{figure*}[!htb]\centering
  \includegraphics[width=0.95\linewidth,height=0.76\textheight,keepaspectratio]{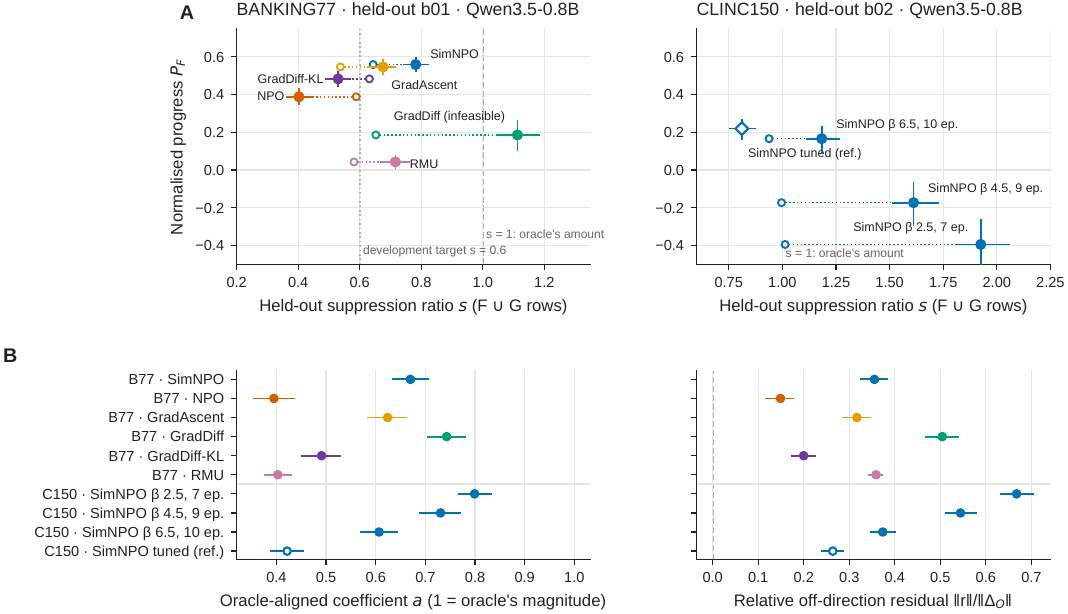}
\caption{Suppression and distributional fidelity. (A) Held-out suppression $s$ against $P_\mathcal{F}$, with open markers locating development suppression for the same configuration and dotted horizontal guides showing drift. The vertical coordinate of each open marker remains the held-out progress; it is not a development-performance measurement. Intervals are conditional row-bootstrap 95\% intervals. (B) Pooled oracle-aligned coefficient and relative residual from per-row projections. Matching is approximate; infeasible configurations remain visible. The projection identity is exact for squared L2, not JS.}
\label{fig:geometry}
  \end{figure*}

Configurations selected to approximately match suppression on development data differ on held-out data (Figure~\ref{fig:geometry}). On BANKING77, SimNPO, GradAscent, and RMU yield $s=$0.78, 0.68, and 0.72, respectively, but $P_\mathcal{F}=$0.56, 0.55, and 0.04. Therefore, similar removal of forgotten-option mass does not ensure similar oracle fidelity. This comparison is approximate, since development matching does not enforce identical held-out suppression.

To describe distributional movement, define per-row changes $\Delta_U=\mathbf p_{M_\mathcal{U}}-\mathbf p_{M_\mathrm{full}}$ and $\Delta_O=\mathbf p_{M_\mathcal{O}}-\mathbf p_{M_\mathrm{full}}$. Their projection gives
\begin{equation}
 \begin{aligned}
a&=\frac{\langle\Delta_U,\Delta_O\rangle}{\|\Delta_O\|_2^2},
 \\ r&=\Delta_U-a\Delta_O,\\
 \|\Delta_U-\Delta_O\|_2^2&=(a-1)^2\|\Delta_O\|_2^2+\|r\|_2^2.
 \end{aligned}
 \label{eq:geometry}
\end{equation}
The identity describes squared Euclidean error, not a decomposition of JS divergence. RMU's pooled aligned coefficient is 0.40, compared with 0.67 for matched SimNPO. These diagnostics support a distinction between suppression and movement toward the oracle, and they do not identify a causal mediator.

On CLINC150, three settings near $s=1$ on development data yield held-out ratios of 1.93, 1.61, and 1.18, with corresponding $P_\mathcal{F}=$-0.39, -0.17, and 0.17. Oversuppression and declining fidelity co-occur in this comparison. Increasing update strength is consequently not a general correction for weak progress.

\subsection{Recovery depends on the starting checkpoint and the metric}
\label{sec:recovery}
Recovery uses BANKING77 bundle 01, training seed 17, with $k\in\{1,5,20\}$ forgotten training examples per class and two adaptation seeds. We adapt the decision head and LoRA parameters \citep{hu2021lora} under matched settings and evaluate at $t\in\{0,1,5,10,25\}$ updates on 160 unseen rows from four forgotten classes. Let $\bar p_M(t)$ be the mean gold probability. Normalized trapezoidal area is $A_M=\sum_t w_t\bar p_M(t)$, with weights $(0.02,0.10,0.18,0.40,0.30)$. Define
\begin{equation}
\begin{aligned}
 G_{\rm abs}&=A_{M_\mathcal{U}}-A_{M_\mathcal{O}},\\
 G_{\rm adj}&=[A_{M_\mathcal{U}}-\bar p_{M_\mathcal{U}}(0)]
              -[A_{M_\mathcal{O}}-\bar p_{M_\mathcal{O}}(0)],\\
 H_M&=\frac{A_M-\bar p_M(0)}{1-\bar p_M(0)},\qquad
 G_H=H_{M_\mathcal{U}}-H_{M_\mathcal{O}}.
\end{aligned}
\label{eq:recovery}
\end{equation}
Hence $G_{\rm abs}=\Delta\bar p(0)+G_{\rm adj}$, where $\Delta\bar p(0)=\bar p_{M_\mathcal{U}}(0)-\bar p_{M_\mathcal{O}}(0)$. Headroom normalization is a secondary sensitivity analysis and does not identify an intrinsic learning rate.

\begin{figure*}[!tbh]\centering
\includegraphics[width=0.98\linewidth,height=0.76\textheight,keepaspectratio]{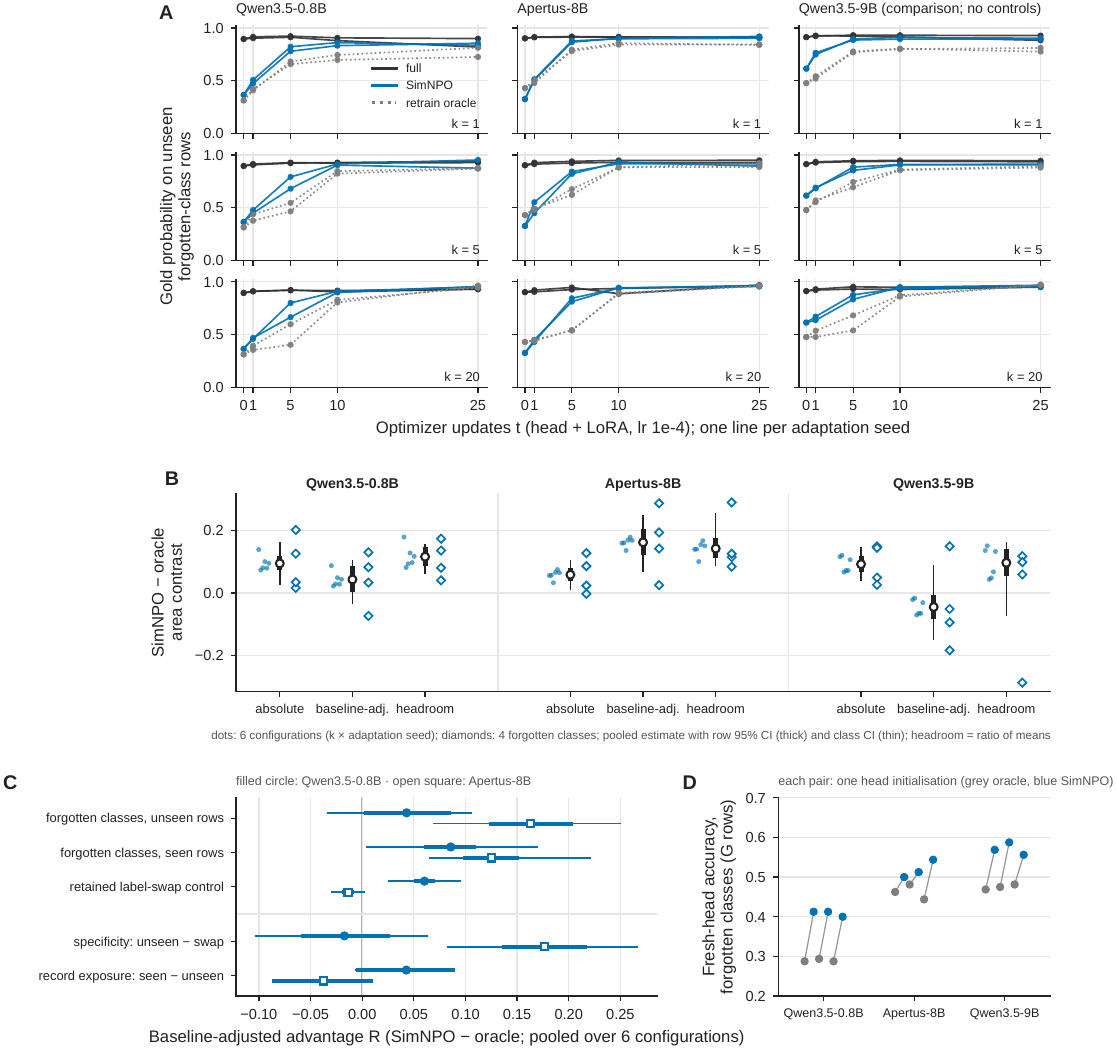}
\caption{Recovery relative to retraining. (A) Gold probability by update, exposure budget, checkpoint, and adaptation seed. (B) Absolute, baseline-adjusted, and headroom-normalized area contrasts; configuration dots and forgotten-class diamonds are descriptive, with pooled row and class intervals. (C) Adjusted recovery on unseen rows, exposure-excluded seen rows, and the retained label-swap control, plus direct contrasts. (D) Three fresh-head initializations per frozen backbone. Repeated evaluations share the same trained checkpoints, so the intervals are conditional on those checkpoints. Qwen3.5-9B has no control arm.}
\label{fig:recovery}
\end{figure*}

Apertus-8B starts below its oracle (0.326 versus 0.429) but has positive baseline-adjusted recovery in every configuration, with pooled $G_{\rm adj}\approx$0.16 [0.12,0.20]. The advantage also survives headroom normalization. Its adjusted recovery exceeds a retained label-swap control by approximately 0.18 [0.14,0.22], with a class-bootstrap interval of [0.08,0.27] (Figure~\ref{fig:recovery}). This specificity is relative to a control that overrides an existing label mapping and starts near zero gold probability.

For Qwen3.5-0.8B, pooled adjusted recovery is approximately 0.04 [0.00,0.09] and its contrast with the label-swap control is -0.02 [-0.06,0.03]. The controls therefore do not establish forgotten-target specificity for this model. Neither control model shows a clear seen-row advantage over unseen rows; this is not an equivalence result.

Qwen3.5-9B starts above its oracle (0.614 versus 0.477). Its positive absolute area contrast coexists with a negative adjusted contrast of approximately -0.05 and a positive headroom-normalized contrast of approximately 0.10. Its apparent recovery advantage therefore depends on the endpoint. No specificity controls were run for this model.

Finally, new heads trained on retained classes atop frozen backbones give higher forgotten-class accuracy point estimates for unlearned classes than oracle backbones. Initialization-level variation and broad class intervals limit this evidence. Because semantic labels remain available, the experiment measures transfer through the frozen representations under this interface, and it does not demonstrate extractable deleted records.

\section{Discussion and Limitations}
\label{sec:discussion}
The results support evaluating unlearning as an approximation to a specified counterfactual, including the decision schema. The oracle may retain substantial recognition through semantic labels; changing the offered alternatives can alter that recognition without changing model weights. Conversely, opaque labels can eliminate oracle recognition while leaving the unlearned model poorly aligned under a reduced option set. Therefore, a fixed-schema score does not establish that unlearning is robust to schema.

Three distinctions prevent stronger conclusions than the evidence supports. Statistical acceptance does not guarantee substantial progress, as illustrated by opaque CLINC150 bundle 06. A broad single-seed screen does not establish reliable transfer across training seeds or deletion targets. Higher recovery after adaptation does not, by itself, indicate faster learning, particularly when initial gold probabilities differ. The probability-geometry analysis is descriptive, and imperfect suppression matching does not isolate a unique mechanism.

The study is limited to two intent-classification datasets and supervised task adaptation from pretrained backbones. It does not establish erasure from pretraining, behavior in open-ended generation, or guarantees against privacy attacks. Granularity comparisons change the membership and organization of deleted rows and do not isolate granularity from every data-composition effect. A single label-code map per dataset cannot separate all consequences of changed label representations and training history. The architecture panel is purposively selected, recovery uses one trained seed per model, and class-bootstrap sensitivity is coarse with four forgotten classes. The retained label-swap control differs from forgotten-target adaptation in its initial difficulty and objective. Finally, checkpoint-level summaries do not recover the distribution over all possible retrained models.

\section{Conclusion}
We introduced \bench{} to examine how deletion changes probabilistic decision models and how well approximate unlearning reproduces those changes. Deletion granularity, label meaning, and candidate composition all affect the retraining target. SimNPO provides reproducible reference-model gains on BANKING77, but those gains do not transfer uniformly across seeds, targets, or datasets. Suppression and recovery diagnostics further show why low forgotten-class accuracy and high post-adaptation performance are insufficient summaries. Reporting the schema-specific oracle, absolute and normalized fidelity, retained utility, and controlled recovery yields a more precise account of what decision-model unlearning achieves.

% Authors are kindly asked to make their submissions as accessible as possible for everyone including people with disabilities and sensory or neurological differences.
% Tips of how to achieve this and what to pay attention to will be provided on the conference website \url{http://icml.cc/}.

% \section*{Software and Data}

% If a paper is accepted, we strongly encourage the publication of software and data with the
% camera-ready version of the paper whenever appropriate. This can be
% done by including a URL in the camera-ready copy. However, \textbf{do not}
% include URLs that reveal your institution or identity in your
% submission for review. Instead, provide an anonymous URL or upload
% the material as ``Supplementary Material'' into the OpenReview reviewing
% system. Note that reviewers are not required to look at this material
% when writing their review.

% % Acknowledgements should only appear in the accepted version.
% \section*{Acknowledgements}

% \textbf{Do not} include acknowledgements in the initial version of
% the paper submitted for blind review.

% If a paper is accepted, the final camera-ready version can (and
% usually should) include acknowledgements.  Such acknowledgements
% should be placed at the end of the section, in an unnumbered section
% that does not count towards the paper page limit. Typically, this will 
% include thanks to reviewers who gave useful comments, to colleagues 
% who contributed to the ideas, and to funding agencies and corporate 
% sponsors that provided financial support.

\section*{Impact Statement}

This work studies how to evaluate the removal of supervised training influence from probabilistic decision models. More accurate evaluation can help avoid unsupported claims that a model has forgotten private information. The experiments use intent-classification benchmarks and do not establish a privacy guarantee, successful record extraction, or the removal of knowledge acquired during pretraining. Recovery analyses are intended to characterize limitations of behavioral evaluation. Any release should respect the licenses of the datasets and base models and distinguish empirical fidelity results from guarantees about individuals' data.

\newpage
\bibliography{reference}
\bibliographystyle{icml2025}

%%%%%%%%%%%%%%%%%%%%%%%%%%%%%%%%%%%%%%%%%%%%%%%%%%%%%%%%%%%%%%%%%%%%%%%%%%%%%%%
%%%%%%%%%%%%%%%%%%%%%%%%%%%%%%%%%%%%%%%%%%%%%%%%%%%%%%%%%%%%%%%%%%%%%%%%%%%%%%%
% APPENDIX
%%%%%%%%%%%%%%%%%%%%%%%%%%%%%%%%%%%%%%%%%%%%%%%%%%%%%%%%%%%%%%%%%%%%%%%%%%%%%%%
%%%%%%%%%%%%%%%%%%%%%%%%%%%%%%%%%%%%%%%%%%%%%%%%%%%%%%%%%%%%%%%%%%%%%%%%%%%%%%%
\newpage
\appendix
\onecolumn
% Appendix fragment: follows the main text and bibliography.

% IMPORTANT: Check all \authorcheck before submission

\appendix
\setcounter{figure}{0}
\setcounter{table}{0}
\renewcommand{\thefigure}{S\arabic{figure}}
\renewcommand{\thetable}{S\arabic{table}}

\section{Notation and Scope of the Retraining Reference}
\label{app:notation}
Throughout the manuscript, $\mathcal{D}_\mathcal{F}$ is deleted training data, $\mathcal{D}_\mathrm{R}$ is retained training data, and $\mathcal{E}_\mathcal{F},\mathcal{E}_\mathcal{G},\mathcal{E}_\mathrm{R}$ are evaluation views. A deletion \emph{bundle} is a fixed collection of target units, not a collection of independently trained models. Training seeds index the construction of full and retain-only checkpoints; adaptation seeds index subsequent recovery runs; head seeds index fresh-head initialization. These sources of variation are not interchangeable.

The full model $M_\mathrm{full}$ and oracle $M_\mathcal{O}$ are both derived from the same pretrained initialization under the task-training procedure. The oracle is a reference for deleting supervised adaptation data. It may retain relevant pretrained information and information inferred from the retained task data or the option descriptions. Its residual recognition is therefore compatible with the specified retraining counterfactual. It is not, by itself, evidence that the oracle was trained on $\mathcal{D}_\mathcal{F}$.

The schema $\mathcal{S}=(\mathcal{C},\ell)$ includes both candidate membership and the textual information supplied about candidates. We suppress $\mathcal{S}$ in formulas when all models being compared use the same schema. An oracle trained with opaque labels is a different checkpoint from the semantic-label oracle. By contrast, changing the candidate set at evaluation reuses checkpoints and supplies them with a different schema.

The supplied plots use $D_0$, $P_\mathcal{F}$, $P_\mathcal{G}$, $s$, $a$, and $G_{\rm adj}$; these symbols have the same meanings here. Source-table strings \texttt{dOD\_F} and \texttt{dOD\_G} are rendered as $\Delta\operatorname{OD}_\mathcal{F}$ and $\Delta\operatorname{OD}_\mathcal{G}$. A change in retained accuracy is reported in percentage points when explicitly marked \emph{pp}; probability and ECE differences are on the $[0,1]$ scale.

\section{Data, Target Construction, and Experimental Coverage}
\label{app:design}
\subsection{Datasets and deletion units}
BANKING77 contains 77 banking intents \citep{casanueva2020efficient}. CLINC150 contains 150 in-scope intents, with an additional out-of-scope option in our interface \citep{larson2019evaluation}. The benchmark uses the existing intent annotations. No inference about a person's identity or private attribute is an evaluation target in this study.

Example deletion removes 5\% of training rows. Association deletion removes sub-intent clusters from selected parent intents, leaving other training examples of those intents. The primary BANKING77 association bundle contains ten clusters, whereas the supplied replication tables contain six and eight clusters in bundles 02 and 03, respectively. Class deletion removes all training rows associated with the selected class. BANKING77 class bundles contain four classes, whereas CLINC150 class bundles contain eight intents.

Target eligibility is evaluated on full-model development predictions. Class targets require accuracy of at least 0.70 and gold probability of at least 0.50, whereas association targets have a 0.65 accuracy threshold. Selection proceeds in hash order among eligible targets at the primary seed. Targets are not reselected to make subsequent seeds favorable. This produces one documented exception, where the semantic Qwen3.5-0.8B at seed 42 has a development accuracy of 0.65 on \texttt{transfer\_timing} in the BANKING77 bundle 02. The corresponding opaque configuration passes its acquisition check.

\subsection{Evaluation populations}
For the primary BANKING77 class bundle, $\mathcal{E}_\mathcal{F}$ contains 428 deleted training rows and $\mathcal{E}_\mathcal{G}$ contains 160 unseen rows. The reference model's second bundle has 382 deleted training rows. Row counts can differ across model-specific bundle-02 records. Therefore, a shared bundle label is not sufficient evidence that every architecture used identical deleted rows. The supplied results preserve the observed counts.

For class deletion, $\mathcal{E}_\mathcal{G}$ comprises unseen rows of the deleted classes. Association-level generalization uses unseen rows of the deleted clusters. At the example level, evaluation rows are associated with deleted training rows through a nearest-training-row relation. Retained utility is computed on retained evaluation rows from the same configuration.

% \authorcheck{The archive does not include dataset split manifests, preprocessing code, the embedding/clustering method used for association targets and nearest-row assignment, or the candidate-neighbor selection rule. Add their exact definitions, seeds, split sizes, and manifest hashes from the original repository. Resolve whether model-specific bundle-02 differences reflect target selection or row filtering. Do not infer these details from shortened bundle labels.}

\subsection{Observed coverage and stages}
The 20-model granularity panel contains 120 full/oracle evaluations across three deletion requests. These are evaluation records, not necessarily 120 separately trained checkpoints, because full models may be reused. The 21st configured model, InternLM3-8B, was excluded due to a checkpoint-reload reproducibility issue.

The class-method confirmation uses Qwen3.5-0.8B, seed 17, bundle 01. Four methods continue to nine seed--bundle cells, with seeds 17, 42, and 99 and class bundles 01--03. The broad frozen-setting panel contains the reference model plus 19 additional model configurations. Four additional models join the reference in the targeted seed/bundle comparison: Apertus-8B, OLMo-3-7B, Qwen3-4B, and Qwen3.5-9B.

The label study contains seven semantic/opaque pairs: BANKING77 bundles 01 and 02 at seeds 17 and 42, and CLINC150 bundles 02, 03, and 06 at seed 17. The opaque bundle-06 result appears in the detailed tables despite a stale not-run entry in the supplied design table. CLINC150 bundle 08 remains reserved. The candidate-set study additionally includes opaque BANKING77 bundles 01 and 02.

The run registry contains 358 displayed records. Its footer reports 1,043 evaluation records in the wider collection, including 646 development-sweep candidates, eight development-reference records, and 31 synthetic fixtures. The omitted records are not independent experimental replications. Table~\ref{tab:S01_run_registry} summarizes the registry, and its CSV companion preserves every displayed row, including shortened keys that are not globally unique. The complete source paths and granularity fields must accompany any public machine-readable release.

\section{Model Training, Methods, and Configuration Reporting}
\label{app:implementation}
The study compares probabilistic decision checkpoints through their complete option distributions. Approximate unlearning begins from the full checkpoint, whereas the oracle is constructed by applying the task-training procedure to the retained data from the shared pretrained initialization. Recovery explicitly updates the decision head and LoRA parameters; fresh-head analysis freezes the corresponding backbone and replaces the head.

GradAscent reverses the supervised objective on deleted examples. GradDiff combines forgetting pressure with retained-data supervision, and GradDiff-KL uses a retained-distribution regularizer. NPO and SimNPO use preference-inspired forgetting objectives \citep{zhang2024npo,fan2025simnpo}. RMU intervenes on representations \citep{li2024wmdp}. Uniformization is an erasure control. These descriptions identify the evaluated method families, and their exact decision-probability losses and implementation-specific normalization must match the released code.

The CLINC150 tuning figure identifies the selected configuration as learning rate $3\times10^{-5}$, $\beta=10$, retained-objective weight $\alpha=0.5$, and 10 epochs. Its development objective is approximately 0.641, versus 0.945 for no-op. The selected point lies on the tested grid boundary. The result is reported as the selected tested setting, without asserting that it is an optimum beyond the grid. Dataset tuning used the development bundle, and the subsequent semantic-label evaluations use frozen settings.

Recovery uses learning rate $10^{-4}$, three exposure budgets, and two adaptation seeds. Conditional seed-42 strength diagnostics use learning rate $6\times10^{-5}$ on Apertus-8B and Qwen3.5-9B. The exact frozen BANKING77 selection and all full/oracle/head-training settings are referenced by a configuration CSV in the original asset notes, but that CSV is absent from this archive.

% \authorcheck{Before submission, supply the exact decision-head architecture and option-scoring rule, loss equations as implemented, model checkpoint identifiers/revisions, trainable parameter sets, LoRA ranks/scaling/target modules, precision or quantization, optimizer and schedule, batch size and gradient accumulation, training duration, checkpoint selection, prompt/schema templates, and retained-data sampling. Add the frozen BANKING77 configurations and fresh-head training recipe. These cannot be reconstructed reliably from outcome tables.}

\section{Metrics and Statistical Analysis}
\label{app:statistics}
\subsection{Distributional metrics}
For distributions $\mathbf p,\mathbf q$ and $\mathbf m=(\mathbf p+\mathbf q)/2$,
\begin{equation}
 \operatorname{JS}_2(\mathbf p,\mathbf q)
 =\tfrac12\operatorname{KL}_2(\mathbf p\|\mathbf m)
 +\tfrac12\operatorname{KL}_2(\mathbf q\|\mathbf m).
\end{equation}
Distances lie between zero and one. Numerical stabilization uses $\epsilon=10^{-12}$. Oracle distance is an average of row divergences. Normalized progress is a ratio of aggregate distances, not an average of row-wise progress ratios. Each schema and label condition uses its own full/oracle denominator.

Retained accuracy uses the model's highest-probability option. ECE compares confidence with empirical correctness within bins. The standard form is
\begin{equation}
 \operatorname{ECE}_\mathrm{R}=\sum_{j=1}^J\frac{|B_j|}{|\mathcal{E}_\mathrm{R}|}
 \left|\operatorname{Acc}(B_j)-\operatorname{conf}(B_j)\right|.
\end{equation}
The specific binning procedure is part of the implementation and must be documented; no new calibration transformation is applied to the JS probabilities for this draft.

\subsection{Within-configuration inference}
Paired resampling keeps full, unlearned, and oracle predictions aligned for the same evaluation rows. The default analysis uses 5000 bootstrap replicates and percentile 95\% intervals. Class- or cluster-level resampling assesses sensitivity to the unit of deletion. With only four forgotten classes in BANKING77, these sensitivity intervals have limited resolution and do not characterize arbitrary new class selections.

Both-view acceptance is an intersection criterion: both $\Delta\operatorname{OD}_\mathcal{F}$ and $\Delta\operatorname{OD}_\mathcal{G}$ must have positive point estimates and positive lower bounds, and the retained-utility constraints must hold. Positive, unconfirmed denotes positive point gains with at least one nonpositive lower bound. No benefit and infeasible are retained as distinct reported outcomes. Statistical acceptance is not an equivalence test and does not require $P_v$ to exceed a practical-effect threshold.

The frozen confirmation family reports a one-sided intersection-union $p$-value formed from the larger of the two view-specific values, followed by Holm adjustment across the candidate family. Screening decisions are uncorrected across model configurations. The reported $p$-values and intervals retain their original roles, and we do not recompute them from rounded table entries.

\subsection{Cross-bundle summaries}
Reference-model Phase 2 uses stored hierarchical resampling over seeds, bundles, and deletion units. A method replicates under the protocol only if the aggregate lower bound is positive on both views, feasibility holds in every observed cell, and the result survives the specified leave-one-seed/bundle-out checks. Positive aggregate distance gains can therefore coexist with a failed replication decision.

For CLINC150, the pooled analysis averages seeds within each intent and considers 48 intents from six bundles. The intent-level estimate is not formed by selectively pooling successful bundles. On $\mathcal{F}$, the reported intent-bootstrap interval is [-0.045,0.090], and the bundle-then-intent interval is [-0.054,0.107] around the same point gain of 0.022. Both retain substantial uncertainty. Mean per-intent progress is omitted because some displacement denominators are small.

% \authorcheck{Add the ECE binning specification, exact one-sided centered-bootstrap algorithm and finite-sample correction, the frozen Holm family membership, and the Phase 2 resampling/leave-one-out implementation from the protocol. The supplied table names these procedures but does not provide enough code to independently reproduce every inferential detail.}

\section{Complete Method and Transfer Results}
\label{app:methods}
Table~\ref{tab:S03_complete_method_results} retains all 150 displayed intervention/control rows, including pilots and infeasible settings. The pilot configurations use default settings and must not replace selected confirmation configurations in the main comparison. In particular, example-deletion pilot results exist even though no example-deletion method advanced through selection to confirmation. The supplied main-table note saying that no method was run is too broad.

The final association-level sensitivity results also require separate treatment. SimNPO has modest positive cross-bundle gains and satisfies the stored replication rule. GradDiff does not establish a positive $\mathcal{G}$ aggregate. Uniformization is reported as a control without a replication decision. Association bundles have different numbers of clusters, and individual seed--bundle results remain visible.

The architecture results in Table~\ref{tab:S04_architecture_replication} include both views, class sensitivity intervals, and observed seed/bundle differences. The reference configuration is included in the total of 20. Thus, 19 accepted configurations correspond to the reference plus 18 of 19 additional configurations. This is a descriptive count for the selected panel, not a probability that an arbitrary model will satisfy the criterion.

Table~\ref{tab:S07_clinc_all_bundles} preserves the CLINC150 comparisons and controls. Its source omits the deletion-granularity field, leaving two visibly identical Qwen3.5-0.8B bundle-01 SimNPO keys with different denominators (0.058 and 0.416). The complete results table distinguishes their unit counts (18 versus 8) and row populations. We retain both records with explicit row identifiers and do not merge them. The main class-deletion transfer result is the eight-intent record with $D_{0,\mathcal{F}}=0.416$.

\section{Label and Candidate-Schema Interventions}
\label{app:schema}
Opaque rendering replaces every option name and description with a fixed code. The original map contains 77 codes for BANKING77 and 151 for CLINC150, including out-of-scope. Input utterances, target identities, option identifiers, and option order are unchanged. Forgotten options remain present, and their logits are not masked. The intervention is therefore a change in the information supplied by labels, together with the corresponding full/oracle training under that rendering.

Table~\ref{tab:S05_semantic_opaque_results} reports acquisition, oracle accuracy, forgotten-option mass, displacement, and progress. Rounded zero accuracy does not imply exactly zero option probability: scientific notation is retained for the mass values. The semantic/opaque effect is strong for oracle recognition, but SimNPO's normalized progress varies with bundle and dataset. A single fixed code map per dataset leaves code-assignment variability untested.

Candidate schemas are fixed within a bundle. $S_0$ supplies all options. $S_1$ supplies the forgotten classes and two retained semantic neighbors per forgotten class, and $S_2$ supplies the forgotten classes with the same number of random retained distractors. Observed subset cardinalities are 12 for BANKING77 and 24 for CLINC150. The exact membership lists and selection code should accompany the release.

All models are rescored under the reduced input schema. This is not slicing and renormalizing a stored full-schema vector. Table~\ref{tab:S06_candidate_sets} reports both views and the new denominator. For example, semantic CLINC150 bundle 02 has $D_{0,\mathcal{F}}=$0.039 with random distractors, compared with 0.298 under the full schema. The smaller denominator changes the interpretation and uncertainty of normalized progress.

\section{Suppression and Probability Geometry}
\label{app:geometry}
The suppression ratio aggregates forgotten-option mass over the relevant rows before dividing. A value $s=1$ matches the oracle's net amount of removed mass; it does not specify which retained options receive that mass or whether individual forgotten classes are suppressed similarly. A value above one indicates greater net mass removal than the oracle.

Development matching uses target $s=0.6$ on BANKING77 and $s=1$ on CLINC150. Table~\ref{tab:S08_suppression_and_geometry} preserves the observed development and held-out values, including configurations whose utility gate fails on held-out data. Uniformization has no feasible development checkpoint near the matched target and is absent from that comparison. The matched SimNPO configuration is distinct from the main confirmation checkpoint, so its $P_\mathcal{F}=$0.56 should not be substituted for the main confirmation value 0.59.

For row $i$ with $\Delta_{O,i}\neq0$, orthogonality gives
\begin{equation}
 \langle r_i,\Delta_{O,i}\rangle=0,\quad
 \|\mathbf p_{U,i}-\mathbf p_{O,i}\|_2^2
 =(a_i-1)^2\|\Delta_{O,i}\|_2^2+\|r_i\|_2^2.
\end{equation}
Summing these per-row identities is exact. The pooled coefficient displayed in the figures is
\begin{equation}
 \bar a=
 \frac{\sum_i\langle\Delta_{U,i},\Delta_{O,i}\rangle}
      {\sum_i\|\Delta_{O,i}\|_2^2},
 \qquad
 \rho_r=\sqrt{\frac{\sum_i\|r_i\|_2^2}
                       {\sum_i\|\Delta_{O,i}\|_2^2}}.
\end{equation}
The pooled $\bar a$ must not replace every $a_i$ in the exact summed decomposition. The matched-suppression table reports identity errors at or below $2.0\times10^{-15}$ and no rows with zero oracle displacement. The wider diagnostic figure includes additional runs with a broader numerical error range.

The magnitude-only and residual-only JS diagnostics use probability vectors that are projected back onto the simplex. This projection and the nonlinear JS metric prevent interpreting them as an additive JS decomposition. Correlations between these diagnostics and progress also share underlying probability vectors. They are descriptive associations, not independent causal evidence about the optimization mechanism.

\section{Recovery, Controls, and Fresh Heads}
\label{app:recovery}
\subsection{Adaptation and numerical summaries}
Full, SimNPO, and retrained checkpoints are adapted using the same exposure examples, update schedule, and evaluation rows within each recovery configuration. Exposure budgets are $k=$1,5,20 rows per forgotten class, and adaptation seeds are 0 and 1. Evaluation occurs at updates 0,1,5,10,25. Trapezoidal integration divided by the horizon of 25 yields weights 0.02,0.10,0.18,0.40,0.30. Consequently, each area has the same probability scale as a time-averaged gold probability.

Table~\ref{tab:S09_recovery_summary} provides all 18 summaries across model, budget, and adaptation seed. Its configuration-level identity residuals are at most approximately $10^{-16}$. The six settings for each model share the same underlying trained checkpoints and four forgotten classes. They are not six independent replications of model training.

Shared row resampling preserves dependence across checkpoints, time points, budgets, and adaptation seeds wherever the same evaluation row is reused. Class intervals condition on the four tested forgotten classes and the reused backbones. Threshold times are the first evaluated update whose mean gold probability reaches the threshold; they are grid-based observations, not interpolated continuous crossing times. An unreached threshold is right-censored at 25 updates.

\subsection{Starting point and headroom}
Absolute area contains both the initial probability difference and the integrated difference in subsequent gains. Subtracting the initial value removes that additive offset. Headroom normalization instead asks what fraction of $1-\bar p(0)$ is recovered. The principal headroom analysis uses a ratio of aggregate means; averaging row-level ratios yields a different estimand and can place large weight on rows starting near one. No equivalence between these aggregations is assumed.

Apertus-8B's adjusted gains are positive across all configurations. Qwen3.5-9B's adjusted gains are negative for every configuration, whereas its headroom point contrasts are positive. This metric dependence is part of the result rather than a basis for selecting one favorable summary. Neither adjustment alone identifies a model-intrinsic learning rate.

\subsection{Specificity and exposure controls}
Controls are available for Qwen3.5-0.8B and Apertus-8B. For an evaluation population $q$, let
\begin{equation}
 R_q=[A_{U,q}-\bar p_{U,q}(0)]-[A_{O,q}-\bar p_{O,q}(0)].
\end{equation}
The specificity contrast is $R_{\rm unseen}-R_{\rm swap}$; the exposure contrast is $R_{\rm seen}-R_{\rm unseen}$. These are direct differences of paired advantages, not comparisons of whether separate intervals include zero.

In the retained label-swap arm, a retained class $X$ is relabeled as $Y$ and adapted using the same budget and schedule. This arm starts with a near-zero target probability and reverses a previously correct mapping. Therefore, a positive specificity contrast establishes a difference relative to this particular control, not exclusion of all forms of general post-unlearning plasticity.

\begin{figure}[tbh]\centering   \includegraphics[width=\linewidth,height=0.79\textheight,keepaspectratio]{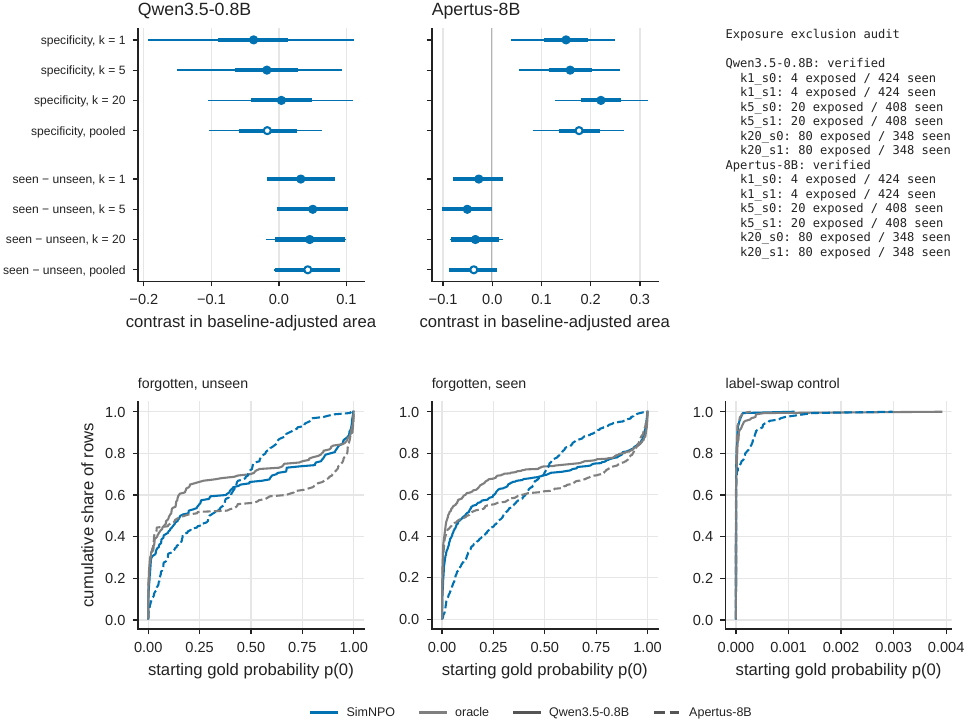}
  \caption{\textbf{Recovery controls and exposure exclusion.} For Qwen3.5-0.8B and Apertus-8B, forgotten-unseen recovery is contrasted directly with retained label-swap recovery, and seen forgotten rows are contrasted with unseen rows. Exposure rows are excluded from seen-row evaluation: the remaining counts are 424, 408, and 348 for $k=1,5,20$. Shared bootstrap draws preserve pairing across conditions. Apertus's adjusted advantage exceeds this control, while Qwen3.5-0.8B's does not clearly do so. The control begins near zero gold probability and changes a correct mapping; it is not matched on starting difficulty.}
  \label{fig:S12}
\end{figure}

Seen evaluation rows belong to the original deletion set but are excluded from the new exposure. Figure~\ref{fig:S12} reports 4 exposed and 424 remaining seen rows for $k=1$, 20 and 408 for $k=5$, and 80 and 348 for $k=20$, in both control models and adaptation seeds. The supplied audit reports verified disjointness. With no clear positive seen-minus-unseen advantage, these experiments do not establish a record-specific recovery effect; they also do not prove the absence of such an effect.

\subsection{Fresh-head transfer}
\begin{figure}[tbh]\centering   \includegraphics[width=\linewidth,height=0.79\textheight,keepaspectratio]{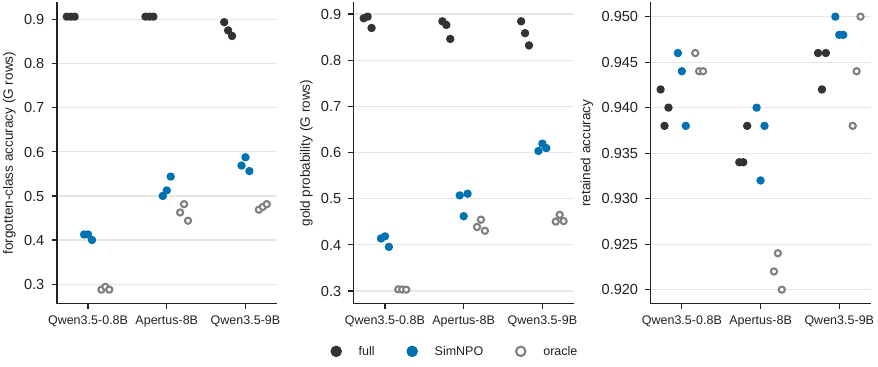}
  \caption{\textbf{Fresh heads on frozen representations.} Forgotten-class accuracy, gold probability, and retained performance are reported for three fresh-head initializations trained on retained-class examples only. Each initialization is shown separately. Across-initialization ranges are not confidence intervals. Reported row-pooled accuracy differences are borderline, and class intervals include zero. Recognition under semantic labels demonstrates transfer from frozen representations under the tested readout; it does not demonstrate extraction of deleted training records.}
  \label{fig:S13}
\end{figure}

The fresh-head study freezes each backbone and trains a newly initialized decision head on retained-class rows, with three matched head initializations. The full backbone is a reference ceiling, and the oracle backbone is the counterfactual comparison. Figure~\ref{fig:S13} reports forgotten-class accuracy, gold probability, and retained accuracy separately for every initialization.

Higher forgotten-class recognition with an unlearned backbone can indicate distinct transfer properties, while semantic option labels remain available. The experiment does not isolate storage of individual deleted records. The supplied caption states that pooled row intervals exclude zero, whereas class intervals do not; the exact pooled fresh-head and control tables are not included in the archive, so the manuscript does not invent additional numerical interval bounds.

\section{Optimization Diagnostics and Conditional Strength Reruns}
\label{app:optimization}
The seed comparison reveals several distinct patterns (Table~\ref{tab:S10_optimization_diagnostics}). For Apertus-8B, suppression on $\mathcal{F}$ falls from 1.07 to 0.39 between seeds 17 and 42; for Qwen3.5-9B it falls from 0.57 to 0.15. Their progress also falls. In OLMo-3-7B, suppression is much more similar (0.38 versus 0.35), while progress decreases from 0.34 to 0.04. For Qwen3-4B, suppression is 0.60 versus 0.61, but progress changes from 0.28 to -0.18. Therefore, insufficient suppression is not a complete explanation for all seed effects.

The recorded training logs provide forget loss, retain loss, and gradient norms. Initial gradient norms are not uniformly smaller for the weaker seed. Parameter-update norms were not recorded, and the figures do not reconstruct a historical trajectory that was never saved.

Conditional stronger-update reruns are exploratory. At learning rate $6\times10^{-5}$, the best displayed Apertus checkpoint reaches approximately $P_\mathcal{F}=0.42$, whereas the Qwen3.5-9B trace reaches only approximately 0.19 and later oversuppresses. The displayed best epochs were chosen after observing these traces; they do not replace the frozen-transfer estimates or establish a validated adjustment rule.

\section{Compute, Integrity, and Reproducibility}
\label{app:integrity}
Table~\ref{tab:S11_compute_and_configuration} reproduces logged GPU-hour summaries by evaluation category. These totals are partial: full-model evaluations record zero and do not include the training time stored with the full checkpoint. The table also does not establish complete recovery-job accounting. We therefore do not report the sum as the study's total training cost or infer an unlearning speedup over full retraining.

The supplied audit reports exact recomputation of oracle distance for 358 evaluations and point-estimate agreement for 252 bootstrap comparisons. It reports identical intervals when the number of bootstrap draws matches, and differences of up to 0.0075 for stored summaries of 2000 versus 5000 draws, without category changes. These are reported upstream checks; the current manuscript package contains figure/table exports, not the raw probabilities needed to repeat them.

The audit also documents repaired filename collisions between semantic and opaque analyses and between training seeds, a reporting-label collision between GradDiff and GradDiff-KL, and recovery reruns after out-of-memory failures. No repaired or post hoc results are treated as additional independent replications. The separate S2-Rank privacy experiment stopped at its acquisition gate and did not test the proposed attack; it provides no evidence of privacy robustness.

% \authorcheck{Complete the release with the frozen protocol, asset manifest and source commit, unique run identifiers including deletion granularity, full probability files or durable access instructions, the missing configuration CSV, exact recovery-control/fresh-head summaries, and complete compute/job accounting. The attached archive alone is insufficient for an independent numerical reproduction.}

\section{Supplementary Figures}
\label{app:figures}
The following figures retain the supplied vector assets. Their captions distinguish supported conclusions from the limits of interpretation. Production identifiers S1--S15 correspond to the original asset grouping.
\begin{figure}[tbh]\centering   \includegraphics[width=\linewidth,height=0.79\textheight,keepaspectratio]{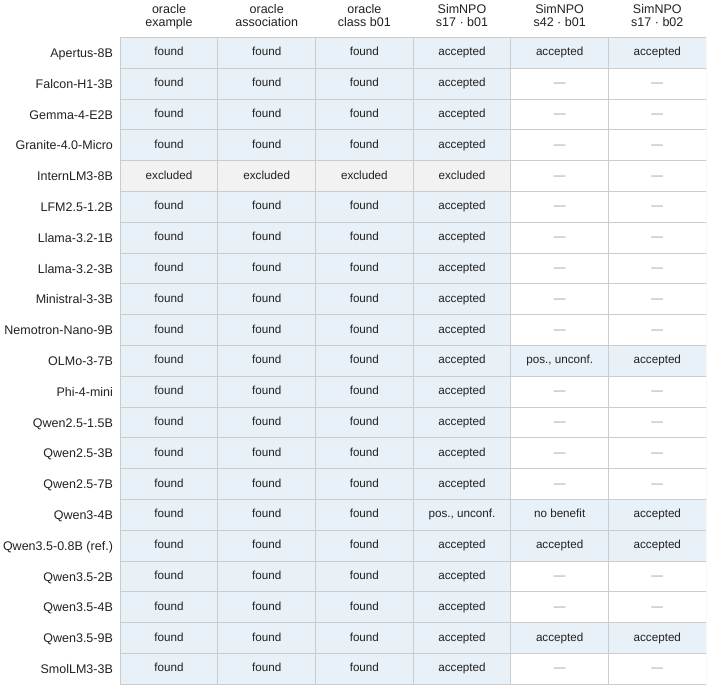}
  \caption{\textbf{Observed experimental coverage.} Cells indicate available evaluations by model and experiment family, including the reference-model replication grid, semantic/opaque comparisons, candidate-set rescoring, and recovery analyses. Coverage is purposive and incomplete across combinations; a filled cell does not imply independent replication of every factor. InternLM3-8B was excluded because checkpoint reload was not reproducible. The registry and machine-readable companion provide the available record keys.}
  \label{fig:S01}
\end{figure}

\begin{figure}[tbh]\centering   \includegraphics[width=\linewidth,height=0.79\textheight,keepaspectratio]{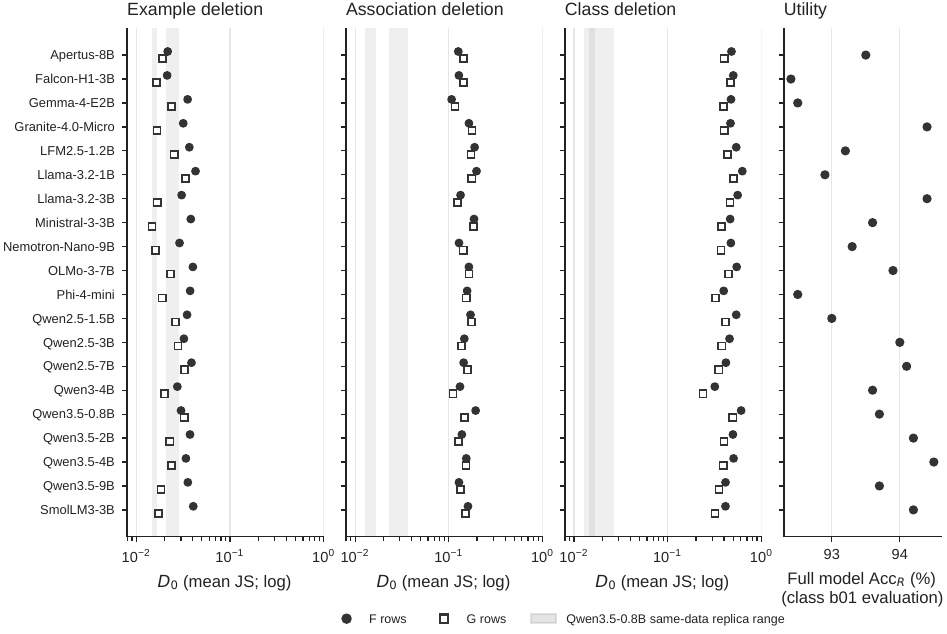}
  \caption{\textbf{Deletion granularity across eligible architectures.} Full-to-oracle displacement on deleted training rows ($\mathrm{F}$), unseen related rows ($\mathrm{G}$), and retained utility is shown for all 20 eligible architectures on BANKING77 at seed 17. Example, association, and class deletion differ in the scope of the target and the amount of task structure removed. The same-data noise reference is available only for Qwen3.5-0.8B. Displacement orders consistently by granularity; this is a statement about the tested decision distributions, not a privacy guarantee.}
  \label{fig:S02}
\end{figure}

\begin{figure}[tbh]\centering   \includegraphics[width=\linewidth,height=0.79\textheight,keepaspectratio]{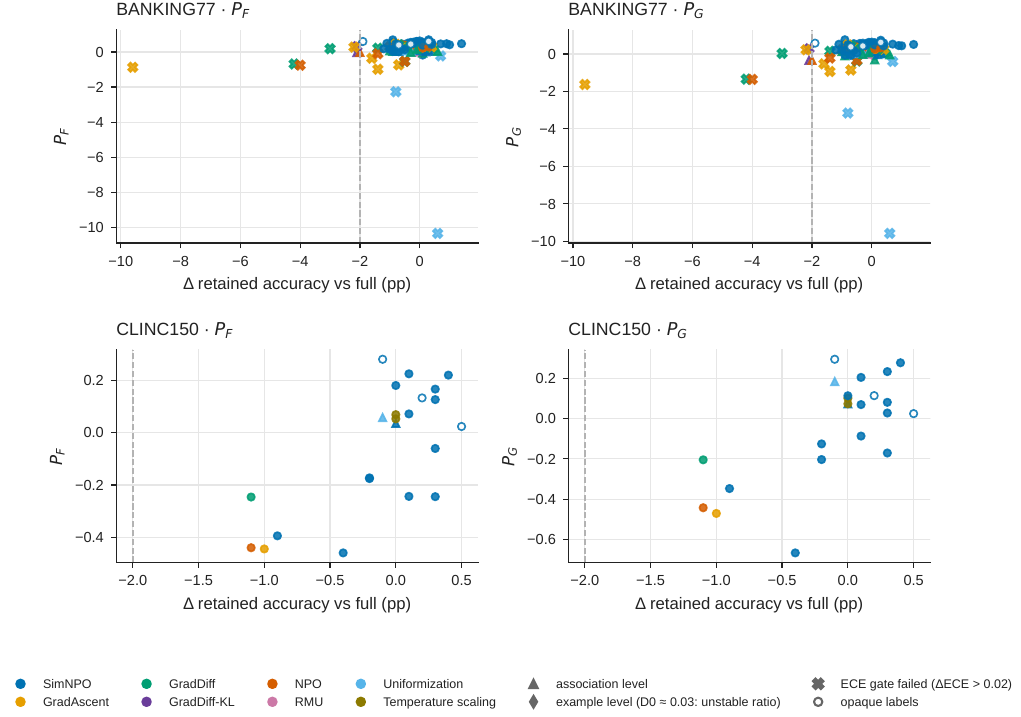}
  \caption{\textbf{Oracle progress and retained utility across evaluated configurations.} Each point is a recorded method configuration, with normalized progress plotted against retained-accuracy change. Calibration is an additional feasibility requirement even where only accuracy is plotted. Exploratory and screening points do not inherit confirmatory status. Extreme utility-loss runs are retained in the display; the complete table separates point estimates, uncertainty, and recorded acceptance category.}
  \label{fig:S03}
\end{figure}

\begin{figure}[tbh]\centering   \includegraphics[width=\linewidth,height=0.79\textheight,keepaspectratio]{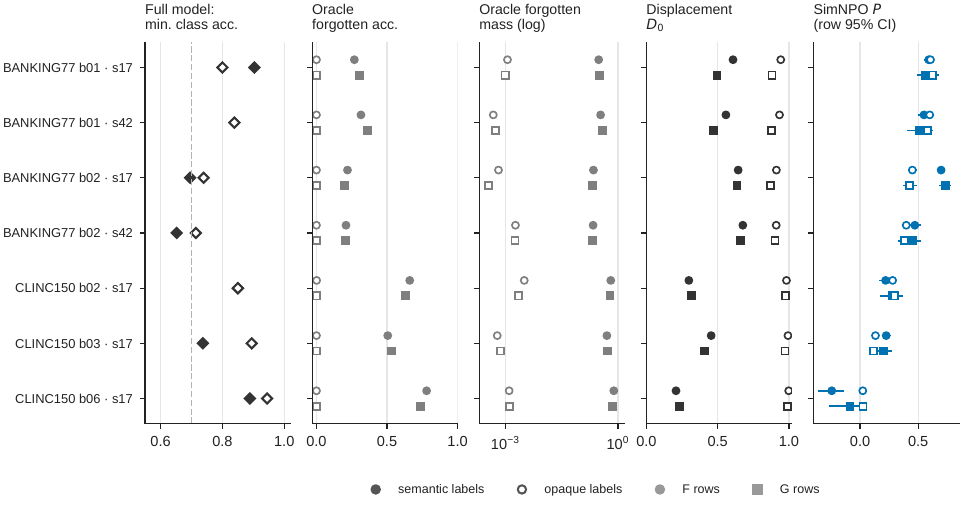}
  \caption{\textbf{Label semantics and the retraining target.} The seven semantic/opaque pairs report acquisition, forgotten-class recognition and probability mass, displacement, and normalized SimNPO progress. All opaque full models pass the acquisition check. Opaque oracles have approximately zero forgotten-class recognition despite successful full-model acquisition. The semantic seed-42 BANKING77 bundle-02 model contains one class below the acquisition threshold; targets were selected at the primary seed and were not reselected. The CLINC150 bundle-06 opaque result is positive but small ($P_\mathrm{F}\simeq0.02$).}
  \label{fig:S04}
\end{figure}

\begin{figure}[tbh]\centering   \includegraphics[width=\linewidth,height=0.79\textheight,keepaspectratio]{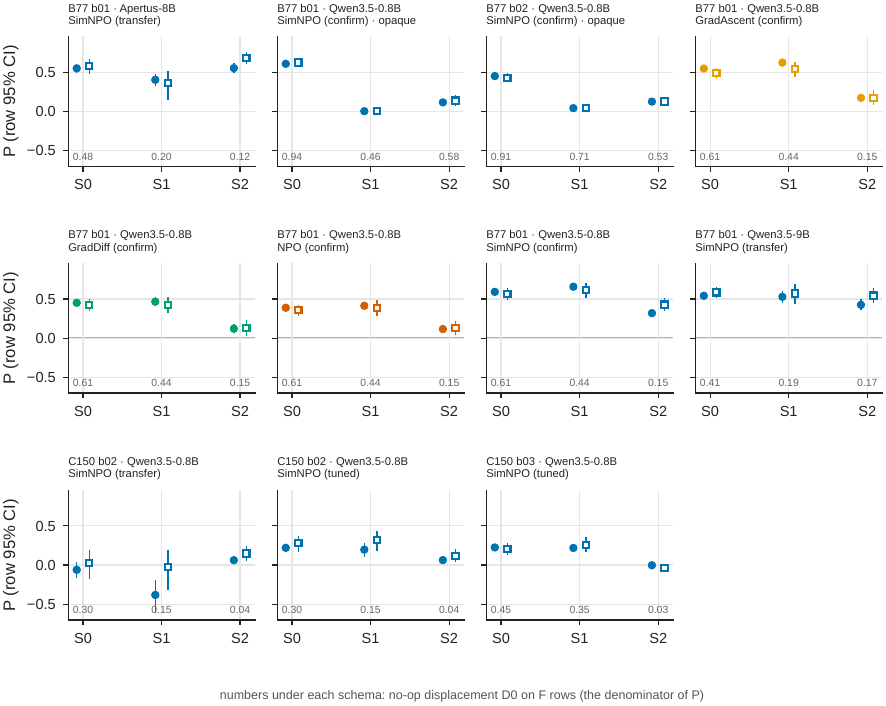}
  \caption{\textbf{All evaluated candidate schemas.} Each checkpoint is re-scored under the original option set ($S_0$), forgotten classes plus nearby retained alternatives ($S_1$), and an equally sized set with random retained alternatives ($S_2$). Oracle recognition, raw displacement, and method progress are shown for the available semantic and opaque conditions. Row-bootstrap intervals are conditional on the fitted checkpoints, and the ratio is recomputed in each replicate. Small displacement can make normalized progress sensitive; the underlying displacement must be read alongside the ratio.}
  \label{fig:S05}
\end{figure}

\begin{figure}[tbh]\centering   \includegraphics[width=\linewidth,height=0.79\textheight,keepaspectratio]{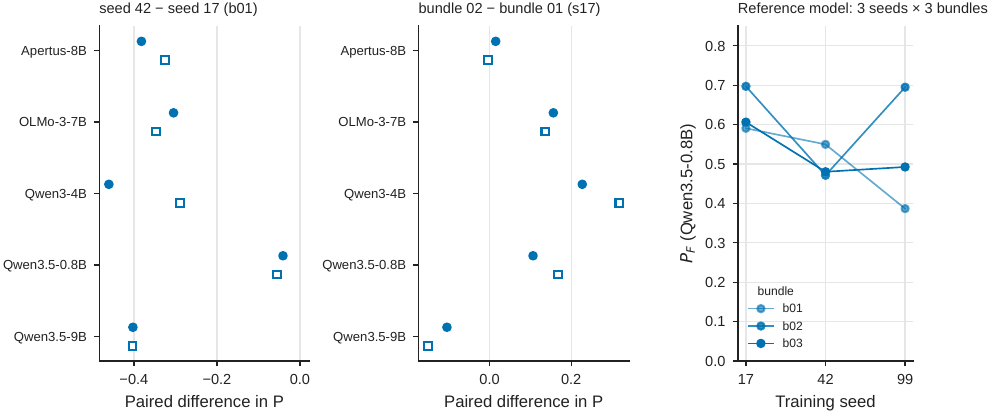}
  \caption{\textbf{Seed and target-bundle sensitivity.} The purposive architecture panel contrasts seed 42 with seed 17 on bundle 01 and bundle 02 with bundle 01 at seed 17. The reference model additionally has a three-seed, three-bundle grid. Guides connect paired comparisons, not a temporal trajectory. Seed labels are identifiers. These sparse contrasts demonstrate sensitivity within an architecture but do not identify a population variance decomposition.}
  \label{fig:S06}
\end{figure}

\begin{figure}[tbh]\centering   \includegraphics[width=\linewidth,height=0.79\textheight,keepaspectratio]{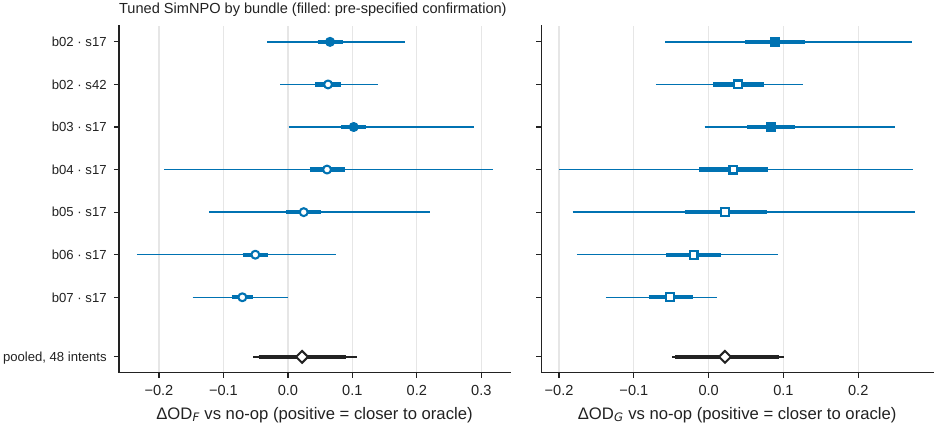}
  \caption{\textbf{CLINC150 transfer and heterogeneity across intent bundles.} Frozen BANKING77 settings and development-tuned SimNPO are distinguished. Tuned effects vary in sign across held-out bundles, and the intent-pooled estimate includes zero. The reported fraction of intents with a positive point estimate does not imply a positive pooled effect. Bundle 08 remains unused. Intervals and the unit of pooling are described in Appendix~\ref{app:statistics}.}
  \label{fig:S07}
\end{figure}

\begin{figure}[tbh]\centering   \includegraphics[width=\linewidth,height=0.79\textheight,keepaspectratio]{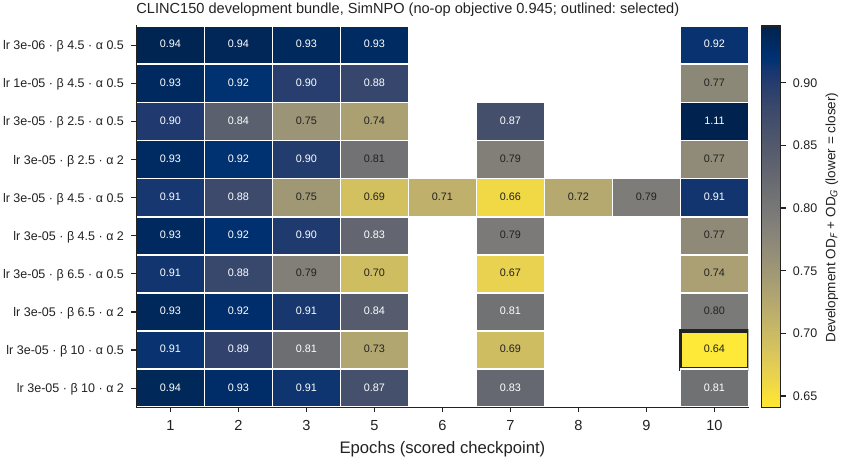}
  \caption{\textbf{CLINC150 development tuning.} The grid displays the development objective and diagnostics over evaluated settings. Selection uses development data, and the chosen setting is on the tested boundary ($\beta=10$, 10 epochs, learning rate $3\times10^{-5}$, retain weight $\alpha=0.5$). The grid was not extended. This landscape is a development diagnostic rather than held-out evidence; unavailable or infeasible cells should be interpreted using the source records.}
  \label{fig:S08}
\end{figure}

\begin{figure}[tbh]\centering   \includegraphics[width=\linewidth,height=0.79\textheight,keepaspectratio]{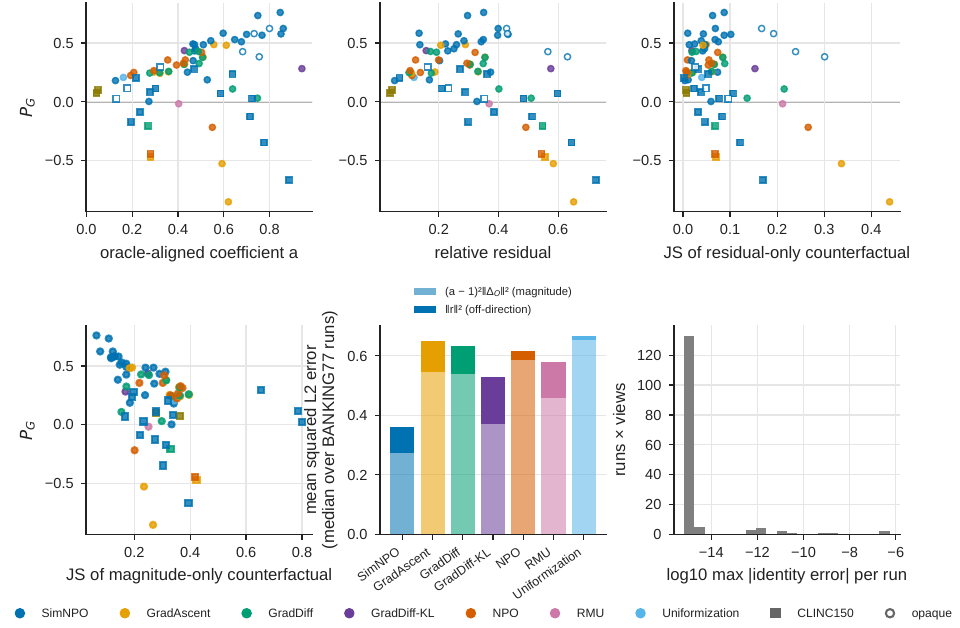}
  \caption{\textbf{Probability-redistribution diagnostics across recorded runs.} The oracle-aligned projection and off-direction residual describe Euclidean changes in the option-probability vector. Per-row orthogonality yields the exact squared-distance identity in Appendix~\ref{app:geometry}. The displayed all-run numerical-error diagnostic is distinct from the matched-suppression subset in Table~\ref{tab:S08_suppression_and_geometry}; its precision range should not be replaced by that subset's $2\times10^{-15}$ bound. Simplex-projected magnitude-only and residual-only diagnostics are not an additive decomposition of Jensen--Shannon divergence.}
  \label{fig:S09}
\end{figure}

\begin{figure}[tbh]\centering   \includegraphics[width=\linewidth,height=0.79\textheight,keepaspectratio]{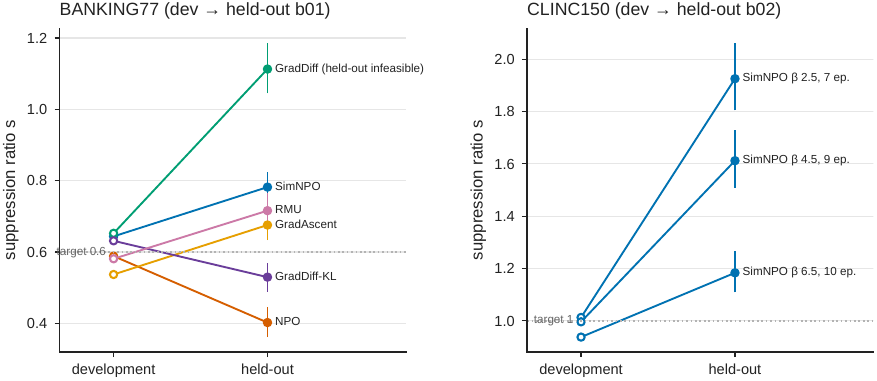}
  \caption{\textbf{Development matching and held-out suppression.} Configurations were chosen on development data to approach suppression targets of 0.6 on BANKING77 and 1 on CLINC150, subject to feasibility. Their held-out suppression can differ substantially. Open development coordinates and held-out coordinates belong to the same selected configuration; matching is approximate. Held-out fidelity cannot be inferred from development suppression alone.}
  \label{fig:S10}
\end{figure}

\begin{figure}[tbh]\centering   \includegraphics[width=\linewidth,height=0.79\textheight,keepaspectratio]{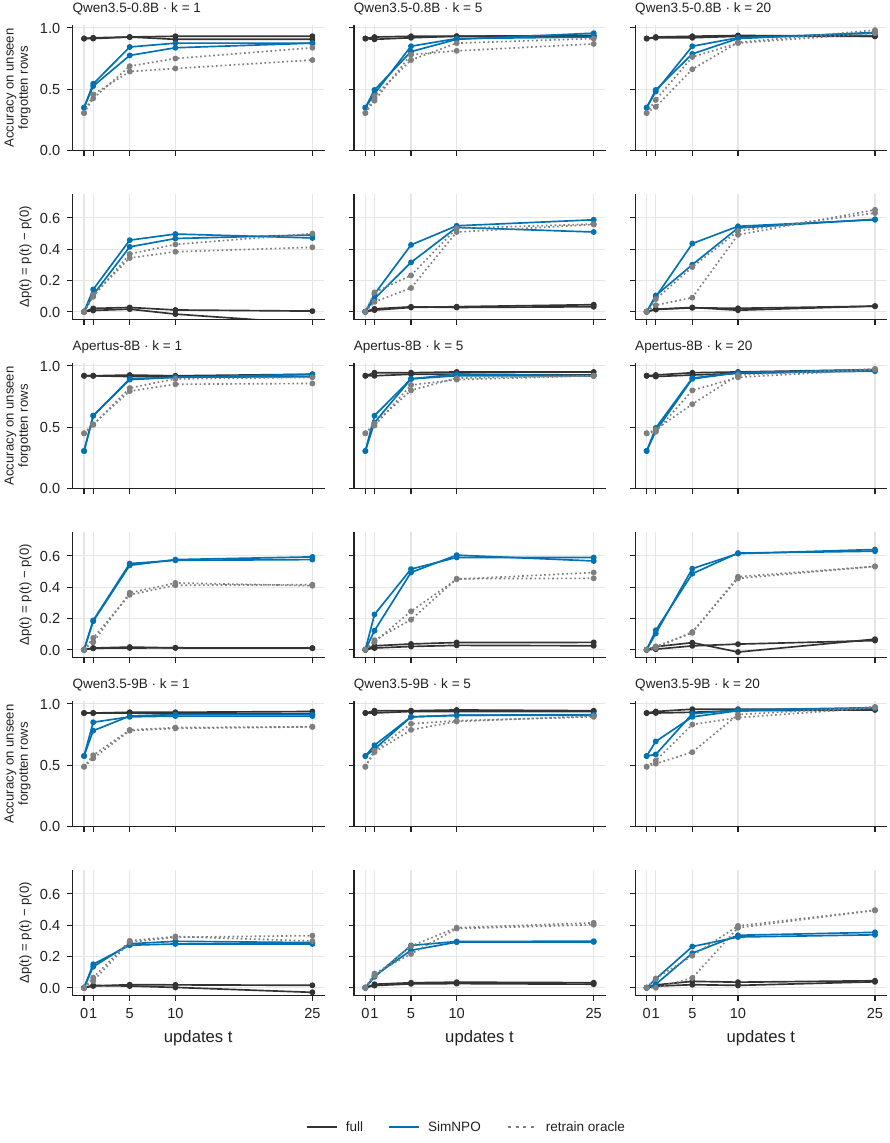}
  \caption{\textbf{Relearning for every exposure budget.} Forgotten-class accuracy and gold-probability changes are shown for $k\in\{1,5,20\}$ exposed deleted-training rows per class and adaptation seeds 0 and 1. Evaluation uses 160 unseen rows from four forgotten classes. Each curve uses one trained backbone seed; repeated budgets and adaptation seeds are not independent backbone replications. Starting probability, absolute recovery, adjusted recovery, and headroom-normalized recovery address different comparisons.}
  \label{fig:S11}
\end{figure}

\begin{figure}[tbh]\centering   \includegraphics[width=\linewidth,height=0.79\textheight,keepaspectratio]{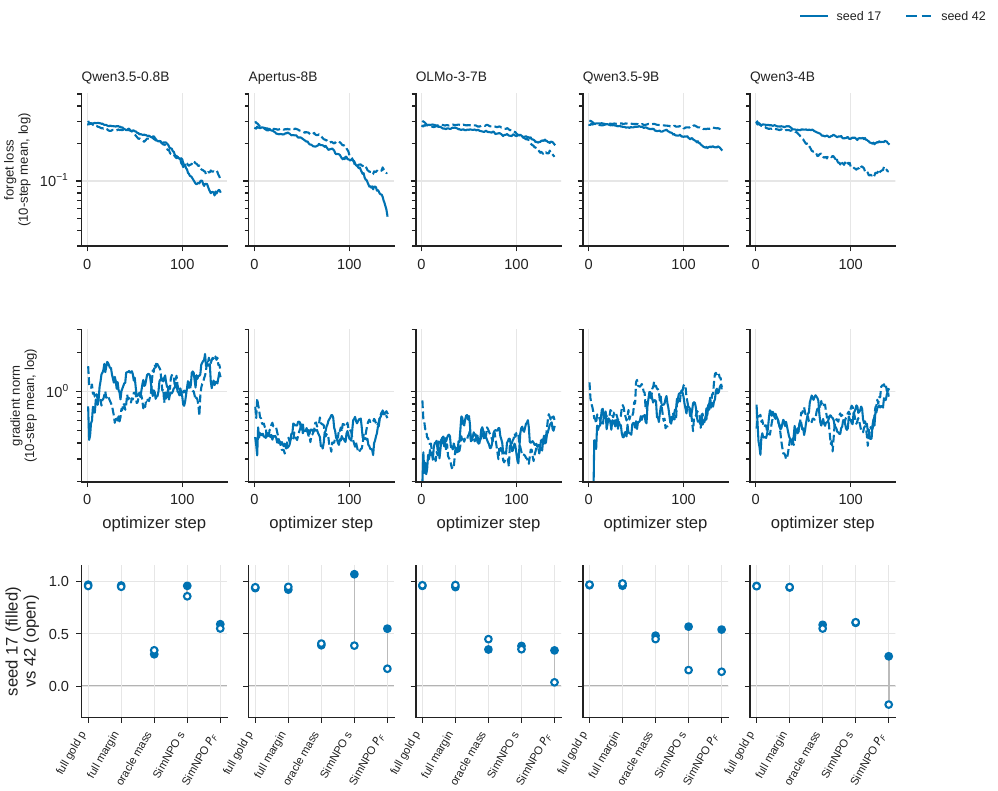}
  \caption{\textbf{Optimization diagnostics for the seed-42 comparison.} Initial decision statistics, forget and retain losses, gradient norms, suppression, and progress are shown for paired model configurations. Apertus-8B and Qwen3.5-9B exhibit weaker suppression at seed 42, but OLMo-3-7B and Qwen3-4B lose fidelity despite similar suppression. No single insufficient-update explanation accounts for all models. Parameter-update norms were not recorded.}
  \label{fig:S14}
\end{figure}

\begin{figure}[tbh]\centering   \includegraphics[width=\linewidth,height=0.79\textheight,keepaspectratio]{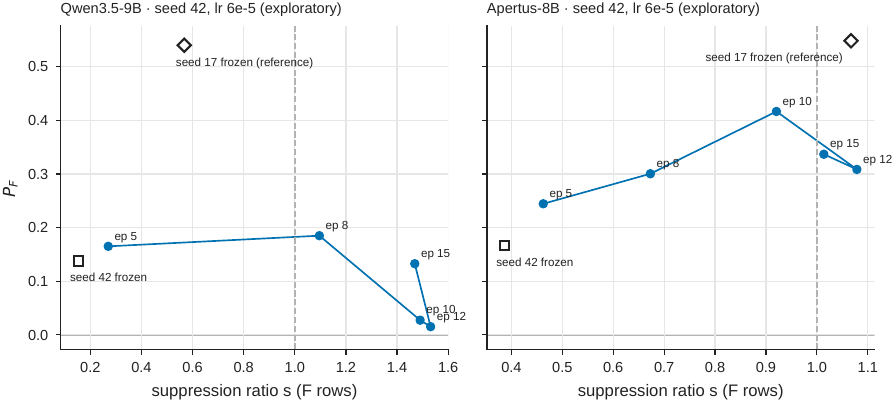}
  \caption{\textbf{Exploratory stronger-update reruns.} Additional traces examine whether changing update strength can recover progress in selected seed-42 failures. Apertus-8B improves at an observed intermediate checkpoint, while Qwen3.5-9B remains limited and later oversuppresses. Best displayed epochs were selected after inspecting these traces. These results motivate a prospective adjustment rule; they do not replace the frozen-setting replication or establish a validated repair.}
  \label{fig:S15}
\end{figure}

\section{Supplementary Tables}
\label{app:tables}
The numerical entries below are transcribed from the supplied LaTeX exports. Formatting and explicit row identifiers have been added; no confidence interval is recalculated from rounded entries. Tables with many columns are divided into readable panels keyed by the same row identifier. The machine-readable transcriptions preserve the original fields.
\clearpage
% \fontsize{8}{10}\selectfont\setlength{\tabcolsep}{3pt}
% \renewcommand{\arraystretch}{1.18}
% \renewcommand{\theHtable}{supp.1.a}
\setcounter{table}{0}
\begin{table*}[tbp]
\centering
\small
\caption{Observed evaluation registry, summarized by dataset, label scheme, and recorded stage.}
\label{tab:S01_run_registry}
\setlength{\tabcolsep}{4pt}
\begin{tabular}{
@{}
>{\raggedright\arraybackslash}p{3cm}
>{\raggedright\arraybackslash}p{3cm}
>{\raggedright\arraybackslash}p{6cm}
>{\raggedright\arraybackslash}p{2cm}
@{}
}
\toprule
\textbf{Dataset} &
\textbf{Labels} &
\textbf{Recorded stage} &
\textbf{Evaluations} \\
\midrule
banking77 & opaque & exploratory & 12 \\
banking77 & semantic & confirmation & 6 \\
banking77 & semantic & confirmation\_\allowbreak{}sensitivity & 29 \\
banking77 & semantic & exploratory & 14 \\
banking77 & semantic & pilot & 15 \\
banking77 & semantic & reference & 170 \\
banking77 & semantic & replication & 40 \\
banking77 & semantic & screening & 19 \\
\midrule
clinc150 & opaque & exploratory & 9 \\
clinc150 & semantic & confirmation & 2 \\
clinc150 & semantic & control & 3 \\
clinc150 & semantic & exploratory & 3 \\
clinc150 & semantic & reference & 24 \\
clinc150 & semantic & replication & 5 \\
clinc150 & semantic & transfer & 7 \\
\bottomrule
\end{tabular}

\par\smallskip
\noindent
\parbox{\textwidth}{\footnotesize
The complete 358-row registry, including model, seed, bundle,
run label, and probability hash, is preserved in
\path{data/S01_run_registry.csv}.
These are evaluations, not independent training replicates.
Several displayed keys lack deletion granularity; source-row
identifiers preserve distinct records.
The upstream audit reports a broader total of 1,043 evaluations,
including 646 development, 8 development-reference, and
31 fixture evaluations outside this registry.
}
\end{table*}
\clearpage

% \clearpage
% {\fontsize{8}{10}\selectfont\setlength{\tabcolsep}{3pt}
% \renewcommand{\arraystretch}{1.18}
% \renewcommand{\theHtable}{supp.2.a}

% ==================================================
% Table S2: Part 1
% ==================================================
\setcounter{table}{1}
\begin{table*}[p]
\centering
\footnotesize
\caption{Metric definitions, gates and uncertainty units.}
\label{tab:S02_metric_definitions}
\setlength{\tabcolsep}{3pt}
\begin{tabular}{
@{}
>{\raggedright\arraybackslash}p{2.8cm}
>{\raggedright\arraybackslash}p{9.4cm}
>{\raggedright\arraybackslash}p{3.6cm}
@{}
}
\toprule
\textbf{Exported quantity} &
\textbf{Definition} &
\textbf{Upstream source (not included)} \\
\midrule

F view (E1, forget\_\allowbreak{}train) &
Rows of the deletion set D\_\allowbreak{}F itself (training rows of the forgotten examples /\allowbreak{} association /\allowbreak{} classes). &
src/\allowbreak{}evaluation/\allowbreak{}views.py \\

G view (E2, forget\_\allowbreak{}gen) &
Unseen evaluation-split rows of the forgotten class /\allowbreak{} association; at example level, test rows whose nearest training row is a forgotten row. &
src/\allowbreak{}evaluation/\allowbreak{}views.py \\

Oracle distance OD\_\allowbreak{}v &
Mean over rows of the view of JS(p\_\allowbreak{}model, p\_\allowbreak{}oracle), base-2 Jensen-Shannon on raw (uncalibrated) option probabilities, eps 1e-12; 0 $\leq$ OD $\leq$ 1. &
src/\allowbreak{}evaluation/\allowbreak{}oracle\_\allowbreak{}distance.py \\

Displacement D0\_\allowbreak{}v &
OD\_\allowbreak{}v of the full (no-op) model against the retrain oracle of the same seed and forget set. &
record.json of 'full' \\

Gain dOD\_\allowbreak{}v &
D0\_\allowbreak{}v - OD\_\allowbreak{}v(method); positive = closer to the oracle than no-op; paired on rows. &
PROTOCOL Amendment 2-3 \\

Normalised progress P\_\allowbreak{}v &
1 - OD\_\allowbreak{}v(method) /\allowbreak{} D0\_\allowbreak{}v from aggregate distances (never a mean of row ratios); recomputed per bootstrap replicate. &
PROTOCOL Amendment 3 \\

Feasibility &
Acc\_\allowbreak{}R $\geq$ Acc\_\allowbreak{}R(full) - 0.02 and ECE $\leq$ ECE(full) + 0.02 on retained rows of the same evaluation. &
PROTOCOL Amendment 1-2 \\

Acceptance (Phase 1 rule) &
accepted: both gains $>$ 0 with row-bootstrap L95 $>$ 0, feasible; positive, unconfirmed: both gains $>$ 0, a bound $\leq$ 0; otherwise no benefit /\allowbreak{} infeasible. &
PROTOCOL Amendment 3 \\

\bottomrule
\end{tabular}
\end{table*}

% ==================================================
% Table S2: Part 2 (continued)
% ==================================================
\begin{table*}[p]
\ContinuedFloat
\centering
\footnotesize
\caption[]{Metric definitions, gates and uncertainty units (continued).}
\setlength{\tabcolsep}{3pt}
\begin{tabular}{
@{}
>{\raggedright\arraybackslash}p{2.8cm}
>{\raggedright\arraybackslash}p{9.4cm}
>{\raggedright\arraybackslash}p{3.6cm}
@{}
}
\toprule
\textbf{Exported quantity} &
\textbf{Definition} &
\textbf{Upstream source (not included)} \\
\midrule

Multiplicity &
Holm-adjusted intersection-union p (max of the two one-sided p) within the frozen candidate family of a confirmation; screening tests per model are uncorrected. &
PROTOCOL Amendment 3; screen record \\

Bootstrap &
5000 paired draws; rows (primary, seeds 1 F /\allowbreak{} 2 G) and forgotten classes or association clusters (sensitivity, seed + 100); percentile 95\% CI; one-sided p from centred resamples. &
scripts/\allowbreak{}confirm\_\allowbreak{}bootstrap.py \\

Cross-bundle replication &
Hierarchical bootstrap: seeds, then bundles, then units within bundle; replicates if aggregate L95 $>$ 0 on both views, feasible in every cell, survives leave-one-bundle/\allowbreak{}seed-out. &
PROTOCOL Phase 2 \\

Noise floor &
JS between same-data full models of different seeds (Qwen3.5-0.8B, seeds 17/\allowbreak{}42/\allowbreak{}99) on the same rows. &
scripts/\allowbreak{}displacement.py \\

Acquisition gate &
Each candidate forgotten class: full-model development accuracy $\geq$ 0.70 and gold probability $\geq$ 0.50 (association: accuracy $\geq$ 0.65); targets selected in hash order among eligible ones at the primary seed. &
src/\allowbreak{}datasets/\allowbreak{}targets.py \\

Suppression ratio s &
Pooled forgotten-class mass removed by the method relative to the oracle over F and G rows: sum(m\_\allowbreak{}full - m\_\allowbreak{}run) /\allowbreak{} sum(m\_\allowbreak{}full - m\_\allowbreak{}oracle). &
scripts/\allowbreak{}matched\_\allowbreak{}suppression.py \\

Projection \(a\), residual \(r\) &
Per row
\(\Delta_U = p_U - p_{\mathrm{full}}\),\allowbreak{}
\(\Delta_O = p_{\mathrm{oracle}} - p_{\mathrm{full}}\),\allowbreak{}
\(a = \frac{\Delta_U \cdot \Delta_O}
{\|\Delta_O\|_2^2}\),\allowbreak{}
\(r = \Delta_U - a\Delta_O\);
pooled
\(a = \frac{\sum \Delta_U \cdot \Delta_O}
{\sum \|\Delta_O\|_2^2}\);\allowbreak{}
relative residual
\(= \sqrt{\frac{\sum \|r\|_2^2}
{\sum \|\Delta_O\|_2^2}}\);\allowbreak{}
exact \(L_2\) identity checked per row. &
\texttt{scripts/\allowbreak{}redistribution\_\allowbreak{}decomposition.py} \\

Recovery areas &
Trapezoid over \(t = 0, 1, 5, 10, 25\) updates,\allowbreak{}
weights normalised to sum to one;\allowbreak{}
absolute
\(A = \sum_t w_t p(t)\);\allowbreak{}
adjusted
\(A_{\mathrm{adj}} = A - p(0)\);\allowbreak{}
headroom
\(H = \frac{A_{\mathrm{adj}}}{1 - p(0)}\)
(ratio of means;\allowbreak{}
mean of row ratios as a second aggregation). &
\texttt{scripts/\allowbreak{}recovery\_\allowbreak{}uncertainty.py} \\

\bottomrule
\end{tabular}

\par\smallskip
\noindent
\parbox{\textwidth}{\footnotesize
The manuscript uses $D_{0,v}$, $\Delta\operatorname{OD}_v$, and $P_v$
consistently for the exported D0, dOD, and P fields.
File paths are provenance pointers from the source table,
not files distributed with this draft.
}
\end{table*}
% }
% \clearpage

% \clearpage
% \begin{landscape}
% {\fontsize{8}{10}\selectfont\setlength{\tabcolsep}{3pt}
% \renewcommand{\arraystretch}{1.18}
% \renewcommand{\theHtable}{supp.3.a}
\setcounter{table}{2}

% Panel 1, row IDs 1--15
\begin{table*}[p]
\centering
\caption{Complete method results: oracle fidelity and recorded status. Row IDs link the utility panel and CSV.}
\label{tab:S03_complete_method_results}
{\footnotesize
\setlength{\tabcolsep}{3pt}
\renewcommand{\arraystretch}{1.12}
\resizebox{\textwidth}{!}{%
\begin{tabular}{
@{}
>{\raggedright\arraybackslash}p{0.65cm}
>{\raggedright\arraybackslash}p{4.3cm}
>{\raggedright\arraybackslash}p{1.6cm}
>{\raggedright\arraybackslash}p{4.4cm}
>{\raggedright\arraybackslash}p{4.4cm}
>{\raggedright\arraybackslash}p{1.5cm}
>{\raggedright\arraybackslash}p{2.5cm}
@{}
}
\toprule
\textbf{ID} & \textbf{Configuration} & \textbf{$D_{0,\mathrm{F}}$ / $D_{0,\mathrm{G}}$} & \textbf{$\Delta\operatorname{OD}_\mathrm{F}$ [rows]; unit CI} & \textbf{$\Delta\operatorname{OD}_\mathrm{G}$ [rows]; unit CI} & \textbf{$P_\mathrm{F}/P_\mathrm{G}$} & \textbf{Status} \\
\midrule
1 & apertus-8b\newline BANKING77 sem. s17 b01\newline SimNPO /\allowbreak{} screen & 0.479\newline 0.400 & +0.263 [+0.226, +0.301]\newline units: [+0.141, +0.434] & +0.231 [+0.164, +0.298]\newline units: [+0.059, +0.464] & 0.55 / 0.58 & accepted \\
2 & apertus-8b\newline BANKING77 sem. s17 b02\newline SimNPO /\allowbreak{} replic. & 0.565\newline 0.552 & +0.318 [+0.289, +0.347]\newline units: [+0.108, +0.516] & +0.316 [+0.272, +0.361]\newline units: [+0.100, +0.500] & 0.56 / 0.57 & accepted \\
3 & apertus-8b\newline BANKING77 sem. s42 b01\newline SimNPO /\allowbreak{} explor. & 0.471\newline 0.405 & +0.196 [+0.153, +0.239]\newline units: [-0.069, +0.514] & +0.184 [+0.111, +0.256]\newline units: [-0.136, +0.498] & 0.42 / 0.45 & accepted \\
4 & apertus-8b\newline BANKING77 sem. s42 b01\newline SimNPO /\allowbreak{} explor. & 0.471\newline 0.405 & +0.145 [+0.096, +0.195]\newline units: [-0.091, +0.533] & +0.137 [+0.051, +0.222]\newline units: [-0.145, +0.494] & 0.31 / 0.34 & accepted \\
5 & apertus-8b\newline BANKING77 sem. s42 b01\newline SimNPO /\allowbreak{} explor. & 0.471\newline 0.405 & +0.115 [+0.085, +0.146]\newline units: [-0.034, +0.293] & +0.121 [+0.071, +0.170]\newline units: [-0.062, +0.240] & 0.24 / 0.30 & accepted \\
6 & apertus-8b\newline BANKING77 sem. s42 b01\newline SimNPO /\allowbreak{} explor. & 0.471\newline 0.405 & +0.141 [+0.098, +0.187]\newline units: [-0.161, +0.487] & +0.170 [+0.095, +0.244]\newline units: [-0.212, +0.513] & 0.30 / 0.42 & accepted \\
7 & apertus-8b\newline BANKING77 sem. s42 b01\newline SimNPO /\allowbreak{} explor. & 0.471\newline 0.405 & +0.159 [+0.110, +0.208]\newline units: [-0.128, +0.524] & +0.140 [+0.056, +0.223]\newline units: [-0.188, +0.510] & 0.34 / 0.35 & accepted \\
8 & apertus-8b\newline BANKING77 sem. s42 b01\newline SimNPO /\allowbreak{} replic. & 0.471\newline 0.405 & +0.078 [+0.047, +0.112]\newline units: [-0.121, +0.316] & +0.102 [+0.048, +0.154]\newline units: [-0.142, +0.276] & 0.17 / 0.25 & accepted \\
9 & falcon-h1-3b\newline BANKING77 sem. s17 b01\newline SimNPO /\allowbreak{} screen & 0.501\newline 0.466 & +0.181 [+0.156, +0.206]\newline units: [+0.068, +0.403] & +0.195 [+0.154, +0.238]\newline units: [+0.074, +0.333] & 0.36 / 0.42 & accepted \\
10 & gemma4-e2b\newline BANKING77 sem. s17 b01\newline SimNPO /\allowbreak{} screen & 0.474\newline 0.394 & +0.195 [+0.167, +0.222]\newline units: [+0.090, +0.381] & +0.172 [+0.128, +0.217]\newline units: [+0.024, +0.321] & 0.41 / 0.44 & accepted \\
11 & granite4-micro\newline BANKING77 sem. s17 b01\newline SimNPO /\allowbreak{} screen & 0.468\newline 0.401 & +0.216 [+0.188, +0.245]\newline units: [+0.109, +0.380] & +0.220 [+0.176, +0.265]\newline units: [+0.091, +0.336] & 0.46 / 0.55 & accepted \\
12 & lfm2.5-1.2b\newline BANKING77 sem. s17 b01\newline SimNPO /\allowbreak{} screen & 0.539\newline 0.434 & +0.160 [+0.116, +0.202]\newline units: [+0.038, +0.308] & +0.098 [+0.027, +0.169]\newline units: [-0.106, +0.318] & 0.30 / 0.23 & accepted \\
13 & llama3.2-1b\newline BANKING77 sem. s17 b01\newline SimNPO /\allowbreak{} screen & 0.626\newline 0.502 & +0.355 [+0.324, +0.386]\newline units: [+0.243, +0.470] & +0.271 [+0.224, +0.320]\newline units: [+0.210, +0.314] & 0.57 / 0.54 & accepted \\
14 & llama3.2-3b\newline BANKING77 sem. s17 b01\newline SimNPO /\allowbreak{} screen & 0.558\newline 0.461 & +0.280 [+0.248, +0.314]\newline units: [+0.168, +0.461] & +0.259 [+0.203, +0.316]\newline units: [+0.113, +0.436] & 0.50 / 0.56 & accepted \\
15 & ministral3-3b\newline BANKING77 sem. s17 b01\newline SimNPO /\allowbreak{} screen & 0.464\newline 0.377 & +0.131 [+0.104, +0.160]\newline units: [+0.055, +0.262] & +0.079 [+0.027, +0.131]\newline units: [+0.013, +0.148] & 0.28 / 0.21 & accepted \\
\bottomrule
\end{tabular}
}% end resizebox
}% end font group
\end{table*}

% Panel 1, row IDs 16--30
\begin{table*}[p]
\ContinuedFloat
\centering
\caption[]{Complete method results (continued): oracle fidelity, IDs 16--30.}
{\footnotesize
\setlength{\tabcolsep}{3pt}
\renewcommand{\arraystretch}{1.12}
\resizebox{\textwidth}{!}{%
\begin{tabular}{
@{}
>{\raggedright\arraybackslash}p{0.65cm}
>{\raggedright\arraybackslash}p{4.3cm}
>{\raggedright\arraybackslash}p{1.6cm}
>{\raggedright\arraybackslash}p{4.4cm}
>{\raggedright\arraybackslash}p{4.4cm}
>{\raggedright\arraybackslash}p{1.5cm}
>{\raggedright\arraybackslash}p{2.5cm}
@{}
}
\toprule
\textbf{ID} & \textbf{Configuration} & \textbf{$D_{0,\mathrm{F}}$ / $D_{0,\mathrm{G}}$} & \textbf{$\Delta\operatorname{OD}_\mathrm{F}$ [rows]; unit CI} & \textbf{$\Delta\operatorname{OD}_\mathrm{G}$ [rows]; unit CI} & \textbf{$P_\mathrm{F}/P_\mathrm{G}$} & \textbf{Status} \\
\midrule
16 & nemotron-nano-9b\newline BANKING77 sem. s17 b01\newline SimNPO /\allowbreak{} screen & 0.472\newline 0.369 & +0.218 [+0.186, +0.250]\newline units: [+0.080, +0.463] & +0.194 [+0.140, +0.248]\newline units: [+0.036, +0.416] & 0.46 / 0.53 & accepted \\
17 & olmo3-7b\newline BANKING77 sem. s17 b01\newline SimNPO /\allowbreak{} screen & 0.545\newline 0.447 & +0.186 [+0.158, +0.215]\newline units: [+0.099, +0.371] & +0.156 [+0.107, +0.204]\newline units: [+0.078, +0.210] & 0.34 / 0.35 & accepted \\
18 & olmo3-7b\newline BANKING77 sem. s17 b02\newline SimNPO /\allowbreak{} replic. & 0.659\newline 0.643 & +0.328 [+0.297, +0.360]\newline units: [+0.071, +0.591] & +0.312 [+0.265, +0.359]\newline units: [+0.078, +0.546] & 0.50 / 0.49 & accepted \\
19 & olmo3-7b\newline BANKING77 sem. s42 b01\newline SimNPO /\allowbreak{} replic. & 0.464\newline 0.388 & +0.017 [-0.021, +0.054]\newline units: [-0.196, +0.301] & +0.001 [-0.055, +0.055]\newline units: [-0.290, +0.206] & 0.04 / 0.00 & positive, unconfirmed \\
20 & phi4-mini\newline BANKING77 sem. s17 b01\newline SimNPO /\allowbreak{} screen & 0.396\newline 0.321 & +0.187 [+0.158, +0.216]\newline units: [+0.069, +0.352] & +0.163 [+0.114, +0.213]\newline units: [+0.037, +0.357] & 0.47 / 0.51 & accepted \\
21 & qwen2.5-1.5b\newline BANKING77 sem. s17 b01\newline SimNPO /\allowbreak{} screen & 0.539\newline 0.417 & +0.243 [+0.214, +0.273]\newline units: [+0.096, +0.402] & +0.227 [+0.181, +0.275]\newline units: [+0.072, +0.374] & 0.45 / 0.54 & accepted \\
22 & qwen2.5-3b\newline BANKING77 sem. s17 b01\newline SimNPO /\allowbreak{} screen & 0.457\newline 0.378 & +0.228 [+0.196, +0.260]\newline units: [+0.083, +0.452] & +0.228 [+0.180, +0.276]\newline units: [+0.091, +0.365] & 0.50 / 0.60 & accepted \\
23 & qwen2.5-7b\newline BANKING77 sem. s17 b01\newline SimNPO /\allowbreak{} screen & 0.418\newline 0.351 & +0.263 [+0.232, +0.295]\newline units: [+0.146, +0.472] & +0.222 [+0.176, +0.268]\newline units: [+0.081, +0.363] & 0.63 / 0.63 & accepted \\
24 & qwen3-4b\newline BANKING77 sem. s17 b01\newline SimNPO /\allowbreak{} screen & 0.318\newline 0.238 & +0.091 [+0.063, +0.117]\newline units: [+0.024, +0.155] & +0.044 [-0.001, +0.091]\newline units: [-0.090, +0.173] & 0.28 / 0.19 & positive, unconfirmed \\
25 & qwen3-4b\newline BANKING77 sem. s17 b02\newline SimNPO /\allowbreak{} replic. & 0.614\newline 0.595 & +0.314 [+0.281, +0.348]\newline units: [+0.094, +0.578] & +0.300 [+0.250, +0.351]\newline units: [+0.089, +0.546] & 0.51 / 0.50 & accepted \\
26 & qwen3-4b\newline BANKING77 sem. s42 b01\newline SimNPO /\allowbreak{} replic. & 0.340\newline 0.298 & -0.060 [-0.109, -0.011]\newline units: [-0.434, +0.272] & -0.031 [-0.106, +0.042]\newline units: [-0.447, +0.258] & -0.18 / -0.10 & no benefit \\
27 & qwen3.5-0.8b\newline BANKING77 opaque s17 b01\newline SimNPO /\allowbreak{} explor. & 0.943\newline 0.881 & +0.571 [+0.539, +0.603]\newline units: [+0.431, +0.732] & +0.549 [+0.495, +0.604]\newline units: [+0.392, +0.706] & 0.61 / 0.62 & accepted \\
28 & qwen3.5-0.8b\newline BANKING77 opaque s17 b02\newline SimNPO /\allowbreak{} explor. & 0.912\newline 0.872 & +0.410 [+0.376, +0.445]\newline units: [+0.234, +0.571] & +0.371 [+0.322, +0.423]\newline units: [+0.207, +0.475] & 0.45 / 0.43 & accepted \\
29 & qwen3.5-0.8b\newline BANKING77 opaque s42 b01\newline SimNPO /\allowbreak{} explor. & 0.934\newline 0.878 & +0.560 [+0.532, +0.587]\newline units: [+0.467, +0.666] & +0.508 [+0.461, +0.554]\newline units: [+0.428, +0.647] & 0.60 / 0.58 & accepted \\
30 & qwen3.5-0.8b\newline BANKING77 opaque s42 b02\newline SimNPO /\allowbreak{} explor. & 0.911\newline 0.902 & +0.362 [+0.326, +0.398]\newline units: [+0.118, +0.611] & +0.345 [+0.293, +0.398]\newline units: [+0.095, +0.594] & 0.40 / 0.38 & accepted \\
\bottomrule
\end{tabular}
}% end resizebox
}% end font group
\end{table*}

% Panel 1, row IDs 31--45
\begin{table*}[p]
\ContinuedFloat
\centering
\caption[]{Complete method results (continued): oracle fidelity, IDs 31--45.}
{\footnotesize
\setlength{\tabcolsep}{3pt}
\renewcommand{\arraystretch}{1.12}
\resizebox{\textwidth}{!}{%
\begin{tabular}{
@{}
>{\raggedright\arraybackslash}p{0.65cm}
>{\raggedright\arraybackslash}p{4.3cm}
>{\raggedright\arraybackslash}p{1.6cm}
>{\raggedright\arraybackslash}p{4.4cm}
>{\raggedright\arraybackslash}p{4.4cm}
>{\raggedright\arraybackslash}p{1.5cm}
>{\raggedright\arraybackslash}p{2.5cm}
@{}
}
\toprule
\textbf{ID} & \textbf{Configuration} & \textbf{$D_{0,\mathrm{F}}$ / $D_{0,\mathrm{G}}$} & \textbf{$\Delta\operatorname{OD}_\mathrm{F}$ [rows]; unit CI} & \textbf{$\Delta\operatorname{OD}_\mathrm{G}$ [rows]; unit CI} & \textbf{$P_\mathrm{F}/P_\mathrm{G}$} & \textbf{Status} \\
\midrule
31 & qwen3.5-0.8b\newline BANKING77 sem. s17 b01\newline GradAscent /\allowbreak{} confirm. & 0.608\newline 0.495 & +0.332 [+0.297, +0.366]\newline units: [+0.224, +0.422] & +0.241 [+0.194, +0.290]\newline units: [+0.192, +0.290] & 0.55 / 0.49 & accepted \\
32 & qwen3.5-0.8b\newline BANKING77 sem. s17 b01\newline GradDiff /\allowbreak{} confirm. & 0.608\newline 0.495 & +0.274 [+0.241, +0.308]\newline units: [+0.183, +0.379] & +0.209 [+0.165, +0.255]\newline units: [+0.145, +0.283] & 0.45 / 0.42 & accepted \\
33 & qwen3.5-0.8b\newline BANKING77 sem. s17 b01\newline GradDiff-KL /\allowbreak{} confirm. & 0.608\newline 0.495 & +0.188 [+0.131, +0.242]\newline units: [+0.006, +0.431] & +0.139 [+0.053, +0.228]\newline units: [-0.092, +0.338] & 0.31 / 0.28 & infeasible \\
34 & qwen3.5-0.8b\newline BANKING77 sem. s17 b01\newline NPO /\allowbreak{} confirm. & 0.608\newline 0.495 & +0.236 [+0.205, +0.266]\newline units: [+0.143, +0.342] & +0.176 [+0.135, +0.220]\newline units: [+0.100, +0.252] & 0.39 / 0.36 & accepted \\
35 & qwen3.5-0.8b\newline BANKING77 sem. s17 b01\newline SimNPO /\allowbreak{} confirm. & 0.608\newline 0.495 & +0.359 [+0.319, +0.396]\newline units: [+0.297, +0.479] & +0.280 [+0.224, +0.336]\newline units: [+0.207, +0.381] & 0.59 / 0.57 & accepted \\
36 & qwen3.5-0.8b\newline BANKING77 sem. s17 b01\newline Uniformization /\allowbreak{} confirm. & 0.608\newline 0.495 & +0.160 [+0.144, +0.175]\newline units: [+0.112, +0.211] & +0.102 [+0.082, +0.124]\newline units: [+0.079, +0.127] & 0.26 / 0.21 & infeasible \\
37 & qwen3.5-0.8b\newline BANKING77 sem. s17 b01\newline GradDiff-KL /\allowbreak{} confirmation\_\allowbreak{}sensitivity & 0.193\newline 0.148 & -0.005 [-0.024, +0.014]\newline units: [-0.076, +0.057] & -0.049 [-0.076, -0.023]\newline units: [-0.125, +0.019] & -0.03 / -0.33 & infeasible \\
38 & qwen3.5-0.8b\newline BANKING77 sem. s17 b01\newline SimNPO /\allowbreak{} confirmation\_\allowbreak{}sensitivity & 0.193\newline 0.148 & +0.040 [+0.033, +0.047]\newline units: [+0.007, +0.083] & +0.044 [+0.031, +0.057]\newline units: [+0.003, +0.091] & 0.21 / 0.29 & accepted \\
39 & qwen3.5-0.8b\newline BANKING77 sem. s17 b01\newline Uniformization /\allowbreak{} confirmation\_\allowbreak{}sensitivity & 0.193\newline 0.148 & +0.030 [+0.025, +0.035]\newline units: [+0.013, +0.045] & +0.023 [+0.017, +0.030]\newline units: [+0.004, +0.044] & 0.16 / 0.16 & accepted \\
40 & qwen3.5-0.8b\newline BANKING77 sem. s17 b01\newline GradDiff /\allowbreak{} confirmation\_\allowbreak{}sensitivity & 0.193\newline 0.148 & +0.007 [-0.007, +0.021]\newline units: [-0.018, +0.038] & -0.014 [-0.033, +0.003]\newline units: [-0.032, +0.004] & 0.04 / -0.10 & no benefit \\
41 & qwen3.5-0.8b\newline BANKING77 sem. s17 b01\newline NPO /\allowbreak{} confirmation\_\allowbreak{}sensitivity & 0.193\newline 0.148 & -0.003 [-0.024, +0.018]\newline units: [-0.097, +0.068] & -0.050 [-0.081, -0.019]\newline units: [-0.160, +0.040] & -0.02 / -0.34 & no benefit \\
42 & qwen3.5-0.8b\newline BANKING77 sem. s17 b01\newline GradAscent /\allowbreak{} pilot & 0.193\newline 0.148 & -0.168 [-0.203, -0.132]\newline units: [-0.303, -0.033] & -0.241 [-0.296, -0.187]\newline units: [-0.420, -0.084] & -0.87 / -1.63 & infeasible \\
43 & qwen3.5-0.8b\newline BANKING77 sem. s17 b01\newline GradAscent /\allowbreak{} pilot & 0.608\newline 0.495 & +0.168 [+0.110, +0.225]\newline units: [-0.011, +0.409] & +0.111 [+0.022, +0.201]\newline units: [-0.125, +0.318] & 0.28 / 0.22 & infeasible \\
44 & qwen3.5-0.8b\newline BANKING77 sem. s17 b01\newline GradDiff /\allowbreak{} pilot & 0.193\newline 0.148 & -0.132 [-0.164, -0.100]\newline units: [-0.249, -0.023] & -0.199 [-0.250, -0.149]\newline units: [-0.373, -0.061] & -0.68 / -1.35 & infeasible \\
45 & qwen3.5-0.8b\newline BANKING77 sem. s17 b01\newline GradDiff /\allowbreak{} pilot & 0.608\newline 0.495 & +0.133 [+0.082, +0.184]\newline units: [-0.003, +0.370] & +0.071 [-0.007, +0.147]\newline units: [-0.120, +0.287] & 0.22 / 0.14 & infeasible \\
\bottomrule
\end{tabular}
}% end resizebox
}% end font group
\end{table*}

% Panel 1, row IDs 46--60
\begin{table*}[p]
\ContinuedFloat
\centering
\caption[]{Complete method results (continued): oracle fidelity, IDs 46--60.}
{\footnotesize
\setlength{\tabcolsep}{3pt}
\renewcommand{\arraystretch}{1.12}
\resizebox{\textwidth}{!}{%
\begin{tabular}{
@{}
>{\raggedright\arraybackslash}p{0.65cm}
>{\raggedright\arraybackslash}p{4.3cm}
>{\raggedright\arraybackslash}p{1.6cm}
>{\raggedright\arraybackslash}p{4.4cm}
>{\raggedright\arraybackslash}p{4.4cm}
>{\raggedright\arraybackslash}p{1.5cm}
>{\raggedright\arraybackslash}p{2.5cm}
@{}
}
\toprule
\textbf{ID} & \textbf{Configuration} & \textbf{$D_{0,\mathrm{F}}$ / $D_{0,\mathrm{G}}$} & \textbf{$\Delta\operatorname{OD}_\mathrm{F}$ [rows]; unit CI} & \textbf{$\Delta\operatorname{OD}_\mathrm{G}$ [rows]; unit CI} & \textbf{$P_\mathrm{F}/P_\mathrm{G}$} & \textbf{Status} \\
\midrule
46 & qwen3.5-0.8b\newline BANKING77 sem. s17 b01\newline GradDiff /\allowbreak{} explor. & 0.608\newline 0.495 & +0.113 [+0.058, +0.167]\newline units: [-0.050, +0.361] & +0.015 [-0.070, +0.099]\newline units: [-0.208, +0.229] & 0.19 / 0.03 & infeasible \\
47 & qwen3.5-0.8b\newline BANKING77 sem. s17 b01\newline GradDiff-KL /\allowbreak{} explor. & 0.608\newline 0.495 & +0.293 [+0.260, +0.325]\newline units: [+0.201, +0.372] & +0.215 [+0.172, +0.261]\newline units: [+0.172, +0.255] & 0.48 / 0.43 & accepted \\
48 & qwen3.5-0.8b\newline BANKING77 sem. s17 b01\newline RMU /\allowbreak{} explor. & 0.608\newline 0.495 & +0.026 [+0.002, +0.048]\newline units: [-0.034, +0.135] & -0.009 [-0.051, +0.033]\newline units: [-0.093, +0.087] & 0.04 / -0.02 & no benefit \\
49 & qwen3.5-0.8b\newline BANKING77 sem. s17 b01\newline SimNPO /\allowbreak{} explor. & 0.608\newline 0.495 & +0.339 [+0.305, +0.372]\newline units: [+0.223, +0.468] & +0.262 [+0.213, +0.312]\newline units: [+0.190, +0.372] & 0.56 / 0.53 & accepted \\
50 & qwen3.5-0.8b\newline BANKING77 sem. s17 b01\newline NPO /\allowbreak{} pilot & 0.193\newline 0.148 & -0.145 [-0.178, -0.111]\newline units: [-0.277, -0.014] & -0.201 [-0.252, -0.151]\newline units: [-0.384, -0.046] & -0.75 / -1.36 & infeasible \\
51 & qwen3.5-0.8b\newline BANKING77 sem. s17 b01\newline NPO /\allowbreak{} pilot & 0.608\newline 0.495 & +0.183 [+0.132, +0.233]\newline units: [+0.028, +0.392] & +0.106 [+0.029, +0.184]\newline units: [-0.075, +0.288] & 0.30 / 0.21 & infeasible \\
52 & qwen3.5-0.8b\newline BANKING77 sem. s17 b01\newline RMU /\allowbreak{} pilot & 0.193\newline 0.148 & -0.000 [-0.001, +0.000]\newline units: [-0.002, +0.001] & -0.000 [-0.001, +0.001]\newline units: [-0.002, +0.001] & -0.00 / -0.00 & no benefit \\
53 & qwen3.5-0.8b\newline BANKING77 sem. s17 b01\newline RMU /\allowbreak{} pilot & 0.608\newline 0.495 & +0.002 [+0.001, +0.003]\newline units: [-0.002, +0.006] & +0.001 [+0.000, +0.003]\newline units: [-0.001, +0.004] & 0.00 / 0.00 & accepted \\
54 & qwen3.5-0.8b\newline BANKING77 sem. s17 b01\newline Uniformization /\allowbreak{} pilot & 0.193\newline 0.148 & -0.437 [-0.466, -0.409]\newline units: [-0.567, -0.303] & -0.466 [-0.508, -0.426]\newline units: [-0.591, -0.337] & -2.26 / -3.16 & infeasible \\
55 & qwen3.5-0.8b\newline BANKING77 sem. s17 b01\newline Uniformization /\allowbreak{} pilot & 0.608\newline 0.495 & -0.135 [-0.173, -0.099]\newline units: [-0.286, -0.008] & -0.198 [-0.255, -0.141]\newline units: [-0.350, -0.119] & -0.22 / -0.40 & infeasible \\
56 & qwen3.5-0.8b\newline BANKING77 sem. s17 b02\newline GradAscent /\allowbreak{} replic. & 0.644\newline 0.636 & +0.185 [+0.152, +0.218]\newline units: [-0.001, +0.374] & +0.162 [+0.117, +0.210]\newline units: [-0.004, +0.327] & 0.29 / 0.25 & infeasible \\
57 & qwen3.5-0.8b\newline BANKING77 sem. s17 b02\newline GradDiff /\allowbreak{} replic. & 0.644\newline 0.636 & +0.185 [+0.153, +0.218]\newline units: [-0.000, +0.375] & +0.164 [+0.119, +0.213]\newline units: [-0.002, +0.330] & 0.29 / 0.26 & infeasible \\
58 & qwen3.5-0.8b\newline BANKING77 sem. s17 b02\newline NPO /\allowbreak{} replic. & 0.644\newline 0.636 & +0.204 [+0.170, +0.237]\newline units: [+0.005, +0.406] & +0.199 [+0.151, +0.249]\newline units: [+0.013, +0.384] & 0.32 / 0.31 & infeasible \\
59 & qwen3.5-0.8b\newline BANKING77 sem. s17 b02\newline SimNPO /\allowbreak{} replic. & 0.644\newline 0.636 & +0.449 [+0.412, +0.487]\newline units: [+0.143, +0.759] & +0.467 [+0.408, +0.524]\newline units: [+0.142, +0.743] & 0.70 / 0.73 & accepted \\
60 & qwen3.5-0.8b\newline BANKING77 sem. s17 b02\newline SimNPO /\allowbreak{} confirmation\_\allowbreak{}sensitivity & 0.225\newline 0.223 & +0.024 [+0.017, +0.032]\newline units: [+0.001, +0.045] & +0.021 [+0.010, +0.035]\newline units: [+0.000, +0.041] & 0.11 / 0.10 & accepted \\
\bottomrule
\end{tabular}
}% end resizebox
}% end font group
\end{table*}

% Panel 1, row IDs 61--75
\begin{table*}[p]
\ContinuedFloat
\centering
\caption[]{Complete method results (continued): oracle fidelity, IDs 61--75.}
{\footnotesize
\setlength{\tabcolsep}{3pt}
\renewcommand{\arraystretch}{1.12}
\resizebox{\textwidth}{!}{%
\begin{tabular}{
@{}
>{\raggedright\arraybackslash}p{0.65cm}
>{\raggedright\arraybackslash}p{4.3cm}
>{\raggedright\arraybackslash}p{1.6cm}
>{\raggedright\arraybackslash}p{4.4cm}
>{\raggedright\arraybackslash}p{4.4cm}
>{\raggedright\arraybackslash}p{1.5cm}
>{\raggedright\arraybackslash}p{2.5cm}
@{}
}
\toprule
\textbf{ID} & \textbf{Configuration} & \textbf{$D_{0,\mathrm{F}}$ / $D_{0,\mathrm{G}}$} & \textbf{$\Delta\operatorname{OD}_\mathrm{F}$ [rows]; unit CI} & \textbf{$\Delta\operatorname{OD}_\mathrm{G}$ [rows]; unit CI} & \textbf{$P_\mathrm{F}/P_\mathrm{G}$} & \textbf{Status} \\
\midrule
61 & qwen3.5-0.8b\newline BANKING77 sem. s17 b02\newline Uniformization /\allowbreak{} confirmation\_\allowbreak{}sensitivity & 0.225\newline 0.223 & +0.016 [+0.012, +0.020]\newline units: [+0.008, +0.023] & +0.018 [+0.011, +0.025]\newline units: [+0.006, +0.027] & 0.07 / 0.08 & accepted \\
62 & qwen3.5-0.8b\newline BANKING77 sem. s17 b02\newline GradDiff /\allowbreak{} confirmation\_\allowbreak{}sensitivity & 0.225\newline 0.223 & +0.082 [+0.061, +0.103]\newline units: [+0.004, +0.185] & +0.055 [+0.029, +0.084]\newline units: [+0.003, +0.122] & 0.36 / 0.25 & accepted \\
63 & qwen3.5-0.8b\newline BANKING77 sem. s17 b03\newline GradAscent /\allowbreak{} replic. & 0.465\newline 0.440 & -0.341 [-0.382, -0.300]\newline units: [-0.582, +0.017] & -0.375 [-0.443, -0.306]\newline units: [-0.638, -0.022] & -0.73 / -0.85 & infeasible \\
64 & qwen3.5-0.8b\newline BANKING77 sem. s17 b03\newline GradDiff /\allowbreak{} replic. & 0.465\newline 0.440 & +0.075 [+0.040, +0.111]\newline units: [-0.146, +0.336] & +0.048 [-0.014, +0.112]\newline units: [-0.213, +0.318] & 0.16 / 0.11 & infeasible \\
65 & qwen3.5-0.8b\newline BANKING77 sem. s17 b03\newline NPO /\allowbreak{} replic. & 0.465\newline 0.440 & +0.159 [+0.135, +0.186]\newline units: [+0.047, +0.318] & +0.156 [+0.111, +0.203]\newline units: [+0.027, +0.298] & 0.34 / 0.36 & accepted \\
66 & qwen3.5-0.8b\newline BANKING77 sem. s17 b03\newline SimNPO /\allowbreak{} replic. & 0.465\newline 0.440 & +0.282 [+0.248, +0.317]\newline units: [+0.089, +0.473] & +0.274 [+0.213, +0.338]\newline units: [+0.081, +0.468] & 0.61 / 0.62 & accepted \\
67 & qwen3.5-0.8b\newline BANKING77 sem. s17 b03\newline SimNPO /\allowbreak{} confirmation\_\allowbreak{}sensitivity & 0.087\newline 0.049 & +0.005 [+0.003, +0.009]\newline units: [+0.000, +0.012] & +0.003 [+0.001, +0.005]\newline units: [+0.000, +0.006] & 0.06 / 0.05 & accepted \\
68 & qwen3.5-0.8b\newline BANKING77 sem. s17 b03\newline Uniformization /\allowbreak{} confirmation\_\allowbreak{}sensitivity & 0.087\newline 0.049 & +0.011 [+0.007, +0.014]\newline units: [+0.003, +0.020] & +0.007 [+0.003, +0.012]\newline units: [+0.001, +0.017] & 0.12 / 0.15 & accepted \\
69 & qwen3.5-0.8b\newline BANKING77 sem. s17 b03\newline GradDiff /\allowbreak{} confirmation\_\allowbreak{}sensitivity & 0.087\newline 0.049 & +0.012 [+0.005, +0.020]\newline units: [-0.000, +0.029] & +0.005 [-0.001, +0.015]\newline units: [-0.001, +0.014] & 0.13 / 0.10 & positive, unconfirmed \\
70 & qwen3.5-0.8b\newline BANKING77 sem. s17 ex\_\allowbreak{}seed17\newline GradAscent /\allowbreak{} pilot & 0.030\newline 0.032 & -0.030 [-0.044, -0.016]\newline units: [-0.044, -0.016] & -0.030 [-0.052, -0.012]\newline units: [-0.053, -0.012] & -0.99 / -0.93 & infeasible \\
71 & qwen3.5-0.8b\newline BANKING77 sem. s17 ex\_\allowbreak{}seed17\newline GradDiff /\allowbreak{} pilot & 0.030\newline 0.032 & -0.016 [-0.024, -0.009]\newline units: [-0.024, -0.009] & -0.012 [-0.023, -0.004]\newline units: [-0.023, -0.004] & -0.53 / -0.37 & infeasible \\
72 & qwen3.5-0.8b\newline BANKING77 sem. s17 ex\_\allowbreak{}seed17\newline NPO /\allowbreak{} pilot & 0.030\newline 0.032 & -0.016 [-0.024, -0.009]\newline units: [-0.024, -0.009] & -0.011 [-0.022, -0.003]\newline units: [-0.022, -0.003] & -0.53 / -0.35 & infeasible \\
73 & qwen3.5-0.8b\newline BANKING77 sem. s17 ex\_\allowbreak{}seed17\newline RMU /\allowbreak{} pilot & 0.030\newline 0.032 & -0.000 [-0.000, +0.000]\newline units: [-0.000, +0.000] & +0.000 [-0.000, +0.000]\newline units: [-0.000, +0.000] & -0.00 / 0.00 & no benefit \\
74 & qwen3.5-0.8b\newline BANKING77 sem. s17 ex\_\allowbreak{}seed17\newline Uniformization /\allowbreak{} pilot & 0.030\newline 0.032 & -0.310 [-0.321, -0.300]\newline units: [-0.320, -0.300] & -0.311 [-0.325, -0.296]\newline units: [-0.325, -0.296] & -10.34 / -9.58 & infeasible \\
75 & qwen3.5-0.8b\newline BANKING77 sem. s42 b01\newline GradAscent /\allowbreak{} replic. & 0.558\newline 0.470 & +0.165 [+0.135, +0.196]\newline units: [+0.066, +0.353] & +0.119 [+0.075, +0.166]\newline units: [+0.006, +0.238] & 0.30 / 0.25 & infeasible \\
\bottomrule
\end{tabular}
}% end resizebox
}% end font group
\end{table*}

% Panel 1, row IDs 76--90
\begin{table*}[p]
\ContinuedFloat
\centering
\caption[]{Complete method results (continued): oracle fidelity, IDs 76--90.}
{\footnotesize
\setlength{\tabcolsep}{3pt}
\renewcommand{\arraystretch}{1.12}
\resizebox{\textwidth}{!}{%
\begin{tabular}{
@{}
>{\raggedright\arraybackslash}p{0.65cm}
>{\raggedright\arraybackslash}p{4.3cm}
>{\raggedright\arraybackslash}p{1.6cm}
>{\raggedright\arraybackslash}p{4.4cm}
>{\raggedright\arraybackslash}p{4.4cm}
>{\raggedright\arraybackslash}p{1.5cm}
>{\raggedright\arraybackslash}p{2.5cm}
@{}
}
\toprule
\textbf{ID} & \textbf{Configuration} & \textbf{$D_{0,\mathrm{F}}$ / $D_{0,\mathrm{G}}$} & \textbf{$\Delta\operatorname{OD}_\mathrm{F}$ [rows]; unit CI} & \textbf{$\Delta\operatorname{OD}_\mathrm{G}$ [rows]; unit CI} & \textbf{$P_\mathrm{F}/P_\mathrm{G}$} & \textbf{Status} \\
\midrule
76 & qwen3.5-0.8b\newline BANKING77 sem. s42 b01\newline GradDiff /\allowbreak{} replic. & 0.558\newline 0.470 & +0.161 [+0.132, +0.191]\newline units: [+0.062, +0.346] & +0.115 [+0.072, +0.161]\newline units: [+0.001, +0.230] & 0.29 / 0.24 & infeasible \\
77 & qwen3.5-0.8b\newline BANKING77 sem. s42 b01\newline NPO /\allowbreak{} replic. & 0.558\newline 0.470 & +0.153 [+0.126, +0.181]\newline units: [+0.072, +0.289] & +0.105 [+0.070, +0.142]\newline units: [+0.021, +0.189] & 0.27 / 0.22 & accepted \\
78 & qwen3.5-0.8b\newline BANKING77 sem. s42 b01\newline SimNPO /\allowbreak{} replic. & 0.558\newline 0.470 & +0.307 [+0.265, +0.347]\newline units: [+0.186, +0.518] & +0.240 [+0.174, +0.304]\newline units: [+0.116, +0.363] & 0.55 / 0.51 & accepted \\
79 & qwen3.5-0.8b\newline BANKING77 sem. s42 b01\newline SimNPO /\allowbreak{} confirmation\_\allowbreak{}sensitivity & 0.208\newline 0.186 & +0.040 [+0.032, +0.049]\newline units: [+0.004, +0.080] & +0.050 [+0.035, +0.066]\newline units: [+0.004, +0.098] & 0.19 / 0.27 & accepted \\
80 & qwen3.5-0.8b\newline BANKING77 sem. s42 b01\newline Uniformization /\allowbreak{} confirmation\_\allowbreak{}sensitivity & 0.208\newline 0.186 & +0.032 [+0.027, +0.037]\newline units: [+0.014, +0.049] & +0.026 [+0.018, +0.034]\newline units: [+0.008, +0.044] & 0.15 / 0.14 & accepted \\
81 & qwen3.5-0.8b\newline BANKING77 sem. s42 b01\newline GradDiff /\allowbreak{} confirmation\_\allowbreak{}sensitivity & 0.208\newline 0.186 & +0.054 [+0.034, +0.074]\newline units: [+0.017, +0.106] & +0.032 [+0.004, +0.061]\newline units: [-0.005, +0.078] & 0.26 / 0.17 & accepted \\
82 & qwen3.5-0.8b\newline BANKING77 sem. s42 b02\newline GradAscent /\allowbreak{} replic. & 0.676\newline 0.660 & +0.257 [+0.222, +0.293]\newline units: [+0.002, +0.518] & +0.250 [+0.200, +0.304]\newline units: [-0.002, +0.502] & 0.38 / 0.38 & infeasible \\
83 & qwen3.5-0.8b\newline BANKING77 sem. s42 b02\newline GradDiff /\allowbreak{} replic. & 0.676\newline 0.660 & +0.256 [+0.221, +0.292]\newline units: [+0.002, +0.516] & +0.249 [+0.198, +0.302]\newline units: [-0.001, +0.499] & 0.38 / 0.38 & infeasible \\
84 & qwen3.5-0.8b\newline BANKING77 sem. s42 b02\newline NPO /\allowbreak{} replic. & 0.676\newline 0.660 & +0.290 [+0.256, +0.326]\newline units: [+0.015, +0.571] & +0.277 [+0.226, +0.331]\newline units: [+0.005, +0.548] & 0.43 / 0.42 & accepted \\
85 & qwen3.5-0.8b\newline BANKING77 sem. s42 b02\newline SimNPO /\allowbreak{} replic. & 0.676\newline 0.660 & +0.319 [+0.283, +0.355]\newline units: [+0.008, +0.636] & +0.297 [+0.245, +0.350]\newline units: [+0.005, +0.589] & 0.47 / 0.45 & accepted \\
86 & qwen3.5-0.8b\newline BANKING77 sem. s42 b02\newline SimNPO /\allowbreak{} confirmation\_\allowbreak{}sensitivity & 0.170\newline 0.133 & +0.019 [+0.014, +0.025]\newline units: [+0.002, +0.035] & +0.016 [+0.009, +0.025]\newline units: [-0.000, +0.030] & 0.11 / 0.12 & accepted \\
87 & qwen3.5-0.8b\newline BANKING77 sem. s42 b02\newline Uniformization /\allowbreak{} confirmation\_\allowbreak{}sensitivity & 0.170\newline 0.133 & +0.018 [+0.014, +0.023]\newline units: [+0.006, +0.029] & +0.013 [+0.007, +0.020]\newline units: [+0.000, +0.024] & 0.11 / 0.10 & accepted \\
88 & qwen3.5-0.8b\newline BANKING77 sem. s42 b02\newline GradDiff /\allowbreak{} confirmation\_\allowbreak{}sensitivity & 0.170\newline 0.133 & -0.009 [-0.032, +0.014]\newline units: [-0.033, +0.010] & -0.041 [-0.080, -0.002]\newline units: [-0.105, +0.009] & -0.06 / -0.31 & no benefit \\
89 & qwen3.5-0.8b\newline BANKING77 sem. s42 b03\newline GradAscent /\allowbreak{} replic. & 0.435\newline 0.398 & -0.156 [-0.192, -0.120]\newline units: [-0.425, +0.119] & -0.210 [-0.272, -0.147]\newline units: [-0.474, +0.068] & -0.36 / -0.53 & infeasible \\
90 & qwen3.5-0.8b\newline BANKING77 sem. s42 b03\newline GradDiff /\allowbreak{} replic. & 0.435\newline 0.398 & +0.147 [+0.121, +0.173]\newline units: [+0.015, +0.316] & +0.129 [+0.083, +0.175]\newline units: [-0.007, +0.267] & 0.34 / 0.32 & accepted \\
\bottomrule
\end{tabular}
}% end resizebox
}% end font group
\end{table*}

% Panel 1, row IDs 91--105
\begin{table*}[p]
\ContinuedFloat
\centering
\caption[]{Complete method results (continued): oracle fidelity, IDs 91--105.}
{\footnotesize
\setlength{\tabcolsep}{3pt}
\renewcommand{\arraystretch}{1.12}
\resizebox{\textwidth}{!}{%
\begin{tabular}{
@{}
>{\raggedright\arraybackslash}p{0.65cm}
>{\raggedright\arraybackslash}p{4.3cm}
>{\raggedright\arraybackslash}p{1.6cm}
>{\raggedright\arraybackslash}p{4.4cm}
>{\raggedright\arraybackslash}p{4.4cm}
>{\raggedright\arraybackslash}p{1.5cm}
>{\raggedright\arraybackslash}p{2.5cm}
@{}
}
\toprule
\textbf{ID} & \textbf{Configuration} & \textbf{$D_{0,\mathrm{F}}$ / $D_{0,\mathrm{G}}$} & \textbf{$\Delta\operatorname{OD}_\mathrm{F}$ [rows]; unit CI} & \textbf{$\Delta\operatorname{OD}_\mathrm{G}$ [rows]; unit CI} & \textbf{$P_\mathrm{F}/P_\mathrm{G}$} & \textbf{Status} \\
\midrule
91 & qwen3.5-0.8b\newline BANKING77 sem. s42 b03\newline NPO /\allowbreak{} replic. & 0.435\newline 0.398 & -0.042 [-0.071, -0.012]\newline units: [-0.261, +0.176] & -0.087 [-0.140, -0.034]\newline units: [-0.314, +0.140] & -0.10 / -0.22 & infeasible \\
92 & qwen3.5-0.8b\newline BANKING77 sem. s42 b03\newline SimNPO /\allowbreak{} replic. & 0.435\newline 0.398 & +0.209 [+0.182, +0.237]\newline units: [+0.057, +0.359] & +0.195 [+0.152, +0.241]\newline units: [+0.078, +0.313] & 0.48 / 0.49 & accepted \\
93 & qwen3.5-0.8b\newline BANKING77 sem. s42 b03\newline SimNPO /\allowbreak{} confirmation\_\allowbreak{}sensitivity & 0.093\newline 0.075 & +0.002 [-0.001, +0.005]\newline units: [-0.002, +0.007] & -0.002 [-0.004, +0.000]\newline units: [-0.004, -0.000] & 0.02 / -0.03 & no benefit \\
94 & qwen3.5-0.8b\newline BANKING77 sem. s42 b03\newline Uniformization /\allowbreak{} confirmation\_\allowbreak{}sensitivity & 0.093\newline 0.075 & +0.007 [+0.004, +0.011]\newline units: [+0.002, +0.014] & +0.006 [+0.002, +0.010]\newline units: [-0.000, +0.013] & 0.08 / 0.08 & accepted \\
95 & qwen3.5-0.8b\newline BANKING77 sem. s42 b03\newline GradDiff /\allowbreak{} confirmation\_\allowbreak{}sensitivity & 0.093\newline 0.075 & +0.004 [-0.003, +0.011]\newline units: [-0.003, +0.013] & -0.003 [-0.009, +0.002]\newline units: [-0.005, -0.001] & 0.04 / -0.04 & no benefit \\
96 & qwen3.5-0.8b\newline BANKING77 sem. s99 b01\newline GradAscent /\allowbreak{} replic. & 0.610\newline 0.495 & +0.136 [+0.109, +0.166]\newline units: [+0.033, +0.291] & +0.119 [+0.076, +0.165]\newline units: [+0.027, +0.217] & 0.22 / 0.24 & accepted \\
97 & qwen3.5-0.8b\newline BANKING77 sem. s99 b01\newline GradDiff /\allowbreak{} replic. & 0.610\newline 0.495 & +0.138 [+0.110, +0.167]\newline units: [+0.034, +0.291] & +0.123 [+0.080, +0.167]\newline units: [+0.028, +0.218] & 0.23 / 0.25 & accepted \\
98 & qwen3.5-0.8b\newline BANKING77 sem. s99 b01\newline NPO /\allowbreak{} replic. & 0.610\newline 0.495 & +0.140 [+0.113, +0.167]\newline units: [+0.037, +0.291] & +0.131 [+0.090, +0.174]\newline units: [+0.034, +0.223] & 0.23 / 0.26 & accepted \\
99 & qwen3.5-0.8b\newline BANKING77 sem. s99 b01\newline SimNPO /\allowbreak{} replic. & 0.610\newline 0.495 & +0.236 [+0.204, +0.267]\newline units: [+0.095, +0.515] & +0.240 [+0.191, +0.290]\newline units: [+0.101, +0.405] & 0.39 / 0.48 & accepted \\
100 & qwen3.5-0.8b\newline BANKING77 sem. s99 b01\newline SimNPO /\allowbreak{} confirmation\_\allowbreak{}sensitivity & 0.230\newline 0.186 & +0.055 [+0.043, +0.069]\newline units: [+0.004, +0.137] & +0.061 [+0.040, +0.084]\newline units: [+0.002, +0.148] & 0.24 / 0.33 & accepted \\
101 & qwen3.5-0.8b\newline BANKING77 sem. s99 b01\newline Uniformization /\allowbreak{} confirmation\_\allowbreak{}sensitivity & 0.230\newline 0.186 & +0.024 [+0.020, +0.028]\newline units: [+0.008, +0.043] & +0.027 [+0.021, +0.034]\newline units: [+0.009, +0.048] & 0.10 / 0.15 & accepted \\
102 & qwen3.5-0.8b\newline BANKING77 sem. s99 b01\newline GradDiff /\allowbreak{} confirmation\_\allowbreak{}sensitivity & 0.230\newline 0.186 & +0.017 [+0.003, +0.032]\newline units: [+0.003, +0.036] & +0.023 [-0.001, +0.048]\newline units: [+0.002, +0.055] & 0.07 / 0.12 & positive, unconfirmed \\
103 & qwen3.5-0.8b\newline BANKING77 sem. s99 b02\newline GradAscent /\allowbreak{} replic. & 0.676\newline 0.648 & +0.185 [+0.154, +0.218]\newline units: [+0.004, +0.370] & +0.207 [+0.158, +0.259]\newline units: [+0.006, +0.408] & 0.27 / 0.32 & accepted \\
104 & qwen3.5-0.8b\newline BANKING77 sem. s99 b02\newline GradDiff /\allowbreak{} replic. & 0.676\newline 0.648 & +0.185 [+0.153, +0.218]\newline units: [+0.004, +0.370] & +0.206 [+0.157, +0.258]\newline units: [+0.006, +0.406] & 0.27 / 0.32 & accepted \\
105 & qwen3.5-0.8b\newline BANKING77 sem. s99 b02\newline NPO /\allowbreak{} replic. & 0.676\newline 0.648 & +0.196 [+0.164, +0.229]\newline units: [+0.005, +0.390] & +0.212 [+0.163, +0.264]\newline units: [+0.010, +0.414] & 0.29 / 0.33 & accepted \\
\bottomrule
\end{tabular}
}% end resizebox
}% end font group
\end{table*}

% Panel 1, row IDs 106--120
\begin{table*}[p]
\ContinuedFloat
\centering
\caption[]{Complete method results (continued): oracle fidelity, IDs 106--120.}
{\footnotesize
\setlength{\tabcolsep}{3pt}
\renewcommand{\arraystretch}{1.12}
\resizebox{\textwidth}{!}{%
\begin{tabular}{
@{}
>{\raggedright\arraybackslash}p{0.65cm}
>{\raggedright\arraybackslash}p{4.3cm}
>{\raggedright\arraybackslash}p{1.6cm}
>{\raggedright\arraybackslash}p{4.4cm}
>{\raggedright\arraybackslash}p{4.4cm}
>{\raggedright\arraybackslash}p{1.5cm}
>{\raggedright\arraybackslash}p{2.5cm}
@{}
}
\toprule
\textbf{ID} & \textbf{Configuration} & \textbf{$D_{0,\mathrm{F}}$ / $D_{0,\mathrm{G}}$} & \textbf{$\Delta\operatorname{OD}_\mathrm{F}$ [rows]; unit CI} & \textbf{$\Delta\operatorname{OD}_\mathrm{G}$ [rows]; unit CI} & \textbf{$P_\mathrm{F}/P_\mathrm{G}$} & \textbf{Status} \\
\midrule
106 & qwen3.5-0.8b\newline BANKING77 sem. s99 b02\newline SimNPO /\allowbreak{} replic. & 0.676\newline 0.648 & +0.469 [+0.432, +0.507]\newline units: [+0.150, +0.707] & +0.492 [+0.433, +0.550]\newline units: [+0.169, +0.740] & 0.69 / 0.76 & accepted \\
107 & qwen3.5-0.8b\newline BANKING77 sem. s99 b02\newline SimNPO /\allowbreak{} confirmation\_\allowbreak{}sensitivity & 0.197\newline 0.183 & +0.016 [+0.010, +0.023]\newline units: [+0.001, +0.030] & +0.019 [+0.010, +0.031]\newline units: [+0.002, +0.032] & 0.08 / 0.10 & accepted \\
108 & qwen3.5-0.8b\newline BANKING77 sem. s99 b02\newline Uniformization /\allowbreak{} confirmation\_\allowbreak{}sensitivity & 0.197\newline 0.183 & +0.014 [+0.011, +0.017]\newline units: [+0.007, +0.022] & +0.012 [+0.006, +0.018]\newline units: [+0.002, +0.021] & 0.07 / 0.06 & accepted \\
109 & qwen3.5-0.8b\newline BANKING77 sem. s99 b02\newline GradDiff /\allowbreak{} confirmation\_\allowbreak{}sensitivity & 0.197\newline 0.183 & +0.031 [+0.019, +0.044]\newline units: [+0.000, +0.075] & +0.011 [-0.007, +0.029]\newline units: [+0.002, +0.020] & 0.15 / 0.06 & positive, unconfirmed \\
110 & qwen3.5-0.8b\newline BANKING77 sem. s99 b03\newline GradAscent /\allowbreak{} replic. & 0.469\newline 0.454 & +0.206 [+0.174, +0.240]\newline units: [+0.004, +0.408] & +0.218 [+0.160, +0.279]\newline units: [-0.006, +0.442] & 0.44 / 0.48 & infeasible \\
111 & qwen3.5-0.8b\newline BANKING77 sem. s99 b03\newline GradDiff /\allowbreak{} replic. & 0.469\newline 0.454 & +0.188 [+0.158, +0.219]\newline units: [+0.012, +0.363] & +0.194 [+0.141, +0.251]\newline units: [+0.006, +0.383] & 0.40 / 0.43 & infeasible \\
112 & qwen3.5-0.8b\newline BANKING77 sem. s99 b03\newline NPO /\allowbreak{} replic. & 0.469\newline 0.454 & +0.119 [+0.096, +0.143]\newline units: [+0.021, +0.239] & +0.113 [+0.079, +0.150]\newline units: [+0.028, +0.198] & 0.25 / 0.25 & accepted \\
113 & qwen3.5-0.8b\newline BANKING77 sem. s99 b03\newline SimNPO /\allowbreak{} replic. & 0.469\newline 0.454 & +0.231 [+0.202, +0.260]\newline units: [+0.051, +0.410] & +0.236 [+0.187, +0.286]\newline units: [+0.055, +0.416] & 0.49 / 0.52 & accepted \\
114 & qwen3.5-0.8b\newline BANKING77 sem. s99 b03\newline SimNPO /\allowbreak{} confirmation\_\allowbreak{}sensitivity & 0.106\newline 0.088 & +0.004 [+0.002, +0.007]\newline units: [+0.001, +0.008] & +0.001 [-0.003, +0.004]\newline units: [-0.002, +0.005] & 0.04 / 0.01 & positive, unconfirmed \\
115 & qwen3.5-0.8b\newline BANKING77 sem. s99 b03\newline Uniformization /\allowbreak{} confirmation\_\allowbreak{}sensitivity & 0.106\newline 0.088 & +0.007 [+0.004, +0.010]\newline units: [+0.001, +0.014] & +0.002 [-0.003, +0.006]\newline units: [-0.004, +0.008] & 0.07 / 0.02 & positive, unconfirmed \\
116 & qwen3.5-0.8b\newline BANKING77 sem. s99 b03\newline GradDiff /\allowbreak{} confirmation\_\allowbreak{}sensitivity & 0.106\newline 0.088 & -0.001 [-0.008, +0.005]\newline units: [-0.011, +0.010] & -0.002 [-0.011, +0.007]\newline units: [-0.012, +0.007] & -0.01 / -0.02 & no benefit \\
117 & qwen3.5-2b\newline BANKING77 sem. s17 b01\newline SimNPO /\allowbreak{} screen & 0.497\newline 0.399 & +0.231 [+0.197, +0.266]\newline units: [+0.107, +0.351] & +0.195 [+0.133, +0.256]\newline units: [+0.053, +0.351] & 0.46 / 0.49 & accepted \\
118 & qwen3.5-4b\newline BANKING77 sem. s17 b01\newline SimNPO /\allowbreak{} screen & 0.505\newline 0.391 & +0.099 [+0.074, +0.126]\newline units: [+0.042, +0.150] & +0.071 [+0.027, +0.115]\newline units: [-0.003, +0.146] & 0.20 / 0.18 & accepted \\
119 & qwen3.5-9b\newline BANKING77 sem. s17 b01\newline SimNPO /\allowbreak{} screen & 0.414\newline 0.353 & +0.223 [+0.194, +0.254]\newline units: [+0.112, +0.422] & +0.206 [+0.164, +0.249]\newline units: [+0.082, +0.329] & 0.54 / 0.58 & accepted \\
120 & qwen3.5-9b\newline BANKING77 sem. s17 b02\newline SimNPO /\allowbreak{} replic. & 0.599\newline 0.591 & +0.261 [+0.222, +0.300]\newline units: [+0.036, +0.538] & +0.255 [+0.203, +0.310]\newline units: [+0.036, +0.507] & 0.44 / 0.43 & accepted \\
\bottomrule
\end{tabular}
}% end resizebox
}% end font group
\end{table*}

% Panel 1, row IDs 121--135
\begin{table*}[p]
\ContinuedFloat
\centering
\caption[]{Complete method results (continued): oracle fidelity, IDs 121--135.}
{\footnotesize
\setlength{\tabcolsep}{3pt}
\renewcommand{\arraystretch}{1.12}
\resizebox{\textwidth}{!}{%
\begin{tabular}{
@{}
>{\raggedright\arraybackslash}p{0.65cm}
>{\raggedright\arraybackslash}p{4.3cm}
>{\raggedright\arraybackslash}p{1.6cm}
>{\raggedright\arraybackslash}p{4.4cm}
>{\raggedright\arraybackslash}p{4.4cm}
>{\raggedright\arraybackslash}p{1.5cm}
>{\raggedright\arraybackslash}p{2.5cm}
@{}
}
\toprule
\textbf{ID} & \textbf{Configuration} & \textbf{$D_{0,\mathrm{F}}$ / $D_{0,\mathrm{G}}$} & \textbf{$\Delta\operatorname{OD}_\mathrm{F}$ [rows]; unit CI} & \textbf{$\Delta\operatorname{OD}_\mathrm{G}$ [rows]; unit CI} & \textbf{$P_\mathrm{F}/P_\mathrm{G}$} & \textbf{Status} \\
\midrule
121 & qwen3.5-9b\newline BANKING77 sem. s42 b01\newline SimNPO /\allowbreak{} explor. & 0.511\newline 0.425 & +0.094 [+0.057, +0.135]\newline units: [+0.015, +0.212] & +0.093 [+0.038, +0.147]\newline units: [-0.035, +0.225] & 0.18 / 0.22 & accepted \\
122 & qwen3.5-9b\newline BANKING77 sem. s42 b01\newline SimNPO /\allowbreak{} explor. & 0.511\newline 0.425 & +0.008 [-0.038, +0.057]\newline units: [-0.085, +0.173] & -0.033 [-0.109, +0.043]\newline units: [-0.191, +0.236] & 0.02 / -0.08 & no benefit \\
123 & qwen3.5-9b\newline BANKING77 sem. s42 b01\newline SimNPO /\allowbreak{} explor. & 0.511\newline 0.425 & +0.014 [-0.030, +0.061]\newline units: [-0.081, +0.149] & -0.027 [-0.094, +0.041]\newline units: [-0.207, +0.190] & 0.03 / -0.06 & no benefit \\
124 & qwen3.5-9b\newline BANKING77 sem. s42 b01\newline SimNPO /\allowbreak{} explor. & 0.511\newline 0.425 & +0.084 [+0.067, +0.103]\newline units: [+0.043, +0.156] & +0.084 [+0.060, +0.110]\newline units: [+0.037, +0.132] & 0.16 / 0.20 & accepted \\
125 & qwen3.5-9b\newline BANKING77 sem. s42 b01\newline SimNPO /\allowbreak{} explor. & 0.511\newline 0.425 & +0.068 [+0.023, +0.117]\newline units: [-0.042, +0.232] & +0.022 [-0.056, +0.099]\newline units: [-0.130, +0.277] & 0.13 / 0.05 & positive, unconfirmed \\
126 & qwen3.5-9b\newline BANKING77 sem. s42 b01\newline SimNPO /\allowbreak{} replic. & 0.511\newline 0.425 & +0.070 [+0.055, +0.087]\newline units: [+0.045, +0.116] & +0.076 [+0.051, +0.103]\newline units: [+0.031, +0.141] & 0.14 / 0.18 & accepted \\
127 & smollm3-3b\newline BANKING77 sem. s17 b01\newline SimNPO /\allowbreak{} screen & 0.414\newline 0.319 & +0.205 [+0.174, +0.237]\newline units: [+0.081, +0.337] & +0.148 [+0.097, +0.199]\newline units: [+0.063, +0.232] & 0.50 / 0.46 & accepted \\
128 & llama3.2-1b\newline CLINC150 sem. s17 b01\newline SimNPO /\allowbreak{} transfer & 0.453\newline 0.404 & -0.079 [-0.125, -0.034]\newline units: [-0.424, +0.281] & -0.082 [-0.159, -0.006]\newline units: [-0.403, +0.241] & -0.18 / -0.20 & no benefit \\
129 & qwen3.5-0.8b\newline CLINC150 opaque s17 b02\newline SimNPO /\allowbreak{} explor. & 0.983\newline 0.976 & +0.275 [+0.247, +0.304]\newline units: [+0.079, +0.538] & +0.287 [+0.237, +0.340]\newline units: [+0.023, +0.560] & 0.28 / 0.29 & accepted \\
130 & qwen3.5-0.8b\newline CLINC150 opaque s17 b03\newline SimNPO /\allowbreak{} explor. & 0.993\newline 0.973 & +0.132 [+0.111, +0.153]\newline units: [+0.007, +0.342] & +0.111 [+0.076, +0.148]\newline units: [+0.002, +0.299] & 0.13 / 0.11 & accepted \\
131 & qwen3.5-0.8b\newline CLINC150 opaque s17 b06\newline SimNPO /\allowbreak{} explor. & 0.999\newline 0.989 & +0.023 [+0.017, +0.031]\newline units: [+0.006, +0.046] & +0.024 [+0.012, +0.039]\newline units: [+0.009, +0.043] & 0.02 / 0.02 & accepted \\
132 & qwen3.5-0.8b\newline CLINC150 sem. s17 b01\newline GradAscent /\allowbreak{} transfer & 0.416\newline 0.356 & -0.185 [-0.217, -0.152]\newline units: [-0.404, -0.001] & -0.168 [-0.228, -0.109]\newline units: [-0.387, +0.021] & -0.44 / -0.47 & no benefit \\
133 & qwen3.5-0.8b\newline CLINC150 sem. s17 b01\newline GradDiff /\allowbreak{} transfer & 0.416\newline 0.356 & -0.102 [-0.130, -0.075]\newline units: [-0.261, -0.005] & -0.073 [-0.121, -0.025]\newline units: [-0.239, +0.027] & -0.25 / -0.20 & no benefit \\
134 & qwen3.5-0.8b\newline CLINC150 sem. s17 b01\newline NPO /\allowbreak{} transfer & 0.416\newline 0.356 & -0.183 [-0.214, -0.151]\newline units: [-0.395, -0.011] & -0.158 [-0.216, -0.099]\newline units: [-0.369, +0.022] & -0.44 / -0.44 & no benefit \\
135 & qwen3.5-0.8b\newline CLINC150 sem. s17 b01\newline SimNPO /\allowbreak{} transfer & 0.058\newline 0.079 & +0.002 [+0.000, +0.005]\newline units: [+0.000, +0.005] & +0.006 [+0.001, +0.011]\newline units: [-0.002, +0.014] & 0.04 / 0.07 & accepted \\
\bottomrule
\end{tabular}
}% end resizebox
}% end font group
\end{table*}

% Panel 1, row IDs 136--150
\begin{table*}[p]
\ContinuedFloat
\centering
\caption[]{Complete method results (continued): oracle fidelity, IDs 136--150.}
{\footnotesize
\setlength{\tabcolsep}{3pt}
\renewcommand{\arraystretch}{1.12}
\resizebox{\textwidth}{!}{%
\begin{tabular}{
@{}
>{\raggedright\arraybackslash}p{0.65cm}
>{\raggedright\arraybackslash}p{4.3cm}
>{\raggedright\arraybackslash}p{1.6cm}
>{\raggedright\arraybackslash}p{4.4cm}
>{\raggedright\arraybackslash}p{4.4cm}
>{\raggedright\arraybackslash}p{1.5cm}
>{\raggedright\arraybackslash}p{2.5cm}
@{}
}
\toprule
\textbf{ID} & \textbf{Configuration} & \textbf{$D_{0,\mathrm{F}}$ / $D_{0,\mathrm{G}}$} & \textbf{$\Delta\operatorname{OD}_\mathrm{F}$ [rows]; unit CI} & \textbf{$\Delta\operatorname{OD}_\mathrm{G}$ [rows]; unit CI} & \textbf{$P_\mathrm{F}/P_\mathrm{G}$} & \textbf{Status} \\
\midrule
136 & qwen3.5-0.8b\newline CLINC150 sem. s17 b01\newline SimNPO /\allowbreak{} transfer & 0.416\newline 0.356 & -0.191 [-0.242, -0.138]\newline units: [-0.577, +0.216] & -0.237 [-0.328, -0.148]\newline units: [-0.628, +0.182] & -0.46 / -0.67 & no benefit \\
137 & qwen3.5-0.8b\newline CLINC150 sem. s17 b01\newline Uniformization /\allowbreak{} transfer & 0.058\newline 0.079 & +0.003 [+0.002, +0.005]\newline units: [+0.001, +0.007] & +0.014 [+0.007, +0.023]\newline units: [+0.002, +0.029] & 0.06 / 0.18 & accepted \\
138 & qwen3.5-0.8b\newline CLINC150 sem. s17 b02\newline SimNPO /\allowbreak{} explor. & 0.298\newline 0.319 & -0.117 [-0.152, -0.083]\newline units: [-0.297, +0.080] & -0.111 [-0.178, -0.044]\newline units: [-0.383, +0.186] & -0.39 / -0.35 & no benefit \\
139 & qwen3.5-0.8b\newline CLINC150 sem. s17 b02\newline SimNPO /\allowbreak{} explor. & 0.298\newline 0.319 & -0.051 [-0.082, -0.020]\newline units: [-0.219, +0.138] & -0.040 [-0.100, +0.022]\newline units: [-0.290, +0.241] & -0.17 / -0.13 & no benefit \\
140 & qwen3.5-0.8b\newline CLINC150 sem. s17 b02\newline SimNPO /\allowbreak{} explor. & 0.298\newline 0.319 & +0.049 [+0.025, +0.074]\newline units: [-0.085, +0.205] & +0.074 [+0.025, +0.124]\newline units: [-0.119, +0.309] & 0.17 / 0.23 & accepted \\
141 & qwen3.5-0.8b\newline CLINC150 sem. s17 b02\newline temperature\_\allowbreak{}control /\allowbreak{} control & 0.298\newline 0.319 & +0.020 [+0.018, +0.023]\newline units: [+0.009, +0.033] & +0.032 [+0.025, +0.041]\newline units: [+0.015, +0.050] & 0.07 / 0.10 & accepted \\
142 & qwen3.5-0.8b\newline CLINC150 sem. s17 b02\newline SimNPO /\allowbreak{} control & 0.298\newline 0.319 & -0.018 [-0.046, +0.011]\newline units: [-0.171, +0.157] & +0.009 [-0.048, +0.067]\newline units: [-0.218, +0.270] & -0.06 / 0.03 & no benefit \\
143 & qwen3.5-0.8b\newline CLINC150 sem. s17 b02\newline SimNPO /\allowbreak{} confirm. & 0.298\newline 0.319 & +0.065 [+0.046, +0.085]\newline units: [-0.032, +0.181] & +0.088 [+0.049, +0.129]\newline units: [-0.058, +0.271] & 0.22 / 0.28 & accepted \\
144 & qwen3.5-0.8b\newline CLINC150 sem. s17 b03\newline temperature\_\allowbreak{}control /\allowbreak{} control & 0.454\newline 0.407 & +0.024 [+0.021, +0.027]\newline units: [+0.007, +0.046] & +0.030 [+0.023, +0.037]\newline units: [+0.013, +0.055] & 0.05 / 0.07 & accepted \\
145 & qwen3.5-0.8b\newline CLINC150 sem. s17 b03\newline SimNPO /\allowbreak{} confirm. & 0.454\newline 0.407 & +0.102 [+0.082, +0.121]\newline units: [+0.001, +0.289] & +0.083 [+0.052, +0.115]\newline units: [-0.006, +0.249] & 0.22 / 0.20 & accepted \\
146 & qwen3.5-0.8b\newline CLINC150 sem. s17 b04\newline SimNPO /\allowbreak{} replic. & 0.481\newline 0.410 & +0.061 [+0.034, +0.089]\newline units: [-0.192, +0.317] & +0.033 [-0.012, +0.080]\newline units: [-0.200, +0.274] & 0.13 / 0.08 & positive, unconfirmed \\
147 & qwen3.5-0.8b\newline CLINC150 sem. s17 b05\newline SimNPO /\allowbreak{} replic. & 0.344\newline 0.322 & +0.025 [-0.003, +0.052]\newline units: [-0.122, +0.220] & +0.022 [-0.032, +0.078]\newline units: [-0.182, +0.276] & 0.07 / 0.07 & positive, unconfirmed \\
148 & qwen3.5-0.8b\newline CLINC150 sem. s17 b06\newline SimNPO /\allowbreak{} replic. & 0.207\newline 0.233 & -0.050 [-0.070, -0.030]\newline units: [-0.234, +0.074] & -0.020 [-0.057, +0.016]\newline units: [-0.176, +0.093] & -0.24 / -0.09 & no benefit \\
149 & qwen3.5-0.8b\newline CLINC150 sem. s17 b07\newline SimNPO /\allowbreak{} replic. & 0.289\newline 0.301 & -0.071 [-0.087, -0.054]\newline units: [-0.148, +0.001] & -0.051 [-0.080, -0.021]\newline units: [-0.137, +0.011] & -0.24 / -0.17 & no benefit \\
150 & qwen3.5-0.8b\newline CLINC150 sem. s42 b02\newline SimNPO /\allowbreak{} replic. & 0.346\newline 0.349 & +0.062 [+0.042, +0.083]\newline units: [-0.012, +0.140] & +0.040 [+0.006, +0.074]\newline units: [-0.071, +0.126] & 0.18 / 0.11 & accepted \\
\bottomrule
\end{tabular}
}% end resizebox
\par\smallskip
\parbox{\textwidth}{\footnotesize Intervals on the first line are paired-row intervals; unit intervals resample forgotten classes or association clusters. Configuration labels are abbreviated; IDs identify the complete original run labels in the CSV. Some source keys omit granularity. Screening status is not a simultaneous guarantee across architectures.}
}% end font group
\end{table*}

% Panel 2, row IDs 1--30
\begin{table*}[p]
\ContinuedFloat
\centering
\caption[]{Complete method results (continued): retained utility and evaluation counts. IDs 1--30.}
{\footnotesize
\setlength{\tabcolsep}{3pt}
\renewcommand{\arraystretch}{1.12}
\begin{tabular}{
@{}
>{\raggedright\arraybackslash}p{1cm}
>{\raggedright\arraybackslash}p{2.9cm}
>{\raggedright\arraybackslash}p{2.5cm}
>{\raggedright\arraybackslash}p{3cm}
>{\raggedright\arraybackslash}p{3cm}
>{\raggedright\arraybackslash}p{2cm}
@{}
}
\toprule
\textbf{ID} & \textbf{$\Delta\operatorname{Acc}_\mathrm{R}$ (pp)} & \textbf{$\operatorname{ECE}_\mathrm{R}$} & \textbf{$\Delta\operatorname{ECE}_\mathrm{R}$} & \textbf{Rows F / G} & \textbf{Units} \\
\midrule
1 & -0.3 & 0.014 & -0.005 & 428/\allowbreak{}160 & 4 \\
2 & -0.1 & 0.021 & -0.001 & 359/\allowbreak{}160 & 4 \\
3 & -0.4 & 0.029 & +0.001 & 428/\allowbreak{}160 & 4 \\
4 & +0.0 & 0.022 & -0.006 & 428/\allowbreak{}160 & 4 \\
5 & -0.3 & 0.034 & +0.006 & 428/\allowbreak{}160 & 4 \\
6 & -0.6 & 0.025 & -0.003 & 428/\allowbreak{}160 & 4 \\
7 & -0.3 & 0.020 & -0.008 & 428/\allowbreak{}160 & 4 \\
8 & +0.0 & 0.028 & -0.000 & 428/\allowbreak{}160 & 4 \\
9 & +0.3 & 0.025 & -0.003 & 428/\allowbreak{}160 & 4 \\
10 & +1.0 & 0.024 & -0.007 & 428/\allowbreak{}160 & 4 \\
11 & -0.6 & 0.033 & +0.004 & 428/\allowbreak{}160 & 4 \\
12 & -0.3 & 0.041 & +0.010 & 428/\allowbreak{}160 & 4 \\
13 & -0.2 & 0.026 & +0.003 & 428/\allowbreak{}160 & 4 \\
14 & -0.4 & 0.031 & +0.003 & 428/\allowbreak{}160 & 4 \\
15 & +0.4 & 0.032 & +0.003 & 428/\allowbreak{}160 & 4 \\
16 & +0.7 & 0.027 & -0.004 & 428/\allowbreak{}160 & 4 \\
17 & +0.1 & 0.018 & +0.000 & 428/\allowbreak{}160 & 4 \\
18 & +0.0 & 0.023 & -0.001 & 382/\allowbreak{}160 & 4 \\
19 & +0.4 & 0.026 & -0.002 & 428/\allowbreak{}160 & 4 \\
20 & +1.4 & 0.020 & -0.007 & 428/\allowbreak{}160 & 4 \\
21 & +0.2 & 0.035 & +0.004 & 428/\allowbreak{}160 & 4 \\
22 & +0.1 & 0.024 & -0.001 & 428/\allowbreak{}160 & 4 \\
23 & +0.0 & 0.034 & +0.006 & 428/\allowbreak{}160 & 4 \\
24 & +0.1 & 0.022 & -0.005 & 428/\allowbreak{}160 & 4 \\
25 & -0.1 & 0.025 & -0.003 & 359/\allowbreak{}160 & 4 \\
26 & +0.1 & 0.016 & -0.007 & 428/\allowbreak{}160 & 4 \\
27 & +0.3 & 0.033 & +0.003 & 428/\allowbreak{}160 & 4 \\
28 & -0.3 & 0.040 & +0.006 & 382/\allowbreak{}160 & 4 \\
29 & -1.9 & 0.028 & -0.001 & 428/\allowbreak{}160 & 4 \\
30 & -0.7 & 0.025 & -0.005 & 382/\allowbreak{}160 & 4 \\
\bottomrule
\end{tabular}
}% end font group
\end{table*}

% Panel 2, row IDs 31--60
\begin{table*}[p]
\ContinuedFloat
\centering
\caption[]{Complete method results (continued): retained utility and evaluation counts. IDs 31--60.}
{\footnotesize
\setlength{\tabcolsep}{3pt}
\renewcommand{\arraystretch}{1.12}
\begin{tabular}{
@{}
>{\raggedright\arraybackslash}p{1cm}
>{\raggedright\arraybackslash}p{2.9cm}
>{\raggedright\arraybackslash}p{2.5cm}
>{\raggedright\arraybackslash}p{3cm}
>{\raggedright\arraybackslash}p{3cm}
>{\raggedright\arraybackslash}p{2cm}
@{}
}
\toprule
\textbf{ID} & \textbf{$\Delta\operatorname{Acc}_\mathrm{R}$ (pp)} & \textbf{$\operatorname{ECE}_\mathrm{R}$} & \textbf{$\Delta\operatorname{ECE}_\mathrm{R}$} & \textbf{Rows F / G} & \textbf{Units} \\
\midrule
31 & -0.1 & 0.039 & +0.013 & 428/\allowbreak{}160 & 4 \\
32 & +0.2 & 0.037 & +0.011 & 428/\allowbreak{}160 & 4 \\
33 & -2.1 & 0.061 & +0.035 & 428/\allowbreak{}160 & 4 \\
34 & +0.2 & 0.029 & +0.003 & 428/\allowbreak{}160 & 4 \\
35 & -0.1 & 0.028 & +0.001 & 428/\allowbreak{}160 & 4 \\
36 & +0.2 & 0.072 & +0.046 & 428/\allowbreak{}160 & 4 \\
37 & -2.1 & 0.047 & +0.019 & 427/\allowbreak{}175 & 10 \\
38 & -0.1 & 0.023 & -0.005 & 427/\allowbreak{}175 & 10 \\
39 & -0.2 & 0.016 & -0.012 & 427/\allowbreak{}175 & 10 \\
40 & -0.9 & 0.045 & +0.017 & 427/\allowbreak{}175 & 10 \\
41 & -2.0 & 0.031 & +0.003 & 427/\allowbreak{}175 & 10 \\
42 & -9.6 & 0.126 & +0.098 & 427/\allowbreak{}175 & 10 \\
43 & -2.2 & 0.064 & +0.038 & 428/\allowbreak{}160 & 4 \\
44 & -4.2 & 0.085 & +0.057 & 427/\allowbreak{}175 & 10 \\
45 & -1.4 & 0.065 & +0.039 & 428/\allowbreak{}160 & 4 \\
46 & -3.0 & 0.072 & +0.046 & 428/\allowbreak{}160 & 4 \\
47 & +0.0 & 0.023 & -0.003 & 428/\allowbreak{}160 & 4 \\
48 & -0.2 & 0.024 & -0.002 & 428/\allowbreak{}160 & 4 \\
49 & +0.1 & 0.027 & +0.000 & 428/\allowbreak{}160 & 4 \\
50 & -4.0 & 0.071 & +0.043 & 427/\allowbreak{}175 & 10 \\
51 & -1.0 & 0.050 & +0.024 & 428/\allowbreak{}160 & 4 \\
52 & -0.2 & 0.028 & -0.000 & 427/\allowbreak{}175 & 10 \\
53 & -0.1 & 0.024 & -0.002 & 428/\allowbreak{}160 & 4 \\
54 & -0.8 & 0.181 & +0.153 & 427/\allowbreak{}175 & 10 \\
55 & +0.7 & 0.079 & +0.053 & 428/\allowbreak{}160 & 4 \\
56 & -0.5 & 0.062 & +0.031 & 382/\allowbreak{}160 & 4 \\
57 & -0.4 & 0.057 & +0.027 & 382/\allowbreak{}160 & 4 \\
58 & -0.4 & 0.052 & +0.022 & 382/\allowbreak{}160 & 4 \\
59 & +0.3 & 0.018 & -0.012 & 382/\allowbreak{}160 & 4 \\
60 & -0.2 & 0.033 & +0.001 & 407/\allowbreak{}146 & 6 \\
\bottomrule
\end{tabular}
}% end font group
\end{table*}

% Panel 2, row IDs 61--90
\begin{table*}[p]
\ContinuedFloat
\centering
\caption[]{Complete method results (continued): retained utility and evaluation counts. IDs 61--90.}
{\footnotesize
\setlength{\tabcolsep}{3pt}
\renewcommand{\arraystretch}{1.12}
\begin{tabular}{
@{}
>{\raggedright\arraybackslash}p{1cm}
>{\raggedright\arraybackslash}p{2.9cm}
>{\raggedright\arraybackslash}p{2.5cm}
>{\raggedright\arraybackslash}p{3cm}
>{\raggedright\arraybackslash}p{3cm}
>{\raggedright\arraybackslash}p{2cm}
@{}
}
\toprule
\textbf{ID} & \textbf{$\Delta\operatorname{Acc}_\mathrm{R}$ (pp)} & \textbf{$\operatorname{ECE}_\mathrm{R}$} & \textbf{$\Delta\operatorname{ECE}_\mathrm{R}$} & \textbf{Rows F / G} & \textbf{Units} \\
\midrule
61 & +0.0 & 0.015 & -0.017 & 407/\allowbreak{}146 & 6 \\
62 & +0.3 & 0.034 & +0.002 & 407/\allowbreak{}146 & 6 \\
63 & -0.7 & 0.638 & +0.609 & 494/\allowbreak{}160 & 4 \\
64 & -0.4 & 0.144 & +0.115 & 494/\allowbreak{}160 & 4 \\
65 & -0.3 & 0.037 & +0.008 & 494/\allowbreak{}160 & 4 \\
66 & -0.1 & 0.039 & +0.010 & 494/\allowbreak{}160 & 4 \\
67 & +0.5 & 0.024 & -0.001 & 386/\allowbreak{}149 & 8 \\
68 & +0.3 & 0.016 & -0.009 & 386/\allowbreak{}149 & 8 \\
69 & +0.5 & 0.037 & +0.012 & 386/\allowbreak{}149 & 8 \\
70 & -1.4 & 0.057 & +0.030 & 386/\allowbreak{}147 & 386 \\
71 & -0.5 & 0.051 & +0.024 & 386/\allowbreak{}147 & 386 \\
72 & -0.5 & 0.051 & +0.024 & 386/\allowbreak{}147 & 386 \\
73 & +0.0 & 0.029 & +0.002 & 386/\allowbreak{}147 & 386 \\
74 & +0.6 & 0.491 & +0.464 & 386/\allowbreak{}147 & 386 \\
75 & +0.4 & 0.050 & +0.024 & 428/\allowbreak{}160 & 4 \\
76 & +0.2 & 0.047 & +0.022 & 428/\allowbreak{}160 & 4 \\
77 & +0.1 & 0.039 & +0.014 & 428/\allowbreak{}160 & 4 \\
78 & -0.9 & 0.021 & -0.004 & 428/\allowbreak{}160 & 4 \\
79 & +0.4 & 0.035 & +0.007 & 427/\allowbreak{}175 & 10 \\
80 & +0.3 & 0.013 & -0.015 & 427/\allowbreak{}175 & 10 \\
81 & +0.5 & 0.022 & -0.006 & 427/\allowbreak{}175 & 10 \\
82 & -0.7 & 0.058 & +0.028 & 382/\allowbreak{}160 & 4 \\
83 & -0.7 & 0.056 & +0.026 & 382/\allowbreak{}160 & 4 \\
84 & -0.6 & 0.039 & +0.010 & 382/\allowbreak{}160 & 4 \\
85 & +0.9 & 0.031 & +0.001 & 382/\allowbreak{}160 & 4 \\
86 & +0.3 & 0.027 & +0.000 & 407/\allowbreak{}146 & 6 \\
87 & +0.0 & 0.010 & -0.017 & 407/\allowbreak{}146 & 6 \\
88 & +0.1 & 0.045 & +0.018 & 407/\allowbreak{}146 & 6 \\
89 & -1.6 & 0.317 & +0.290 & 494/\allowbreak{}160 & 4 \\
90 & -0.4 & 0.040 & +0.013 & 494/\allowbreak{}160 & 4 \\
\bottomrule
\end{tabular}
}% end font group
\end{table*}

% Panel 2, row IDs 91--120
\begin{table*}[p]
\ContinuedFloat
\centering
\caption[]{Complete method results (continued): retained utility and evaluation counts. IDs 91--120.}
{\footnotesize
\setlength{\tabcolsep}{3pt}
\renewcommand{\arraystretch}{1.12}
\begin{tabular}{
@{}
>{\raggedright\arraybackslash}p{1cm}
>{\raggedright\arraybackslash}p{2.9cm}
>{\raggedright\arraybackslash}p{2.5cm}
>{\raggedright\arraybackslash}p{3cm}
>{\raggedright\arraybackslash}p{3cm}
>{\raggedright\arraybackslash}p{2cm}
@{}
}
\toprule
\textbf{ID} & \textbf{$\Delta\operatorname{Acc}_\mathrm{R}$ (pp)} & \textbf{$\operatorname{ECE}_\mathrm{R}$} & \textbf{$\Delta\operatorname{ECE}_\mathrm{R}$} & \textbf{Rows F / G} & \textbf{Units} \\
\midrule
91 & -1.4 & 0.218 & +0.191 & 494/\allowbreak{}160 & 4 \\
92 & -0.1 & 0.036 & +0.009 & 494/\allowbreak{}160 & 4 \\
93 & +0.2 & 0.030 & +0.006 & 386/\allowbreak{}149 & 8 \\
94 & -0.1 & 0.011 & -0.013 & 386/\allowbreak{}149 & 8 \\
95 & +0.6 & 0.032 & +0.008 & 386/\allowbreak{}149 & 8 \\
96 & +0.1 & 0.055 & +0.020 & 428/\allowbreak{}160 & 4 \\
97 & +0.0 & 0.049 & +0.014 & 428/\allowbreak{}160 & 4 \\
98 & +0.1 & 0.032 & -0.002 & 428/\allowbreak{}160 & 4 \\
99 & +0.1 & 0.015 & -0.020 & 428/\allowbreak{}160 & 4 \\
100 & +0.0 & 0.033 & +0.000 & 427/\allowbreak{}175 & 10 \\
101 & -0.5 & 0.022 & -0.010 & 427/\allowbreak{}175 & 10 \\
102 & -1.0 & 0.049 & +0.016 & 427/\allowbreak{}175 & 10 \\
103 & -0.5 & 0.050 & +0.013 & 382/\allowbreak{}160 & 4 \\
104 & -0.4 & 0.050 & +0.013 & 382/\allowbreak{}160 & 4 \\
105 & +0.3 & 0.043 & +0.006 & 382/\allowbreak{}160 & 4 \\
106 & -0.9 & 0.030 & -0.007 & 382/\allowbreak{}160 & 4 \\
107 & +0.0 & 0.038 & +0.001 & 407/\allowbreak{}146 & 6 \\
108 & -0.3 & 0.020 & -0.017 & 407/\allowbreak{}146 & 6 \\
109 & -0.1 & 0.042 & +0.005 & 407/\allowbreak{}146 & 6 \\
110 & -0.8 & 0.069 & +0.034 & 494/\allowbreak{}160 & 4 \\
111 & -0.7 & 0.059 & +0.025 & 494/\allowbreak{}160 & 4 \\
112 & -0.9 & 0.041 & +0.006 & 494/\allowbreak{}160 & 4 \\
113 & -1.1 & 0.031 & -0.004 & 494/\allowbreak{}160 & 4 \\
114 & -0.7 & 0.031 & -0.000 & 386/\allowbreak{}149 & 8 \\
115 & -0.5 & 0.021 & -0.010 & 386/\allowbreak{}149 & 8 \\
116 & -0.3 & 0.045 & +0.014 & 386/\allowbreak{}149 & 8 \\
117 & -0.8 & 0.040 & +0.016 & 428/\allowbreak{}160 & 4 \\
118 & -0.6 & 0.026 & +0.003 & 428/\allowbreak{}160 & 4 \\
119 & +0.4 & 0.026 & +0.001 & 428/\allowbreak{}160 & 4 \\
120 & +0.4 & 0.021 & -0.007 & 382/\allowbreak{}160 & 4 \\
\bottomrule
\end{tabular}
}% end font group
\end{table*}

% Panel 2, row IDs 121--150
\begin{table*}[p]
\ContinuedFloat
\centering
\caption[]{Complete method results (continued): retained utility and evaluation counts. IDs 121--150.}
{\footnotesize
\setlength{\tabcolsep}{3pt}
\renewcommand{\arraystretch}{1.12}
\begin{tabular}{
@{}
>{\raggedright\arraybackslash}p{1cm}
>{\raggedright\arraybackslash}p{2.9cm}
>{\raggedright\arraybackslash}p{2.5cm}
>{\raggedright\arraybackslash}p{3cm}
>{\raggedright\arraybackslash}p{3cm}
>{\raggedright\arraybackslash}p{2cm}
@{}
}
\toprule
\textbf{ID} & \textbf{$\Delta\operatorname{Acc}_\mathrm{R}$ (pp)} & \textbf{$\operatorname{ECE}_\mathrm{R}$} & \textbf{$\Delta\operatorname{ECE}_\mathrm{R}$} & \textbf{Rows F / G} & \textbf{Units} \\
\midrule
121 & -1.2 & 0.028 & +0.001 & 428/\allowbreak{}160 & 4 \\
122 & -0.7 & 0.022 & -0.005 & 428/\allowbreak{}160 & 4 \\
123 & -0.9 & 0.022 & -0.005 & 428/\allowbreak{}160 & 4 \\
124 & -1.0 & 0.023 & -0.004 & 428/\allowbreak{}160 & 4 \\
125 & -0.5 & 0.023 & -0.004 & 428/\allowbreak{}160 & 4 \\
126 & -0.9 & 0.021 & -0.005 & 428/\allowbreak{}160 & 4 \\
127 & -0.2 & 0.030 & +0.000 & 428/\allowbreak{}160 & 4 \\
128 & -0.2 & 0.010 & -0.011 & 785/\allowbreak{}240 & 8 \\
129 & -0.1 & 0.022 & -0.005 & 791/\allowbreak{}240 & 8 \\
130 & +0.2 & 0.021 & -0.008 & 780/\allowbreak{}240 & 8 \\
131 & +0.5 & 0.015 & -0.015 & 789/\allowbreak{}240 & 8 \\
132 & -1.0 & 0.041 & +0.017 & 785/\allowbreak{}240 & 8 \\
133 & -1.1 & 0.040 & +0.015 & 785/\allowbreak{}240 & 8 \\
134 & -1.1 & 0.040 & +0.016 & 785/\allowbreak{}240 & 8 \\
135 & +0.0 & 0.025 & +0.001 & 768/\allowbreak{}236 & 18 \\
136 & -0.4 & 0.015 & -0.009 & 785/\allowbreak{}240 & 8 \\
137 & -0.1 & 0.021 & -0.003 & 768/\allowbreak{}236 & 18 \\
138 & -0.9 & 0.022 & -0.001 & 791/\allowbreak{}240 & 8 \\
139 & -0.2 & 0.024 & +0.001 & 791/\allowbreak{}240 & 8 \\
140 & +0.3 & 0.015 & -0.009 & 791/\allowbreak{}240 & 8 \\
141 & +0.0 & 0.026 & +0.002 & 791/\allowbreak{}240 & 8 \\
142 & +0.3 & 0.013 & -0.010 & 791/\allowbreak{}240 & 8 \\
143 & +0.4 & 0.011 & -0.012 & 791/\allowbreak{}240 & 8 \\
144 & +0.0 & 0.027 & +0.003 & 780/\allowbreak{}240 & 8 \\
145 & +0.1 & 0.021 & -0.002 & 780/\allowbreak{}240 & 8 \\
146 & +0.3 & 0.014 & -0.010 & 778/\allowbreak{}240 & 8 \\
147 & +0.1 & 0.019 & -0.001 & 790/\allowbreak{}240 & 8 \\
148 & +0.1 & 0.020 & -0.005 & 789/\allowbreak{}240 & 8 \\
149 & +0.3 & 0.018 & -0.005 & 784/\allowbreak{}240 & 8 \\
150 & +0.0 & 0.017 & -0.004 & 791/\allowbreak{}240 & 8 \\
\bottomrule
\end{tabular}
\par\smallskip
\parbox{\textwidth}{\footnotesize IDs match the preceding panel; no estimate is recomputed from rounded entries.}
}% end font group
\end{table*}
% }
% \clearpage

% \clearpage
% \begin{landscape}
% {\fontsize{8}{10}\selectfont\setlength{\tabcolsep}{3pt}
% \renewcommand{\arraystretch}{1.18}
% \renewcommand{\theHtable}{supp.4.a}
\setcounter{table}{3}
\begin{table*}[p]
\centering
\scriptsize
\setlength{\tabcolsep}{3pt}
\renewcommand{\arraystretch}{1.13}
\caption{Architecture screening and seed / bundle replication of frozen class SimNPO (BANKING77).}
\label{tab:S04_architecture_replication}
\begin{tabular}{
@{}
>{\raggedright\arraybackslash}p{2.45cm}
>{\raggedright\arraybackslash}p{1.65cm}
>{\raggedright\arraybackslash}p{3.15cm}
>{\raggedright\arraybackslash}p{2.55cm}
>{\raggedright\arraybackslash}p{2.70cm}
>{\centering\arraybackslash}p{1.35cm}
>{\raggedright\arraybackslash}p{\dimexpr\textwidth-13.85cm-12\tabcolsep\relax}
@{}
}
\toprule
\textbf{Model} &
\textbf{Seed / bundle; stage} &
\textbf{$P_\mathrm{F}$ [rows]; class CI} &
\textbf{$P_\mathrm{G}$ [rows]} &
\textbf{$\Delta P_\mathrm{F}/\Delta P_\mathrm{G}$ vs. seed 17, b01} &
\textbf{$\Delta\operatorname{Acc}_\mathrm{R}$ (pp)} &
\textbf{Status} \\
\midrule
Apertus-8B & 17 / b01\newline screening & 0.55 [0.50, 0.60]\newline classes: [0.32, 0.74] & 0.58 [0.47, 0.66] & --- / --- & -0.3 & accepted \\
Apertus-8B & 42 / b01\newline replication & 0.17 [0.10, 0.23]\newline classes: [-0.33, 0.61] & 0.25 [0.13, 0.35] & -0.38 / -0.33 & +0.0 & accepted \\
Apertus-8B & 17 / b02\newline replication & 0.56 [0.53, 0.59]\newline classes: [0.47, 0.62] & 0.57 [0.52, 0.62] & +0.02 / -0.00 & -0.1 & accepted \\
Falcon-H1-3B & 17 / b01\newline screening & 0.36 [0.32, 0.41]\newline classes: [0.16, 0.69] & 0.42 [0.35, 0.49] & --- / --- & +0.3 & accepted \\
Gemma-4-E2B & 17 / b01\newline screening & 0.41 [0.36, 0.46]\newline classes: [0.25, 0.59] & 0.44 [0.35, 0.52] & --- / --- & +1.0 & accepted \\
Granite-4.0-Micro & 17 / b01\newline screening & 0.46 [0.42, 0.51]\newline classes: [0.34, 0.59] & 0.55 [0.48, 0.62] & --- / --- & -0.6 & accepted \\
LFM2.5-1.2B & 17 / b01\newline screening & 0.30 [0.23, 0.36]\newline classes: [0.09, 0.43] & 0.23 [0.07, 0.36] & --- / --- & -0.3 & accepted \\
Llama-3.2-1B & 17 / b01\newline screening & 0.57 [0.53, 0.60]\newline classes: [0.46, 0.71] & 0.54 [0.47, 0.61] & --- / --- & -0.2 & accepted \\
Llama-3.2-3B & 17 / b01\newline screening & 0.50 [0.46, 0.55]\newline classes: [0.31, 0.70] & 0.56 [0.48, 0.63] & --- / --- & -0.4 & accepted \\
Ministral-3-3B & 17 / b01\newline screening & 0.28 [0.23, 0.34]\newline classes: [0.13, 0.47] & 0.21 [0.08, 0.33] & --- / --- & +0.4 & accepted \\
Nemotron-Nano-9B & 17 / b01\newline screening & 0.46 [0.41, 0.51]\newline classes: [0.21, 0.72] & 0.53 [0.42, 0.62] & --- / --- & +0.7 & accepted \\
OLMo-3-7B & 17 / b01\newline screening & 0.34 [0.30, 0.39]\newline classes: [0.18, 0.61] & 0.35 [0.26, 0.44] & --- / --- & +0.1 & accepted \\
OLMo-3-7B & 42 / b01\newline replication & 0.04 [-0.05, 0.11]\newline classes: [-0.62, 0.51] & 0.00 [-0.16, 0.13] & -0.30 / -0.35 & +0.4 & positive, unconfirmed \\
OLMo-3-7B & 17 / b02\newline replication & 0.50 [0.46, 0.54]\newline classes: [0.14, 0.73] & 0.49 [0.42, 0.55] & +0.16 / +0.14 & +0.0 & accepted \\
Phi-4-mini & 17 / b01\newline screening & 0.47 [0.42, 0.52]\newline classes: [0.28, 0.68] & 0.51 [0.40, 0.60] & --- / --- & +1.4 & accepted \\
\bottomrule
\end{tabular}
\end{table*}

\begin{table*}[p]
\ContinuedFloat
\centering
\scriptsize
\setlength{\tabcolsep}{3pt}
\renewcommand{\arraystretch}{1.13}
\caption[]{Architecture screening and seed / bundle replication (continued).}
\begin{tabular}{
@{}
>{\raggedright\arraybackslash}p{2.45cm}
>{\raggedright\arraybackslash}p{1.65cm}
>{\raggedright\arraybackslash}p{3.15cm}
>{\raggedright\arraybackslash}p{2.55cm}
>{\raggedright\arraybackslash}p{2.70cm}
>{\centering\arraybackslash}p{1.35cm}
>{\raggedright\arraybackslash}p{\dimexpr\textwidth-13.85cm-12\tabcolsep\relax}
@{}
}
\toprule
\textbf{Model} &
\textbf{Seed / bundle; stage} &
\textbf{$P_\mathrm{F}$ [rows]; class CI} &
\textbf{$P_\mathrm{G}$ [rows]} &
\textbf{$\Delta P_\mathrm{F}/\Delta P_\mathrm{G}$ vs. seed 17, b01} &
\textbf{$\Delta\operatorname{Acc}_\mathrm{R}$ (pp)} &
\textbf{Status} \\
\midrule
Qwen2.5-1.5B & 17 / b01\newline screening & 0.45 [0.41, 0.50]\newline classes: [0.18, 0.68] & 0.54 [0.47, 0.62] & --- / --- & +0.2 & accepted \\
Qwen2.5-3B & 17 / b01\newline screening & 0.50 [0.45, 0.55]\newline classes: [0.25, 0.74] & 0.60 [0.52, 0.68] & --- / --- & +0.1 & accepted \\
Qwen2.5-7B & 17 / b01\newline screening & 0.63 [0.58, 0.67]\newline classes: [0.51, 0.79] & 0.63 [0.56, 0.70] & --- / --- & +0.0 & accepted \\
Qwen3-4B & 17 / b01\newline screening & 0.28 [0.21, 0.35]\newline classes: [0.10, 0.41] & 0.19 [-0.01, 0.34] & --- / --- & +0.1 & positive, unconfirmed \\
Qwen3-4B & 42 / b01\newline replication & -0.18 [-0.34, -0.03]\newline classes: [-1.90, 0.58] & -0.10 [-0.41, 0.13] & -0.46 / -0.29 & +0.1 & no benefit \\
Qwen3-4B & 17 / b02\newline replication & 0.51 [0.47, 0.55]\newline classes: [0.35, 0.64] & 0.50 [0.44, 0.56] & +0.23 / +0.32 & -0.1 & accepted \\
Qwen3.5-0.8B & 17 / b01\newline confirmation & 0.59 [0.55, 0.63]\newline classes: [0.49, 0.76] & 0.57 [0.49, 0.64] & --- / --- & -0.1 & accepted \\
Qwen3.5-0.8B & 42 / b01\newline replication & 0.55 [0.50, 0.60]\newline classes: [0.40, 0.76] & 0.51 [0.40, 0.60] & -0.04 / -0.06 & -0.9 & accepted \\
Qwen3.5-0.8B & 17 / b02\newline replication & 0.70 [0.66, 0.73]\newline classes: [0.44, 0.83] & 0.73 [0.68, 0.78] & +0.11 / +0.17 & +0.3 & accepted \\
Qwen3.5-2B & 17 / b01\newline screening & 0.46 [0.41, 0.51]\newline classes: [0.23, 0.62] & 0.49 [0.37, 0.59] & --- / --- & -0.8 & accepted \\
Qwen3.5-4B & 17 / b01\newline screening & 0.20 [0.15, 0.24]\newline classes: [0.10, 0.32] & 0.18 [0.07, 0.28] & --- / --- & -0.6 & accepted \\
Qwen3.5-9B & 17 / b01\newline screening & 0.54 [0.49, 0.59]\newline classes: [0.35, 0.76] & 0.58 [0.51, 0.65] & --- / --- & +0.4 & accepted \\
Qwen3.5-9B & 42 / b01\newline replication & 0.14 [0.11, 0.17]\newline classes: [0.07, 0.25] & 0.18 [0.13, 0.24] & -0.40 / -0.40 & -0.9 & accepted \\
Qwen3.5-9B & 17 / b02\newline replication & 0.44 [0.38, 0.49]\newline classes: [0.07, 0.78] & 0.43 [0.36, 0.51] & -0.10 / -0.15 & +0.4 & accepted \\
SmolLM3-3B & 17 / b01\newline screening & 0.50 [0.44, 0.55]\newline classes: [0.24, 0.66] & 0.46 [0.34, 0.57] & --- / --- & -0.2 & accepted \\
\bottomrule
\end{tabular}
\end{table*}
% }
% \end{landscape}
% \clearpage

% \clearpage
% \begin{landscape}
% {\fontsize{8}{10}\selectfont\setlength{\tabcolsep}{3pt}
% \renewcommand{\arraystretch}{1.18}
% \renewcommand{\theHtable}{supp.5.a}

% ==================================================
% Table S5: Part 1 — BANKING77
% ==================================================
\setcounter{table}{4}
\begin{table*}[p]
\centering
\scriptsize
\caption{Semantic versus opaque decision labels: acquisition, oracle recognition, displacement and SimNPO progress.}
\label{tab:S05_semantic_opaque_results}
\setlength{\tabcolsep}{2pt}
\begin{tabular}{
@{}
>{\raggedright\arraybackslash}p{1.7cm}
>{\raggedright\arraybackslash}p{1.3cm}
>{\raggedright\arraybackslash}p{1.8cm}
>{\raggedright\arraybackslash}p{1.8cm}
>{\raggedright\arraybackslash}p{2.4cm}
>{\raggedright\arraybackslash}p{1.75cm}
>{\raggedright\arraybackslash}p{1.35cm}
>{\raggedright\arraybackslash}p{\dimexpr\textwidth-12.1cm-14\tabcolsep\relax}
@{}
}
\toprule
\textbf{Dataset / bundle / seed} &
\textbf{Labels; status} &
\textbf{Full accuracy F / G} &
\textbf{Oracle accuracy F / G} &
\textbf{Oracle forgotten mass F / G} &
\textbf{$D_{0,\mathrm{F}}/D_{0,\mathrm{G}}$} &
\textbf{$P_\mathrm{F}/P_\mathrm{G}$} &
\textbf{Opaque acquisition} \\
\midrule

banking77\newline b01, seed 17 &
semantic\newline found &
0.981 /\allowbreak{} 0.912 &
0.2687 /\allowbreak{} 0.3063 &
3.043e-01 /\allowbreak{} 3.262e-01 &
0.608 /\allowbreak{} 0.495 &
0.59 /\allowbreak{} 0.57 &
--- \\

banking77\newline b01, seed 17 &
opaque\newline found &
0.953 /\allowbreak{} 0.906 &
0.0000 /\allowbreak{} 0.0000 &
1.126e-03 /\allowbreak{} 9.742e-04 &
0.943 /\allowbreak{} 0.881 &
0.61 /\allowbreak{} 0.62 &
pass; min class acc. 0.80 (semantic 0.90), min gold p 0.77 \\

banking77\newline b01, seed 42 &
semantic\newline found &
0.974 /\allowbreak{} 0.906 &
0.3154 /\allowbreak{} 0.3625 &
3.419e-01 /\allowbreak{} 3.789e-01 &
0.558 /\allowbreak{} 0.470 &
0.55 /\allowbreak{} 0.51 &
--- \\

banking77\newline b01, seed 42 &
opaque\newline found &
0.960 /\allowbreak{} 0.894 &
0.0000 /\allowbreak{} 0.0000 &
4.706e-04 /\allowbreak{} 5.307e-04 &
0.934 /\allowbreak{} 0.878 &
0.60 /\allowbreak{} 0.58 &
pass; min class acc. 0.84 (semantic 0.84), min gold p 0.83 \\

banking77\newline b02, seed 17 &
semantic\newline found &
0.940 /\allowbreak{} 0.925 &
0.2199 /\allowbreak{} 0.2000 &
2.209e-01 /\allowbreak{} 2.108e-01 &
0.644 /\allowbreak{} 0.636 &
0.70 /\allowbreak{} 0.73 &
--- \\

banking77\newline b02, seed 17 &
opaque\newline found &
0.966 /\allowbreak{} 0.931 &
0.0000 /\allowbreak{} 0.0000 &
6.479e-04 /\allowbreak{} 3.491e-04 &
0.912 /\allowbreak{} 0.872 &
0.45 /\allowbreak{} 0.43 &
pass; min class acc. 0.74 (semantic 0.70), min gold p 0.69 \\

banking77\newline b02, seed 42 &
semantic\newline found &
0.958 /\allowbreak{} 0.956 &
0.2094 /\allowbreak{} 0.2062 &
2.160e-01 /\allowbreak{} 2.124e-01 &
0.676 /\allowbreak{} 0.660 &
0.47 /\allowbreak{} 0.45 &
--- \\

banking77\newline b02, seed 42 &
opaque\newline found &
0.955 /\allowbreak{} 0.950 &
0.0000 /\allowbreak{} 0.0000 &
1.836e-03 /\allowbreak{} 1.780e-03 &
0.911 /\allowbreak{} 0.902 &
0.40 /\allowbreak{} 0.38 &
pass; min class acc. 0.71 (semantic 0.65), min gold p 0.70 \\

\bottomrule
\end{tabular}
\end{table*}

% ==================================================
% Table S5: Part 2 — CLINC150
% ==================================================
\begin{table*}[p]
\ContinuedFloat
\centering
\scriptsize
\caption[]{Semantic versus opaque decision labels (continued): CLINC150.}
\setlength{\tabcolsep}{2pt}
\begin{tabular}{
@{}
>{\raggedright\arraybackslash}p{1.7cm}
>{\raggedright\arraybackslash}p{1.3cm}
>{\raggedright\arraybackslash}p{1.8cm}
>{\raggedright\arraybackslash}p{1.8cm}
>{\raggedright\arraybackslash}p{2.4cm}
>{\raggedright\arraybackslash}p{1.75cm}
>{\raggedright\arraybackslash}p{1.35cm}
>{\raggedright\arraybackslash}p{\dimexpr\textwidth-12.1cm-14\tabcolsep\relax}
@{}
}
\toprule
\textbf{Dataset / bundle / seed} &
\textbf{Labels; status} &
\textbf{Full accuracy F / G} &
\textbf{Oracle accuracy F / G} &
\textbf{Oracle forgotten mass F / G} &
\textbf{$D_{0,\mathrm{F}}/D_{0,\mathrm{G}}$} &
\textbf{$P_\mathrm{F}/P_\mathrm{G}$} &
\textbf{Opaque acquisition} \\
\midrule

clinc150\newline b02, seed 17 &
semantic\newline found &
0.995 /\allowbreak{} 0.983 &
0.6612 /\allowbreak{} 0.6333 &
6.400e-01 /\allowbreak{} 6.079e-01 &
0.298 /\allowbreak{} 0.319 &
0.22 /\allowbreak{} 0.28 &
--- \\

clinc150\newline b02, seed 17 &
opaque\newline found &
0.992 /\allowbreak{} 0.971 &
0.0013 /\allowbreak{} 0.0000 &
3.153e-03 /\allowbreak{} 2.200e-03 &
0.983 /\allowbreak{} 0.976 &
0.28 /\allowbreak{} 0.29 &
pass; min class acc. 0.85 (semantic 0.85), min gold p 0.85 \\

clinc150\newline b03, seed 17 &
semantic\newline found &
0.996 /\allowbreak{} 0.983 &
0.5051 /\allowbreak{} 0.5333 &
5.047e-01 /\allowbreak{} 5.316e-01 &
0.454 /\allowbreak{} 0.407 &
0.22 /\allowbreak{} 0.20 &
--- \\

clinc150\newline b03, seed 17 &
opaque\newline found &
0.994 /\allowbreak{} 0.971 &
0.0000 /\allowbreak{} 0.0000 &
6.001e-04 /\allowbreak{} 7.282e-04 &
0.993 /\allowbreak{} 0.973 &
0.13 /\allowbreak{} 0.11 &
pass; min class acc. 0.89 (semantic 0.74), min gold p 0.87 \\

clinc150\newline b06, seed 17 &
semantic\newline found &
0.999 /\allowbreak{} 0.979 &
0.7807 /\allowbreak{} 0.7375 &
7.705e-01 /\allowbreak{} 7.241e-01 &
0.207 /\allowbreak{} 0.233 &
-0.24 /\allowbreak{} -0.09 &
--- \\

clinc150\newline b06, seed 17 &
opaque\newline found &
0.996 /\allowbreak{} 0.983 &
0.0000 /\allowbreak{} 0.0000 &
1.247e-03 /\allowbreak{} 1.292e-03 &
0.999 /\allowbreak{} 0.989 &
0.02 /\allowbreak{} 0.02 &
pass; min class acc. 0.94 (semantic 0.89), min gold p 0.93 \\

\bottomrule
\end{tabular}

\par\smallskip
\noindent
\parbox{\textwidth}{\footnotesize
All seven opaque acquisition checks pass. The semantic seed-42
b02 condition includes one primary-seed-selected class below
the acquisition threshold at replication. Masses use the
precision available in the source export.
}
\end{table*}
% }
% \end{landscape}
% \clearpage

% \clearpage
% \begin{landscape}
% {\fontsize{8}{10}\selectfont\setlength{\tabcolsep}{3pt}
% \renewcommand{\arraystretch}{1.18}
% \renewcommand{\theHtable}{supp.6.a}
\setcounter{table}{5}
\begin{table*}[p]
\centering
\scriptsize
\setlength{\tabcolsep}{3pt}
\caption{Candidate-schema results with paired uncertainty.}
\label{tab:S06_candidate_sets}
\begin{tabular}{@{}
>{\raggedright\arraybackslash}p{0.65cm}
>{\raggedright\arraybackslash}p{3.45cm}
>{\raggedright\arraybackslash}p{2.45cm}
>{\raggedright\arraybackslash}p{1.12cm}
>{\raggedright\arraybackslash}p{3.15cm}
>{\raggedright\arraybackslash}p{2.55cm}
>{\raggedright\arraybackslash}p{1.65cm}
@{}}
\toprule
\textbf{ID} & \textbf{Configuration} & \textbf{Schema / view} & \textbf{$D_{0,v}$} & \textbf{$P_v$ [rows]} & \textbf{Class CI} & \textbf{Oracle accuracy} \\
\midrule
1 & apertus-8b\newline BANKING77 sem. \_\allowbreak{}01\newline SimNPO & S0\_\allowbreak{}original\newline 77 options; F & 0.479 & 0.55 [0.50, 0.60] & [0.31, 0.72] & 0.395 \\
2 & apertus-8b\newline BANKING77 sem. \_\allowbreak{}01\newline SimNPO & S0\_\allowbreak{}original\newline 77 options; G & 0.400 & 0.58 [0.48, 0.66] & [0.15, 0.77] & 0.450 \\
3 & apertus-8b\newline BANKING77 sem. \_\allowbreak{}01\newline SimNPO & S1\_\allowbreak{}neighbors\newline 12 options; F & 0.204 & 0.40 [0.32, 0.47] & [0.22, 0.61] & 0.682 \\
4 & apertus-8b\newline BANKING77 sem. \_\allowbreak{}01\newline SimNPO & S1\_\allowbreak{}neighbors\newline 12 options; G & 0.132 & 0.36 [0.14, 0.52] & [0.09, 0.57] & 0.775 \\
5 & apertus-8b\newline BANKING77 sem. \_\allowbreak{}01\newline SimNPO & S2\_\allowbreak{}distractors\newline 12 options; F & 0.122 & 0.55 [0.48, 0.61] & [-0.12, 0.68] & 0.827 \\
6 & apertus-8b\newline BANKING77 sem. \_\allowbreak{}01\newline SimNPO & S2\_\allowbreak{}distractors\newline 12 options; G & 0.150 & 0.68 [0.60, 0.75] & [-0.45, 0.77] & 0.775 \\
7 & qwen3.5-0.8b\newline BANKING77 sem. \_\allowbreak{}01\newline SimNPO & S0\_\allowbreak{}original\newline 77 options; F & 0.608 & 0.59 [0.55, 0.63] & [0.39, 0.73] & 0.269 \\
8 & qwen3.5-0.8b\newline BANKING77 sem. \_\allowbreak{}01\newline NPO & S0\_\allowbreak{}original\newline 77 options; F & 0.608 & 0.39 [0.34, 0.43] & [0.21, 0.59] & 0.269 \\
9 & qwen3.5-0.8b\newline BANKING77 sem. \_\allowbreak{}01\newline GradAscent & S0\_\allowbreak{}original\newline 77 options; F & 0.608 & 0.55 [0.50, 0.59] & [0.41, 0.62] & 0.269 \\
10 & qwen3.5-0.8b\newline BANKING77 sem. \_\allowbreak{}01\newline GradDiff & S0\_\allowbreak{}original\newline 77 options; F & 0.608 & 0.45 [0.41, 0.50] & [0.31, 0.60] & 0.269 \\
11 & qwen3.5-0.8b\newline BANKING77 sem. \_\allowbreak{}01\newline SimNPO & S0\_\allowbreak{}original\newline 77 options; G & 0.495 & 0.57 [0.49, 0.64] & [0.40, 0.71] & 0.306 \\
12 & qwen3.5-0.8b\newline BANKING77 sem. \_\allowbreak{}01\newline NPO & S0\_\allowbreak{}original\newline 77 options; G & 0.495 & 0.36 [0.28, 0.43] & [0.18, 0.52] & 0.306 \\
13 & qwen3.5-0.8b\newline BANKING77 sem. \_\allowbreak{}01\newline GradAscent & S0\_\allowbreak{}original\newline 77 options; G & 0.495 & 0.49 [0.41, 0.56] & [0.38, 0.58] & 0.306 \\
14 & qwen3.5-0.8b\newline BANKING77 sem. \_\allowbreak{}01\newline GradDiff & S0\_\allowbreak{}original\newline 77 options; G & 0.495 & 0.42 [0.35, 0.50] & [0.27, 0.57] & 0.306 \\
\bottomrule
\end{tabular}
\end{table*}

\begin{table*}[p]
\ContinuedFloat
\centering
\scriptsize
\setlength{\tabcolsep}{3pt}
\caption[]{Candidate-schema results (continued): BANKING77: alternative candidates.}
\begin{tabular}{@{}
>{\raggedright\arraybackslash}p{0.65cm}
>{\raggedright\arraybackslash}p{3.45cm}
>{\raggedright\arraybackslash}p{2.45cm}
>{\raggedright\arraybackslash}p{1.12cm}
>{\raggedright\arraybackslash}p{3.15cm}
>{\raggedright\arraybackslash}p{2.55cm}
>{\raggedright\arraybackslash}p{1.65cm}
@{}}
\toprule
\textbf{ID} & \textbf{Configuration} & \textbf{Schema / view} & \textbf{$D_{0,v}$} & \textbf{$P_v$ [rows]} & \textbf{Class CI} & \textbf{Oracle accuracy} \\
\midrule
15 & qwen3.5-0.8b\newline BANKING77 sem. \_\allowbreak{}01\newline SimNPO & S1\_\allowbreak{}neighbors\newline 12 options; F & 0.438 & 0.66 [0.61, 0.70] & [0.50, 0.74] & 0.460 \\
16 & qwen3.5-0.8b\newline BANKING77 sem. \_\allowbreak{}01\newline NPO & S1\_\allowbreak{}neighbors\newline 12 options; F & 0.438 & 0.41 [0.36, 0.47] & [0.14, 0.65] & 0.460 \\
17 & qwen3.5-0.8b\newline BANKING77 sem. \_\allowbreak{}01\newline GradAscent & S1\_\allowbreak{}neighbors\newline 12 options; F & 0.438 & 0.62 [0.57, 0.67] & [0.50, 0.67] & 0.460 \\
18 & qwen3.5-0.8b\newline BANKING77 sem. \_\allowbreak{}01\newline GradDiff & S1\_\allowbreak{}neighbors\newline 12 options; F & 0.438 & 0.47 [0.41, 0.52] & [0.24, 0.66] & 0.460 \\
19 & qwen3.5-0.8b\newline BANKING77 sem. \_\allowbreak{}01\newline SimNPO & S1\_\allowbreak{}neighbors\newline 12 options; G & 0.339 & 0.61 [0.51, 0.70] & [0.41, 0.71] & 0.519 \\
20 & qwen3.5-0.8b\newline BANKING77 sem. \_\allowbreak{}01\newline NPO & S1\_\allowbreak{}neighbors\newline 12 options; G & 0.339 & 0.38 [0.28, 0.48] & [0.07, 0.60] & 0.519 \\
21 & qwen3.5-0.8b\newline BANKING77 sem. \_\allowbreak{}01\newline GradAscent & S1\_\allowbreak{}neighbors\newline 12 options; G & 0.339 & 0.54 [0.44, 0.63] & [0.35, 0.67] & 0.519 \\
22 & qwen3.5-0.8b\newline BANKING77 sem. \_\allowbreak{}01\newline GradDiff & S1\_\allowbreak{}neighbors\newline 12 options; G & 0.339 & 0.43 [0.33, 0.52] & [0.14, 0.61] & 0.519 \\
23 & qwen3.5-0.8b\newline BANKING77 sem. \_\allowbreak{}01\newline SimNPO & S2\_\allowbreak{}distractors\newline 12 options; F & 0.148 & 0.32 [0.27, 0.37] & [0.13, 0.45] & 0.799 \\
24 & qwen3.5-0.8b\newline BANKING77 sem. \_\allowbreak{}01\newline NPO & S2\_\allowbreak{}distractors\newline 12 options; F & 0.148 & 0.12 [0.07, 0.17] & [0.02, 0.20] & 0.799 \\
25 & qwen3.5-0.8b\newline BANKING77 sem. \_\allowbreak{}01\newline GradAscent & S2\_\allowbreak{}distractors\newline 12 options; F & 0.148 & 0.17 [0.12, 0.23] & [0.11, 0.23] & 0.799 \\
26 & qwen3.5-0.8b\newline BANKING77 sem. \_\allowbreak{}01\newline GradDiff & S2\_\allowbreak{}distractors\newline 12 options; F & 0.148 & 0.12 [0.07, 0.18] & [0.03, 0.19] & 0.799 \\
27 & qwen3.5-0.8b\newline BANKING77 sem. \_\allowbreak{}01\newline SimNPO & S2\_\allowbreak{}distractors\newline 12 options; G & 0.133 & 0.43 [0.35, 0.51] & [0.16, 0.56] & 0.756 \\
28 & qwen3.5-0.8b\newline BANKING77 sem. \_\allowbreak{}01\newline NPO & S2\_\allowbreak{}distractors\newline 12 options; G & 0.133 & 0.13 [0.03, 0.22] & [-0.10, 0.21] & 0.756 \\
29 & qwen3.5-0.8b\newline BANKING77 sem. \_\allowbreak{}01\newline GradAscent & S2\_\allowbreak{}distractors\newline 12 options; G & 0.133 & 0.17 [0.09, 0.27] & [-0.06, 0.26] & 0.756 \\
30 & qwen3.5-0.8b\newline BANKING77 sem. \_\allowbreak{}01\newline GradDiff & S2\_\allowbreak{}distractors\newline 12 options; G & 0.133 & 0.13 [0.03, 0.23] & [-0.12, 0.20] & 0.756 \\
\bottomrule
\end{tabular}
\end{table*}

\begin{table*}[p]
\ContinuedFloat
\centering
\scriptsize
\setlength{\tabcolsep}{3pt}
\caption[]{Candidate-schema results (continued): BANKING77: opaque labels and 9B.}
\begin{tabular}{@{}
>{\raggedright\arraybackslash}p{0.65cm}
>{\raggedright\arraybackslash}p{3.45cm}
>{\raggedright\arraybackslash}p{2.45cm}
>{\raggedright\arraybackslash}p{1.12cm}
>{\raggedright\arraybackslash}p{3.15cm}
>{\raggedright\arraybackslash}p{2.55cm}
>{\raggedright\arraybackslash}p{1.65cm}
@{}}
\toprule
\textbf{ID} & \textbf{Configuration} & \textbf{Schema / view} & \textbf{$D_{0,v}$} & \textbf{$P_v$ [rows]} & \textbf{Class CI} & \textbf{Oracle accuracy} \\
\midrule
31 & qwen3.5-0.8b\newline BANKING77 opaque \_\allowbreak{}01\newline SimNPO & S0\_\allowbreak{}original\newline 77 options; F & 0.943 & 0.61 [0.57, 0.64] & [0.44, 0.76] & 0.000 \\
32 & qwen3.5-0.8b\newline BANKING77 opaque \_\allowbreak{}01\newline SimNPO & S0\_\allowbreak{}original\newline 77 options; G & 0.881 & 0.62 [0.57, 0.68] & [0.44, 0.75] & 0.000 \\
33 & qwen3.5-0.8b\newline BANKING77 opaque \_\allowbreak{}01\newline SimNPO & S1\_\allowbreak{}neighbors\newline 12 options; F & 0.464 & 0.00 [-0.03, 0.03] & [-0.15, 0.15] & 0.016 \\
34 & qwen3.5-0.8b\newline BANKING77 opaque \_\allowbreak{}01\newline SimNPO & S1\_\allowbreak{}neighbors\newline 12 options; G & 0.489 & 0.01 [-0.03, 0.04] & [-0.09, 0.09] & 0.050 \\
35 & qwen3.5-0.8b\newline BANKING77 opaque \_\allowbreak{}01\newline SimNPO & S2\_\allowbreak{}distractors\newline 12 options; F & 0.583 & 0.12 [0.07, 0.16] & [-0.08, 0.46] & 0.000 \\
36 & qwen3.5-0.8b\newline BANKING77 opaque \_\allowbreak{}01\newline SimNPO & S2\_\allowbreak{}distractors\newline 12 options; G & 0.553 & 0.14 [0.07, 0.20] & [-0.11, 0.48] & 0.000 \\
37 & qwen3.5-0.8b\newline BANKING77 opaque \_\allowbreak{}02\newline SimNPO & S0\_\allowbreak{}original\newline 77 options; F & 0.913 & 0.45 [0.41, 0.49] & [0.26, 0.71] & 0.000 \\
38 & qwen3.5-0.8b\newline BANKING77 opaque \_\allowbreak{}02\newline SimNPO & S0\_\allowbreak{}original\newline 77 options; G & 0.872 & 0.43 [0.37, 0.49] & [0.25, 0.61] & 0.000 \\
39 & qwen3.5-0.8b\newline BANKING77 opaque \_\allowbreak{}02\newline SimNPO & S1\_\allowbreak{}neighbors\newline 12 options; F & 0.714 & 0.04 [0.03, 0.05] & [0.01, 0.12] & 0.000 \\
40 & qwen3.5-0.8b\newline BANKING77 opaque \_\allowbreak{}02\newline SimNPO & S1\_\allowbreak{}neighbors\newline 12 options; G & 0.691 & 0.04 [0.03, 0.06] & [0.01, 0.09] & 0.000 \\
41 & qwen3.5-0.8b\newline BANKING77 opaque \_\allowbreak{}02\newline SimNPO & S2\_\allowbreak{}distractors\newline 12 options; F & 0.530 & 0.12 [0.11, 0.14] & [0.01, 0.17] & 0.005 \\
42 & qwen3.5-0.8b\newline BANKING77 opaque \_\allowbreak{}02\newline SimNPO & S2\_\allowbreak{}distractors\newline 12 options; G & 0.507 & 0.13 [0.10, 0.15] & [0.03, 0.17] & 0.019 \\
43 & qwen3.5-9b\newline BANKING77 sem. \_\allowbreak{}01\newline SimNPO & S0\_\allowbreak{}original\newline 77 options; F & 0.414 & 0.54 [0.49, 0.59] & [0.33, 0.72] & 0.484 \\
44 & qwen3.5-9b\newline BANKING77 sem. \_\allowbreak{}01\newline SimNPO & S0\_\allowbreak{}original\newline 77 options; G & 0.352 & 0.58 [0.51, 0.66] & [0.37, 0.70] & 0.487 \\
45 & qwen3.5-9b\newline BANKING77 sem. \_\allowbreak{}01\newline SimNPO & S1\_\allowbreak{}neighbors\newline 12 options; F & 0.187 & 0.53 [0.45, 0.60] & [0.16, 0.75] & 0.745 \\
46 & qwen3.5-9b\newline BANKING77 sem. \_\allowbreak{}01\newline SimNPO & S1\_\allowbreak{}neighbors\newline 12 options; G & 0.139 & 0.57 [0.43, 0.69] & [0.12, 0.73] & 0.787 \\
47 & qwen3.5-9b\newline BANKING77 sem. \_\allowbreak{}01\newline SimNPO & S2\_\allowbreak{}distractors\newline 12 options; F & 0.167 & 0.43 [0.36, 0.50] & [0.03, 0.59] & 0.771 \\
48 & qwen3.5-9b\newline BANKING77 sem. \_\allowbreak{}01\newline SimNPO & S2\_\allowbreak{}distractors\newline 12 options; G & 0.174 & 0.54 [0.45, 0.63] & [0.09, 0.63] & 0.738 \\
\bottomrule
\end{tabular}
\end{table*}

\begin{table*}[p]
\ContinuedFloat
\centering
\scriptsize
\setlength{\tabcolsep}{3pt}
\caption[]{Candidate-schema results (continued): CLINC150.}
\begin{tabular}{@{}
>{\raggedright\arraybackslash}p{0.65cm}
>{\raggedright\arraybackslash}p{3.45cm}
>{\raggedright\arraybackslash}p{2.45cm}
>{\raggedright\arraybackslash}p{1.12cm}
>{\raggedright\arraybackslash}p{3.15cm}
>{\raggedright\arraybackslash}p{2.55cm}
>{\raggedright\arraybackslash}p{1.65cm}
@{}}
\toprule
\textbf{ID} & \textbf{Configuration} & \textbf{Schema / view} & \textbf{$D_{0,v}$} & \textbf{$P_v$ [rows]} & \textbf{Class CI} & \textbf{Oracle accuracy} \\
\midrule
49 & qwen3.5-0.8b\newline CLINC150 sem. \_\allowbreak{}02\newline SimNPO & S0\_\allowbreak{}original\newline 151 options; F & 0.298 & 0.22 [0.16, 0.27] & [-0.15, 0.43] & 0.661 \\
50 & qwen3.5-0.8b\newline CLINC150 sem. \_\allowbreak{}02\newline SimNPO & S0\_\allowbreak{}original\newline 151 options; F & 0.298 & -0.06 [-0.17, 0.04] & [-0.91, 0.35] & 0.661 \\
51 & qwen3.5-0.8b\newline CLINC150 sem. \_\allowbreak{}02\newline SimNPO & S0\_\allowbreak{}original\newline 151 options; G & 0.319 & 0.28 [0.17, 0.37] & [-0.29, 0.51] & 0.633 \\
52 & qwen3.5-0.8b\newline CLINC150 sem. \_\allowbreak{}02\newline SimNPO & S0\_\allowbreak{}original\newline 151 options; G & 0.319 & 0.03 [-0.17, 0.19] & [-1.44, 0.48] & 0.633 \\
53 & qwen3.5-0.8b\newline CLINC150 sem. \_\allowbreak{}02\newline SimNPO & S1\_\allowbreak{}neighbors\newline 24 options; F & 0.151 & 0.20 [0.10, 0.28] & [-0.54, 0.45] & 0.827 \\
54 & qwen3.5-0.8b\newline CLINC150 sem. \_\allowbreak{}02\newline SimNPO & S1\_\allowbreak{}neighbors\newline 24 options; F & 0.151 & -0.38 [-0.62, -0.19] & [-2.09, 0.28] & 0.827 \\
55 & qwen3.5-0.8b\newline CLINC150 sem. \_\allowbreak{}02\newline SimNPO & S1\_\allowbreak{}neighbors\newline 24 options; G & 0.199 & 0.32 [0.18, 0.43] & [-0.56, 0.59] & 0.762 \\
56 & qwen3.5-0.8b\newline CLINC150 sem. \_\allowbreak{}02\newline SimNPO & S1\_\allowbreak{}neighbors\newline 24 options; G & 0.199 & -0.02 [-0.32, 0.19] & [-2.59, 0.56] & 0.762 \\
57 & qwen3.5-0.8b\newline CLINC150 sem. \_\allowbreak{}02\newline SimNPO & S2\_\allowbreak{}distractors\newline 24 options; F & 0.039 & 0.06 [0.03, 0.10] & [0.02, 0.14] & 0.957 \\
58 & qwen3.5-0.8b\newline CLINC150 sem. \_\allowbreak{}02\newline SimNPO & S2\_\allowbreak{}distractors\newline 24 options; F & 0.039 & 0.06 [0.03, 0.10] & [0.02, 0.14] & 0.957 \\
59 & qwen3.5-0.8b\newline CLINC150 sem. \_\allowbreak{}02\newline SimNPO & S2\_\allowbreak{}distractors\newline 24 options; G & 0.029 & 0.12 [0.04, 0.21] & [0.04, 0.24] & 0.963 \\
60 & qwen3.5-0.8b\newline CLINC150 sem. \_\allowbreak{}02\newline SimNPO & S2\_\allowbreak{}distractors\newline 24 options; G & 0.029 & 0.15 [0.05, 0.25] & [0.04, 0.29] & 0.963 \\
61 & qwen3.5-0.8b\newline CLINC150 sem. \_\allowbreak{}03\newline SimNPO & S0\_\allowbreak{}original\newline 151 options; F & 0.454 & 0.22 [0.19, 0.26] & [0.01, 0.54] & 0.505 \\
62 & qwen3.5-0.8b\newline CLINC150 sem. \_\allowbreak{}03\newline SimNPO & S0\_\allowbreak{}original\newline 151 options; G & 0.407 & 0.20 [0.13, 0.28] & [-0.02, 0.55] & 0.533 \\
63 & qwen3.5-0.8b\newline CLINC150 sem. \_\allowbreak{}03\newline SimNPO & S1\_\allowbreak{}neighbors\newline 24 options; F & 0.348 & 0.22 [0.18, 0.26] & [0.01, 0.51] & 0.601 \\
64 & qwen3.5-0.8b\newline CLINC150 sem. \_\allowbreak{}03\newline SimNPO & S1\_\allowbreak{}neighbors\newline 24 options; G & 0.272 & 0.26 [0.16, 0.35] & [-0.02, 0.60] & 0.683 \\
65 & qwen3.5-0.8b\newline CLINC150 sem. \_\allowbreak{}03\newline SimNPO & S2\_\allowbreak{}distractors\newline 24 options; F & 0.027 & -0.00 [-0.01, 0.00] & [-0.01, 0.00] & 0.971 \\
66 & qwen3.5-0.8b\newline CLINC150 sem. \_\allowbreak{}03\newline SimNPO & S2\_\allowbreak{}distractors\newline 24 options; G & 0.038 & -0.03 [-0.09, 0.01] & [-0.10, 0.02] & 0.963 \\
\bottomrule
\end{tabular}
\par\smallskip
\noindent\parbox{\textwidth}{\footnotesize
Candidates are re-scored by forward passes. The source bundle label \texttt{\_01} is retained as exported; it denotes the bundle-01 comparison. Ratios are recomputed within each bootstrap, rather than averaging row-level ratios.
}
\end{table*}
% }
% \end{landscape}
\clearpage

% \clearpage
% \begin{landscape}
% {\fontsize{8}{10}\selectfont\setlength{\tabcolsep}{3pt}
% \renewcommand{\arraystretch}{1.18}
% \renewcommand{\theHtable}{supp.7.a}
\setcounter{table}{6}
\newcommand{\SSevenHeader}{%
\toprule
\textbf{ID} & \textbf{Configuration} & \textbf{$D_{0,\mathrm{F}}$} &
\textbf{$\Delta\operatorname{OD}_\mathrm{F}$ [rows]; intent CI} &
\textbf{$\Delta\operatorname{OD}_\mathrm{G}$ [rows]; intent CI} &
\textbf{$P_\mathrm{F}/P_\mathrm{G}$} &
\textbf{$\Delta\operatorname{Acc}_\mathrm{R}$ (pp)} &
\textbf{Status} \\
\midrule
}
% Sum of first seven column widths = 13.85cm; last fills text width.

% Part 1: IDs 1--7.
\begin{table*}[p]
\centering\scriptsize\setlength{\tabcolsep}{2pt}
\caption{CLINC150: every observed bundle, frozen transfer, tuned confirmation and replication.}
\label{tab:S07_clinc_all_bundles}
\begin{tabular}{@{}
>{\raggedright\arraybackslash}p{0.55cm}
>{\raggedright\arraybackslash}p{2.8cm}
>{\raggedright\arraybackslash}p{1.0cm}
>{\raggedright\arraybackslash}p{3.5cm}
>{\raggedright\arraybackslash}p{3.5cm}
>{\raggedright\arraybackslash}p{1.25cm}
>{\raggedright\arraybackslash}p{1.25cm}
>{\raggedright\arraybackslash}p{\dimexpr\textwidth-13.85cm-14\tabcolsep\relax}
@{}
}
\SSevenHeader
1 & llama3.2-1b\newline seed 17, b01\newline transfer\_\allowbreak{}\_\allowbreak{}simnpo\newline transfer & 0.453 & -0.079 [-0.125, -0.034]\newline intents: [-0.424, +0.281] & -0.082 [-0.159, -0.006]\newline intents: [-0.403, +0.241] & -0.18 /\allowbreak{} -0.20 & -0.2 & no benefit \\
2 & qwen3.5-0.8b\newline seed 17, b01\newline transfer\_\allowbreak{}\_\allowbreak{}simnpo\newline transfer & 0.058 & +0.002 [+0.000, +0.005]\newline intents: [+0.000, +0.005] & +0.006 [+0.001, +0.011]\newline intents: [-0.002, +0.014] & 0.04 /\allowbreak{} 0.07 & +0.0 & accepted \\
3 & qwen3.5-0.8b\newline seed 17, b01\newline transfer\_\allowbreak{}\_\allowbreak{}uniformization\newline transfer & 0.058 & +0.003 [+0.002, +0.005]\newline intents: [+0.001, +0.007] & +0.014 [+0.007, +0.023]\newline intents: [+0.002, +0.029] & 0.06 /\allowbreak{} 0.18 & -0.1 & accepted \\
4 & qwen3.5-0.8b\newline seed 17, b01\newline transfer\_\allowbreak{}\_\allowbreak{}grad\_\allowbreak{}ascent\newline transfer & 0.416 & -0.185 [-0.217, -0.152]\newline intents: [-0.404, -0.001] & -0.168 [-0.228, -0.109]\newline intents: [-0.387, +0.021] & -0.44 /\allowbreak{} -0.47 & -1.0 & no benefit \\
5 & qwen3.5-0.8b\newline seed 17, b01\newline transfer\_\allowbreak{}\_\allowbreak{}grad\_\allowbreak{}diff\newline transfer & 0.416 & -0.102 [-0.130, -0.075]\newline intents: [-0.261, -0.005] & -0.073 [-0.121, -0.025]\newline intents: [-0.239, +0.027] & -0.25 /\allowbreak{} -0.20 & -1.1 & no benefit \\
6 & qwen3.5-0.8b\newline seed 17, b01\newline transfer\_\allowbreak{}\_\allowbreak{}npo\newline transfer & 0.416 & -0.183 [-0.214, -0.151]\newline intents: [-0.395, -0.011] & -0.158 [-0.216, -0.099]\newline intents: [-0.369, +0.022] & -0.44 /\allowbreak{} -0.44 & -1.1 & no benefit \\
7 & qwen3.5-0.8b\newline seed 17, b01\newline transfer\_\allowbreak{}\_\allowbreak{}simnpo\newline transfer & 0.416 & -0.191 [-0.242, -0.138]\newline intents: [-0.577, +0.216] & -0.237 [-0.328, -0.148]\newline intents: [-0.628, +0.182] & -0.46 /\allowbreak{} -0.67 & -0.4 & no benefit \\
\bottomrule
\end{tabular}
\end{table*}

% Part 2: IDs 8--13.
\begin{table*}[p]
\ContinuedFloat
\centering\scriptsize\setlength{\tabcolsep}{2pt}
\caption[]{CLINC150 results (continued): bundle-02 tuning and controls.}
\begin{tabular}{@{}
>{\raggedright\arraybackslash}p{0.55cm}
>{\raggedright\arraybackslash}p{2.8cm}
>{\raggedright\arraybackslash}p{1.0cm}
>{\raggedright\arraybackslash}p{3.5cm}
>{\raggedright\arraybackslash}p{3.5cm}
>{\raggedright\arraybackslash}p{1.25cm}
>{\raggedright\arraybackslash}p{1.25cm}
>{\raggedright\arraybackslash}p{\dimexpr\textwidth-13.85cm-14\tabcolsep\relax}
@{}
}
\SSevenHeader
8 & qwen3.5-0.8b\newline seed 17, b02\newline matched\_\allowbreak{}s1\_\allowbreak{}\_\allowbreak{}simnpo\_\allowbreak{}b2.5\_\allowbreak{}ep7\newline exploratory & 0.298 & -0.117 [-0.152, -0.083]\newline intents: [-0.297, +0.080] & -0.111 [-0.178, -0.044]\newline intents: [-0.383, +0.186] & -0.39 /\allowbreak{} -0.35 & -0.9 & no benefit \\
9 & qwen3.5-0.8b\newline seed 17, b02\newline matched\_\allowbreak{}s1\_\allowbreak{}\_\allowbreak{}simnpo\_\allowbreak{}b4.5\_\allowbreak{}ep9\newline exploratory & 0.298 & -0.051 [-0.082, -0.020]\newline intents: [-0.219, +0.138] & -0.040 [-0.100, +0.022]\newline intents: [-0.290, +0.241] & -0.17 /\allowbreak{} -0.13 & -0.2 & no benefit \\
10 & qwen3.5-0.8b\newline seed 17, b02\newline matched\_\allowbreak{}s1\_\allowbreak{}\_\allowbreak{}simnpo\_\allowbreak{}b6.5\_\allowbreak{}ep10\newline exploratory & 0.298 & +0.049 [+0.025, +0.074]\newline intents: [-0.085, +0.205] & +0.074 [+0.025, +0.124]\newline intents: [-0.119, +0.309] & 0.17 /\allowbreak{} 0.23 & +0.3 & accepted \\
11 & qwen3.5-0.8b\newline seed 17, b02\newline temperature\_\allowbreak{}control\newline control & 0.298 & +0.020 [+0.018, +0.023]\newline intents: [+0.009, +0.033] & +0.032 [+0.025, +0.041]\newline intents: [+0.015, +0.050] & 0.07 /\allowbreak{} 0.10 & +0.0 & accepted \\
12 & qwen3.5-0.8b\newline seed 17, b02\newline transfer\_\allowbreak{}\_\allowbreak{}simnpo\newline control & 0.298 & -0.018 [-0.046, +0.011]\newline intents: [-0.171, +0.157] & +0.009 [-0.048, +0.067]\newline intents: [-0.218, +0.270] & -0.06 /\allowbreak{} 0.03 & +0.3 & no benefit \\
13 & qwen3.5-0.8b\newline seed 17, b02\newline tuned\_\allowbreak{}\_\allowbreak{}simnpo\newline confirmation & 0.298 & +0.065 [+0.046, +0.085]\newline intents: [-0.032, +0.181] & +0.088 [+0.049, +0.129]\newline intents: [-0.058, +0.271] & 0.22 /\allowbreak{} 0.28 & +0.4 & accepted \\
\bottomrule
\end{tabular}
\end{table*}

% Part 3: IDs 14--20.
\begin{table*}[p]
\ContinuedFloat
\centering\scriptsize\setlength{\tabcolsep}{2pt}
\caption[]{CLINC150 results (continued): confirmation and replication.}
\begin{tabular}{@{}
>{\raggedright\arraybackslash}p{0.55cm}
>{\raggedright\arraybackslash}p{2.8cm}
>{\raggedright\arraybackslash}p{1.0cm}
>{\raggedright\arraybackslash}p{3.5cm}
>{\raggedright\arraybackslash}p{3.5cm}
>{\raggedright\arraybackslash}p{1.25cm}
>{\raggedright\arraybackslash}p{1.25cm}
>{\raggedright\arraybackslash}p{\dimexpr\textwidth-13.85cm-14\tabcolsep\relax}
@{}
}
\SSevenHeader
14 & qwen3.5-0.8b\newline seed 17, b03\newline temperature\_\allowbreak{}control\newline control & 0.454 & +0.024 [+0.021, +0.027]\newline intents: [+0.007, +0.046] & +0.030 [+0.023, +0.037]\newline intents: [+0.013, +0.055] & 0.05 /\allowbreak{} 0.07 & +0.0 & accepted \\
15 & qwen3.5-0.8b\newline seed 17, b03\newline tuned\_\allowbreak{}\_\allowbreak{}simnpo\newline confirmation & 0.454 & +0.102 [+0.082, +0.121]\newline intents: [+0.001, +0.289] & +0.083 [+0.052, +0.115]\newline intents: [-0.006, +0.249] & 0.22 /\allowbreak{} 0.20 & +0.1 & accepted \\
16 & qwen3.5-0.8b\newline seed 17, b04\newline tuned\_\allowbreak{}\_\allowbreak{}simnpo\newline replication & 0.481 & +0.061 [+0.034, +0.089]\newline intents: [-0.192, +0.317] & +0.033 [-0.012, +0.080]\newline intents: [-0.200, +0.274] & 0.13 /\allowbreak{} 0.08 & +0.3 & positive, unconfirmed \\
17 & qwen3.5-0.8b\newline seed 17, b05\newline tuned\_\allowbreak{}\_\allowbreak{}simnpo\newline replication & 0.344 & +0.025 [-0.003, +0.052]\newline intents: [-0.122, +0.220] & +0.022 [-0.032, +0.078]\newline intents: [-0.182, +0.276] & 0.07 /\allowbreak{} 0.07 & +0.1 & positive, unconfirmed \\
18 & qwen3.5-0.8b\newline seed 17, b06\newline tuned\_\allowbreak{}\_\allowbreak{}simnpo\newline replication & 0.207 & -0.050 [-0.070, -0.030]\newline intents: [-0.234, +0.074] & -0.020 [-0.057, +0.016]\newline intents: [-0.176, +0.093] & -0.24 /\allowbreak{} -0.09 & +0.1 & no benefit \\
19 & qwen3.5-0.8b\newline seed 17, b07\newline tuned\_\allowbreak{}\_\allowbreak{}simnpo\newline replication & 0.289 & -0.071 [-0.087, -0.054]\newline intents: [-0.148, +0.001] & -0.051 [-0.080, -0.021]\newline intents: [-0.137, +0.011] & -0.24 /\allowbreak{} -0.17 & +0.3 & no benefit \\
20 & qwen3.5-0.8b\newline seed 42, b02\newline tuned\_\allowbreak{}\_\allowbreak{}simnpo\newline replication & 0.346 & +0.062 [+0.042, +0.083]\newline intents: [-0.012, +0.140] & +0.040 [+0.006, +0.074]\newline intents: [-0.071, +0.126] & 0.18 /\allowbreak{} 0.11 & +0.0 & accepted \\
\bottomrule
\end{tabular}
\par\smallskip\noindent\parbox{\textwidth}{\footnotesize
Some visible source keys repeat with different displacement and row counts; the export omits a distinguishing granularity field. They are retained as separate source rows, without assigning a missing label. The main-text class result is the bundle-01 row with $D_{0,\mathrm{F}}=0.416$.
}
\end{table*}
% }
% \end{landscape}
% \clearpage

% \clearpage
% \begin{landscape}
% {\fontsize{8}{10}\selectfont\setlength{\tabcolsep}{3pt}
% \renewcommand{\arraystretch}{1.18}
% \renewcommand{\theHtable}{supp.8.a}
\setcounter{table}{7}
\newcommand{\SEightHeader}{%
\toprule
\textbf{ID} & \textbf{Dataset / run} & \textbf{View} &
\textbf{Dev. $s$} & \textbf{Held-out $s$ [rows]; class CI} &
\textbf{Aligned $a$ [rows]} & \textbf{Relative residual [rows]} &
\textbf{$P_\mathrm{F}$; gate} & \textbf{Identity error; zero directions} \\
\midrule
}

% Part 1: IDs 1--9; BANKING77: SimNPO, NPO and GradAscent
\begin{table*}[p]
\centering
\footnotesize
\setlength{\tabcolsep}{2pt}
\caption{Matched suppression (development vs held-out) and oracle-direction geometry.}
\label{tab:S08_suppression_and_geometry}
\begin{tabular}{@{}
>{\raggedright\arraybackslash}p{0.50cm}
>{\raggedright\arraybackslash}p{2.70cm}
>{\raggedright\arraybackslash}p{0.90cm}
>{\raggedright\arraybackslash}p{0.78cm}
>{\raggedright\arraybackslash}p{2.72cm}
>{\raggedright\arraybackslash}p{2.50cm}
>{\raggedright\arraybackslash}p{2.50cm}
>{\raggedright\arraybackslash}p{1.35cm}
>{\raggedright\arraybackslash}p{\dimexpr\textwidth-13.95cm-16\tabcolsep\relax}
@{}}
\SEightHeader
1 & banking77\newline matched\_\allowbreak{}\_\allowbreak{}simnpo & FG & 0.64 & 0.78 [0.74, 0.83]\newline classes: [0.48, 1.07] & 0.67 [0.63, 0.71] & 0.36 [0.32, 0.39] & 0.56\newline feasible: True & 1.1e-15\newline zero rows: 0 \\
2 & banking77\newline matched\_\allowbreak{}\_\allowbreak{}simnpo & F & 0.64 & 0.79 [0.74, 0.84]\newline classes: [0.48, 1.07] & 0.68 [0.63, 0.72] & 0.36 [0.32, 0.39] & 0.56\newline feasible: True & 1.1e-15\newline zero rows: 0 \\
3 & banking77\newline matched\_\allowbreak{}\_\allowbreak{}simnpo & G & 0.64 & 0.77 [0.68, 0.86]\newline classes: [0.45, 1.09] & 0.65 [0.57, 0.73] & 0.35 [0.29, 0.41] & 0.56\newline feasible: True & 1.1e-15\newline zero rows: 0 \\
4 & banking77\newline confirm\_\allowbreak{}\_\allowbreak{}npo & FG & 0.59 & 0.40 [0.36, 0.44]\newline classes: [0.21, 0.60] & 0.39 [0.35, 0.44] & 0.15 [0.12, 0.18] & 0.39\newline feasible: True & 1.3e-15\newline zero rows: 0 \\
5 & banking77\newline confirm\_\allowbreak{}\_\allowbreak{}npo & F & 0.59 & 0.42 [0.37, 0.47]\newline classes: [0.22, 0.64] & 0.41 [0.36, 0.46] & 0.16 [0.12, 0.19] & 0.39\newline feasible: True & 8.9e-16\newline zero rows: 0 \\
6 & banking77\newline confirm\_\allowbreak{}\_\allowbreak{}npo & G & 0.59 & 0.36 [0.28, 0.44]\newline classes: [0.17, 0.57] & 0.36 [0.27, 0.44] & 0.12 [0.05, 0.19] & 0.39\newline feasible: True & 1.3e-15\newline zero rows: 0 \\
7 & banking77\newline confirm\_\allowbreak{}\_\allowbreak{}grad\_\allowbreak{}ascent & FG & 0.54 & 0.68 [0.63, 0.72]\newline classes: [0.48, 0.86] & 0.62 [0.58, 0.66] & 0.32 [0.28, 0.35] & 0.55\newline feasible: True & 1.1e-15\newline zero rows: 0 \\
8 & banking77\newline confirm\_\allowbreak{}\_\allowbreak{}grad\_\allowbreak{}ascent & F & 0.54 & 0.70 [0.65, 0.75]\newline classes: [0.49, 0.86] & 0.64 [0.60, 0.69] & 0.32 [0.29, 0.36] & 0.55\newline feasible: True & 1.1e-15\newline zero rows: 0 \\
9 & banking77\newline confirm\_\allowbreak{}\_\allowbreak{}grad\_\allowbreak{}ascent & G & 0.54 & 0.61 [0.52, 0.70]\newline classes: [0.45, 0.88] & 0.56 [0.47, 0.64] & 0.29 [0.22, 0.35] & 0.55\newline feasible: True & 8.9e-16\newline zero rows: 0 \\
\bottomrule
\end{tabular}
\end{table*}

% Part 2: IDs 10--18; BANKING77: GradDiff, GradDiff-KL and RMU
\begin{table*}[p]
\ContinuedFloat
\centering
\footnotesize
\setlength{\tabcolsep}{2pt}
\caption[]{Matched suppression and oracle-direction geometry (continued).}
\begin{tabular}{@{}
>{\raggedright\arraybackslash}p{0.50cm}
>{\raggedright\arraybackslash}p{2.70cm}
>{\raggedright\arraybackslash}p{0.90cm}
>{\raggedright\arraybackslash}p{0.78cm}
>{\raggedright\arraybackslash}p{2.72cm}
>{\raggedright\arraybackslash}p{2.50cm}
>{\raggedright\arraybackslash}p{2.50cm}
>{\raggedright\arraybackslash}p{1.35cm}
>{\raggedright\arraybackslash}p{\dimexpr\textwidth-13.95cm-16\tabcolsep\relax}
@{}}
\SEightHeader
10 & banking77\newline matched\_\allowbreak{}\_\allowbreak{}grad\_\allowbreak{}diff & FG & 0.65 & 1.11 [1.04, 1.19]\newline classes: [0.60, 1.60] & 0.74 [0.70, 0.78] & 0.50 [0.47, 0.54] & 0.19\newline feasible: False & 2.0e-15\newline zero rows: 0 \\
11 & banking77\newline matched\_\allowbreak{}\_\allowbreak{}grad\_\allowbreak{}diff & F & 0.65 & 1.10 [1.02, 1.18]\newline classes: [0.61, 1.55] & 0.74 [0.70, 0.78] & 0.50 [0.46, 0.54] & 0.19\newline feasible: False & 2.0e-15\newline zero rows: 0 \\
12 & banking77\newline matched\_\allowbreak{}\_\allowbreak{}grad\_\allowbreak{}diff & G & 0.65 & 1.16 [1.01, 1.33]\newline classes: [0.59, 1.81] & 0.75 [0.66, 0.83] & 0.51 [0.43, 0.59] & 0.19\newline feasible: False & 1.3e-15\newline zero rows: 0 \\
13 & banking77\newline matched\_\allowbreak{}\_\allowbreak{}grad\_\allowbreak{}diff\_\allowbreak{}kl & FG & 0.63 & 0.53 [0.49, 0.57]\newline classes: [0.36, 0.72] & 0.49 [0.45, 0.53] & 0.20 [0.17, 0.23] & 0.48\newline feasible: True & 8.9e-16\newline zero rows: 0 \\
14 & banking77\newline matched\_\allowbreak{}\_\allowbreak{}grad\_\allowbreak{}diff\_\allowbreak{}kl & F & 0.63 & 0.55 [0.50, 0.60]\newline classes: [0.37, 0.73] & 0.51 [0.46, 0.56] & 0.21 [0.18, 0.24] & 0.48\newline feasible: True & 8.9e-16\newline zero rows: 0 \\
15 & banking77\newline matched\_\allowbreak{}\_\allowbreak{}grad\_\allowbreak{}diff\_\allowbreak{}kl & G & 0.63 & 0.46 [0.38, 0.54]\newline classes: [0.32, 0.71] & 0.43 [0.35, 0.51] & 0.16 [0.10, 0.22] & 0.48\newline feasible: True & 8.9e-16\newline zero rows: 0 \\
16 & banking77\newline matched\_\allowbreak{}\_\allowbreak{}rmu & FG & 0.58 & 0.72 [0.67, 0.76]\newline classes: [0.43, 0.95] & 0.40 [0.38, 0.43] & 0.36 [0.34, 0.38] & 0.04\newline feasible: True & 1.1e-15\newline zero rows: 0 \\
17 & banking77\newline matched\_\allowbreak{}\_\allowbreak{}rmu & F & 0.58 & 0.72 [0.66, 0.77]\newline classes: [0.41, 0.95] & 0.40 [0.37, 0.43] & 0.36 [0.34, 0.37] & 0.04\newline feasible: True & 1.1e-15\newline zero rows: 0 \\
18 & banking77\newline matched\_\allowbreak{}\_\allowbreak{}rmu & G & 0.58 & 0.71 [0.62, 0.82]\newline classes: [0.51, 0.95] & 0.40 [0.35, 0.46] & 0.37 [0.33, 0.41] & 0.04\newline feasible: True & 8.9e-16\newline zero rows: 0 \\
\bottomrule
\end{tabular}
\end{table*}

% Part 3: IDs 19--24; CLINC150: matched SimNPO, ep7 and ep9
\begin{table*}[p]
\ContinuedFloat
\centering
\footnotesize
\setlength{\tabcolsep}{2pt}
\caption[]{Matched suppression and oracle-direction geometry (continued).}
\begin{tabular}{@{}
>{\raggedright\arraybackslash}p{0.50cm}
>{\raggedright\arraybackslash}p{2.70cm}
>{\raggedright\arraybackslash}p{0.90cm}
>{\raggedright\arraybackslash}p{0.78cm}
>{\raggedright\arraybackslash}p{2.72cm}
>{\raggedright\arraybackslash}p{2.50cm}
>{\raggedright\arraybackslash}p{2.50cm}
>{\raggedright\arraybackslash}p{1.35cm}
>{\raggedright\arraybackslash}p{\dimexpr\textwidth-13.95cm-16\tabcolsep\relax}
@{}}
\SEightHeader
19 & clinc150\newline matched\_\allowbreak{}s1\_\allowbreak{}\_\allowbreak{}simnpo\_\allowbreak{}b2.5\_\allowbreak{}ep7 & FG & 1.01 & 1.93 [1.80, 2.06]\newline classes: [1.32, 3.30] & 0.80 [0.77, 0.83] & 0.67 [0.63, 0.71] & -0.39\newline feasible: True & 1.6e-15\newline zero rows: 0 \\
20 & clinc150\newline matched\_\allowbreak{}s1\_\allowbreak{}\_\allowbreak{}simnpo\_\allowbreak{}b2.5\_\allowbreak{}ep7 & F & 1.01 & 1.94 [1.80, 2.10]\newline classes: [1.38, 3.13] & 0.81 [0.77, 0.85] & 0.67 [0.63, 0.72] & -0.39\newline feasible: True & 1.3e-15\newline zero rows: 0 \\
21 & clinc150\newline matched\_\allowbreak{}s1\_\allowbreak{}\_\allowbreak{}simnpo\_\allowbreak{}b2.5\_\allowbreak{}ep7 & G & 1.01 & 1.87 [1.63, 2.18]\newline classes: [1.21, 3.74] & 0.78 [0.71, 0.84] & 0.64 [0.57, 0.73] & -0.39\newline feasible: True & 1.6e-15\newline zero rows: 0 \\
22 & clinc150\newline matched\_\allowbreak{}s1\_\allowbreak{}\_\allowbreak{}simnpo\_\allowbreak{}b4.5\_\allowbreak{}ep9 & FG & 1.00 & 1.61 [1.51, 1.73]\newline classes: [0.99, 2.89] & 0.73 [0.69, 0.77] & 0.54 [0.51, 0.58] & -0.17\newline feasible: True & 1.6e-15\newline zero rows: 0 \\
23 & clinc150\newline matched\_\allowbreak{}s1\_\allowbreak{}\_\allowbreak{}simnpo\_\allowbreak{}b4.5\_\allowbreak{}ep9 & F & 1.00 & 1.63 [1.50, 1.77]\newline classes: [1.03, 2.78] & 0.74 [0.69, 0.79] & 0.55 [0.51, 0.60] & -0.17\newline feasible: True & 1.6e-15\newline zero rows: 0 \\
24 & clinc150\newline matched\_\allowbreak{}s1\_\allowbreak{}\_\allowbreak{}simnpo\_\allowbreak{}b4.5\_\allowbreak{}ep9 & G & 1.00 & 1.55 [1.34, 1.81]\newline classes: [0.91, 3.16] & 0.71 [0.62, 0.81] & 0.51 [0.44, 0.59] & -0.17\newline feasible: True & 1.1e-15\newline zero rows: 0 \\
\bottomrule
\end{tabular}
\end{table*}

% Part 4: IDs 25--30; CLINC150: matched SimNPO ep10 and tuned SimNPO
\begin{table*}[p]
\ContinuedFloat
\centering
\footnotesize
\setlength{\tabcolsep}{2pt}
\caption[]{Matched suppression and oracle-direction geometry (continued).}
\begin{tabular}{@{}
>{\raggedright\arraybackslash}p{0.50cm}
>{\raggedright\arraybackslash}p{2.70cm}
>{\raggedright\arraybackslash}p{0.90cm}
>{\raggedright\arraybackslash}p{0.78cm}
>{\raggedright\arraybackslash}p{2.72cm}
>{\raggedright\arraybackslash}p{2.50cm}
>{\raggedright\arraybackslash}p{2.50cm}
>{\raggedright\arraybackslash}p{1.35cm}
>{\raggedright\arraybackslash}p{\dimexpr\textwidth-13.95cm-16\tabcolsep\relax}
@{}}
\SEightHeader
25 & clinc150\newline matched\_\allowbreak{}s1\_\allowbreak{}\_\allowbreak{}simnpo\_\allowbreak{}b6.5\_\allowbreak{}ep10 & FG & 0.94 & 1.18 [1.11, 1.27]\newline classes: [0.74, 2.00] & 0.61 [0.57, 0.64] & 0.37 [0.34, 0.40] & 0.17\newline feasible: True & 8.9e-16\newline zero rows: 0 \\
26 & clinc150\newline matched\_\allowbreak{}s1\_\allowbreak{}\_\allowbreak{}simnpo\_\allowbreak{}b6.5\_\allowbreak{}ep10 & F & 0.94 & 1.18 [1.09, 1.28]\newline classes: [0.76, 1.91] & 0.60 [0.55, 0.64] & 0.38 [0.35, 0.41] & 0.17\newline feasible: True & 8.9e-16\newline zero rows: 0 \\
27 & clinc150\newline matched\_\allowbreak{}s1\_\allowbreak{}\_\allowbreak{}simnpo\_\allowbreak{}b6.5\_\allowbreak{}ep10 & G & 0.94 & 1.19 [1.04, 1.37]\newline classes: [0.74, 2.24] & 0.64 [0.55, 0.72] & 0.36 [0.29, 0.43] & 0.17\newline feasible: True & 6.7e-16\newline zero rows: 0 \\
28 & clinc150\newline tuned\_\allowbreak{}\_\allowbreak{}simnpo & FG & --- & 0.81 [0.75, 0.88]\newline classes: [0.46, 1.46] & 0.42 [0.39, 0.46] & 0.26 [0.24, 0.29] & 0.22\newline feasible: True & 1.1e-15\newline zero rows: 0 \\
29 & clinc150\newline tuned\_\allowbreak{}\_\allowbreak{}simnpo & F & --- & 0.80 [0.73, 0.87]\newline classes: [0.46, 1.39] & 0.41 [0.37, 0.44] & 0.26 [0.23, 0.29] & 0.22\newline feasible: True & 1.1e-15\newline zero rows: 0 \\
30 & clinc150\newline tuned\_\allowbreak{}\_\allowbreak{}simnpo & G & --- & 0.86 [0.73, 1.00]\newline classes: [0.51, 1.66] & 0.47 [0.39, 0.55] & 0.27 [0.21, 0.33] & 0.22\newline feasible: True & 8.9e-16\newline zero rows: 0 \\
\bottomrule
\end{tabular}

\par\smallskip
\noindent\parbox{\textwidth}{\footnotesize
The largest identity error in this table is $2\times10^{-15}$.
This precision statement applies to these matched-suppression rows;
the broader all-run diagnostic figure has a different displayed numerical range.
}
\end{table*}
% }
% \end{landscape}
% \clearpage

% \clearpage
% \begin{landscape}
% {\fontsize{8}{10}\selectfont\setlength{\tabcolsep}{3pt}
% \renewcommand{\arraystretch}{1.18}
% \renewcommand{\theHtable}{supp.9.a}
\setcounter{table}{8}
\begin{table*}[p]
\centering
\footnotesize
\setlength{\tabcolsep}{3pt}
\renewcommand{\arraystretch}{1.13}
\caption{Recovery contrasts for every exposure budget and adaptation seed.}
\label{tab:S09_recovery_summary}
\begin{tabular}{@{}
>{\raggedright\arraybackslash}p{0.5cm}
>{\raggedright\arraybackslash}p{2.35cm}
>{\raggedright\arraybackslash}p{1.65cm}
>{\raggedright\arraybackslash}p{3.4cm}
>{\raggedright\arraybackslash}p{4.6cm}
>{\raggedright\arraybackslash}p{3.1cm}@{}}
\toprule
\textbf{ID} & \textbf{Model / configuration} & \textbf{$p(0)$: U / O} & \textbf{$G_{\rm abs}$ [rows]} & \textbf{$G_{\rm adj}$ [rows]; class CI} & \textbf{$G_H$ [rows]} \\
\midrule
1 & qwen3.5-0.8b\newline k1\_\allowbreak{}s0 & 0.364 /\allowbreak{} 0.313 & +0.139 [+0.092, +0.187] & +0.087 [+0.043, +0.131]\newline classes: [+0.039, +0.139] & +0.178 [+0.119, +0.237] \\
2 & qwen3.5-0.8b\newline k1\_\allowbreak{}s1 & 0.364 /\allowbreak{} 0.313 & +0.073 [+0.039, +0.107] & +0.022 [-0.029, +0.073]\newline classes: [-0.049, +0.088] & +0.081 [+0.031, +0.130] \\
3 & qwen3.5-0.8b\newline k5\_\allowbreak{}s0 & 0.364 /\allowbreak{} 0.313 & +0.080 [+0.059, +0.104] & +0.029 [-0.016, +0.073]\newline classes: [-0.087, +0.120] & +0.093 [+0.062, +0.127] \\
4 & qwen3.5-0.8b\newline k5\_\allowbreak{}s1 & 0.364 /\allowbreak{} 0.313 & +0.100 [+0.077, +0.126] & +0.049 [+0.001, +0.099]\newline classes: [-0.074, +0.136] & +0.128 [+0.095, +0.164] \\
5 & qwen3.5-0.8b\newline k20\_\allowbreak{}s0 & 0.364 /\allowbreak{} 0.313 & +0.079 [+0.057, +0.103] & +0.028 [-0.017, +0.075]\newline classes: [-0.090, +0.122] & +0.097 [+0.064, +0.132] \\
6 & qwen3.5-0.8b\newline k20\_\allowbreak{}s1 & 0.364 /\allowbreak{} 0.313 & +0.095 [+0.073, +0.116] & +0.043 [+0.002, +0.087]\newline classes: [-0.045, +0.129] & +0.117 [+0.086, +0.148] \\
\midrule
7 & apertus-8b\newline k1\_\allowbreak{}s0 & 0.326 /\allowbreak{} 0.429 & +0.056 [+0.024, +0.090] & +0.159 [+0.114, +0.205]\newline classes: [+0.093, +0.225] & +0.139 [+0.089, +0.195] \\
8 & apertus-8b\newline k1\_\allowbreak{}s1 & 0.326 /\allowbreak{} 0.429 & +0.057 [+0.034, +0.080] & +0.160 [+0.114, +0.207]\newline classes: [+0.028, +0.272] & +0.140 [+0.100, +0.184] \\
9 & apertus-8b\newline k5\_\allowbreak{}s0 & 0.326 /\allowbreak{} 0.429 & +0.033 [+0.009, +0.058] & +0.136 [+0.094, +0.178]\newline classes: [+0.052, +0.219] & +0.100 [+0.064, +0.139] \\
10 & apertus-8b\newline k5\_\allowbreak{}s1 & 0.326 /\allowbreak{} 0.429 & +0.066 [+0.044, +0.089] & +0.169 [+0.127, +0.213]\newline classes: [+0.061, +0.279] & +0.154 [+0.118, +0.194] \\
11 & apertus-8b\newline k20\_\allowbreak{}s0 & 0.326 /\allowbreak{} 0.429 & +0.075 [+0.056, +0.096] & +0.178 [+0.140, +0.217]\newline classes: [+0.078, +0.278] & +0.167 [+0.142, +0.196] \\
12 & apertus-8b\newline k20\_\allowbreak{}s1 & 0.326 /\allowbreak{} 0.429 & +0.065 [+0.043, +0.087] & +0.168 [+0.127, +0.210]\newline classes: [+0.087, +0.249] & +0.150 [+0.121, +0.184] \\
\midrule
13 & qwen3.5-9b\newline k1\_\allowbreak{}s0 & 0.614 /\allowbreak{} 0.477 & +0.116 [+0.070, +0.162] & -0.021 [-0.057, +0.015]\newline classes: [-0.104, +0.068] & +0.135 [+0.061, +0.210] \\
14 & qwen3.5-9b\newline k1\_\allowbreak{}s1 & 0.614 /\allowbreak{} 0.477 & +0.120 [+0.080, +0.166] & -0.016 [-0.068, +0.032]\newline classes: [-0.180, +0.167] & +0.151 [+0.074, +0.229] \\
15 & qwen3.5-9b\newline k5\_\allowbreak{}s0 & 0.614 /\allowbreak{} 0.477 & +0.066 [+0.044, +0.092] & -0.070 [-0.115, -0.029]\newline classes: [-0.202, +0.063] & +0.043 [-0.002, +0.090] \\
16 & qwen3.5-9b\newline k5\_\allowbreak{}s1 & 0.614 /\allowbreak{} 0.477 & +0.072 [+0.044, +0.102] & -0.065 [-0.112, -0.021]\newline classes: [-0.194, +0.100] & +0.048 [-0.013, +0.110] \\
17 & qwen3.5-9b\newline k20\_\allowbreak{}s0 & 0.614 /\allowbreak{} 0.477 & +0.072 [+0.050, +0.093] & -0.065 [-0.108, -0.025]\newline classes: [-0.165, +0.053] & +0.068 [+0.025, +0.108] \\
18 & qwen3.5-9b\newline k20\_\allowbreak{}s1 & 0.614 /\allowbreak{} 0.477 & +0.106 [+0.083, +0.129] & -0.031 [-0.070, +0.008]\newline classes: [-0.118, +0.097] & +0.132 [+0.089, +0.174] \\
\bottomrule
\end{tabular}

\par\smallskip\noindent
\parbox{\textwidth}{\footnotesize
Each interval is conditional on the two adaptation seeds and the reused trained backbones. Four-class intervals are coarse. The ratio-of-means headroom definition is used here.
}
\end{table*}

% Table S9, continued: component areas, threshold crossings, residuals.
\begin{table*}[p]
\ContinuedFloat
\centering
\footnotesize
\setlength{\tabcolsep}{3pt}
\renewcommand{\arraystretch}{1.16}
\caption[]{Recovery summaries (continued): component areas, threshold crossings, and algebraic residual.}
\begin{tabular}{@{}
>{\raggedright\arraybackslash}p{0.5cm}
>{\raggedright\arraybackslash}p{2.45cm}
>{\raggedright\arraybackslash}p{2.45cm}
>{\raggedright\arraybackslash}p{3.6cm}
>{\raggedright\arraybackslash}p{4.6cm}
>{\raggedright\arraybackslash}p{1.5cm}@{}}
\toprule
\textbf{ID} & \textbf{$A_{\rm abs}$: U / O} & \textbf{$A_{\rm adj}$: U / O} & \textbf{First $t$: $p\geq0.5$, U / O} & \textbf{First $t$: $p\geq0.8$, U / O} & \textbf{Identity residual} \\
\midrule
1 & 0.801 /\allowbreak{} 0.662 & 0.437 /\allowbreak{} 0.349 & 1 /\allowbreak{} 5 & 5 /\allowbreak{} not reached by t = 25 (censored) & 6e-17 \\
2 & 0.784 /\allowbreak{} 0.711 & 0.420 /\allowbreak{} 0.398 & 5 /\allowbreak{} 5 & 10 /\allowbreak{} 25 & 6e-17 \\
3 & 0.798 /\allowbreak{} 0.717 & 0.433 /\allowbreak{} 0.404 & 5 /\allowbreak{} 10 & 10 /\allowbreak{} 10 & 6e-17 \\
4 & 0.848 /\allowbreak{} 0.748 & 0.484 /\allowbreak{} 0.435 & 5 /\allowbreak{} 5 & 10 /\allowbreak{} 10 & 1e-16 \\
5 & 0.847 /\allowbreak{} 0.768 & 0.483 /\allowbreak{} 0.455 & 5 /\allowbreak{} 5 & 10 /\allowbreak{} 10 & 1e-16 \\
6 & 0.820 /\allowbreak{} 0.725 & 0.456 /\allowbreak{} 0.412 & 5 /\allowbreak{} 10 & 10 /\allowbreak{} 10 & 8e-17 \\
\midrule
7 & 0.845 /\allowbreak{} 0.789 & 0.519 /\allowbreak{} 0.360 & 1 /\allowbreak{} 1 & 5 /\allowbreak{} 10 & 6e-17 \\
8 & 0.850 /\allowbreak{} 0.793 & 0.524 /\allowbreak{} 0.364 & 1 /\allowbreak{} 5 & 5 /\allowbreak{} 10 & 6e-17 \\
9 & 0.839 /\allowbreak{} 0.806 & 0.513 /\allowbreak{} 0.377 & 5 /\allowbreak{} 5 & 5 /\allowbreak{} 10 & 6e-17 \\
10 & 0.854 /\allowbreak{} 0.788 & 0.528 /\allowbreak{} 0.359 & 1 /\allowbreak{} 5 & 5 /\allowbreak{} 10 & 6e-17 \\
11 & 0.867 /\allowbreak{} 0.793 & 0.541 /\allowbreak{} 0.363 & 5 /\allowbreak{} 5 & 5 /\allowbreak{} 10 & 6e-17 \\
12 & 0.862 /\allowbreak{} 0.797 & 0.536 /\allowbreak{} 0.368 & 5 /\allowbreak{} 5 & 5 /\allowbreak{} 10 & 6e-17 \\
\midrule
13 & 0.873 /\allowbreak{} 0.758 & 0.260 /\allowbreak{} 0.281 & 0 /\allowbreak{} 1 & 5 /\allowbreak{} 10 & 0e+00 \\
14 & 0.883 /\allowbreak{} 0.763 & 0.269 /\allowbreak{} 0.286 & 0 /\allowbreak{} 1 & 5 /\allowbreak{} 25 & 0e+00 \\
15 & 0.877 /\allowbreak{} 0.810 & 0.263 /\allowbreak{} 0.333 & 0 /\allowbreak{} 1 & 5 /\allowbreak{} 10 & 0e+00 \\
16 & 0.869 /\allowbreak{} 0.797 & 0.255 /\allowbreak{} 0.320 & 0 /\allowbreak{} 1 & 5 /\allowbreak{} 10 & 0e+00 \\
17 & 0.898 /\allowbreak{} 0.826 & 0.284 /\allowbreak{} 0.349 & 0 /\allowbreak{} 1 & 5 /\allowbreak{} 10 & 0e+00 \\
18 & 0.896 /\allowbreak{} 0.790 & 0.282 /\allowbreak{} 0.313 & 0 /\allowbreak{} 5 & 5 /\allowbreak{} 10 & 0e+00 \\
\bottomrule
\end{tabular}

\par\smallskip\noindent
\parbox{\textwidth}{\footnotesize
Unreached thresholds are censored at $t=25$; no extrapolated crossing time is assigned.
}
\end{table*}
% }
% \end{landscape}
% \clearpage

% \clearpage
% \begin{landscape}
% {\fontsize{8}{10}\selectfont\setlength{\tabcolsep}{3pt}
% \renewcommand{\arraystretch}{1.18}
% \renewcommand{\theHtable}{supp.10.a}
\setcounter{table}{9}
\begin{table*}[tbp]
\centering
\footnotesize
\caption{Seed-17 and seed-42 optimisation diagnostics
(frozen class SimNPO, BANKING77 bundle 01).}
\label{tab:S10_optimization_diagnostics}

\setlength{\tabcolsep}{2.5pt}
\renewcommand{\arraystretch}{1.12}

\begin{tabular}{
@{}
>{\raggedright\arraybackslash}p{1.9cm}
>{\raggedright\arraybackslash}p{1.45cm}
>{\raggedright\arraybackslash}p{2.5cm}
>{\raggedright\arraybackslash}p{1.35cm}
>{\raggedright\arraybackslash}p{2.3cm}
>{\raggedright\arraybackslash}p{2.3cm}
>{\raggedright\arraybackslash}p{2.35cm}
>{\raggedright\arraybackslash}p{\dimexpr\textwidth-14.15cm-14\tabcolsep\relax}
@{}
}
\toprule
\textbf{Model / seed} &
\textbf{$P_\mathrm{F}/s_\mathrm{F}$} &
\textbf{Full gold p / confidence / margin} &
\textbf{Oracle accuracy} &
\textbf{Forget loss: first / last 10} &
\textbf{Retain loss: first / last 10} &
\textbf{Gradient norm: first / last 10} &
\textbf{Source} \\
\midrule

qwen3.5-0.8b\newline seed 17 &
+0.59 / 0.96 &
0.967 /\allowbreak{} 0.977 /\allowbreak{} 0.959 &
0.27 &
0.291 /\allowbreak{} 0.082 &
0.057 /\allowbreak{} 0.146 &
0.822 /\allowbreak{} 1.314 &
recorded training log \\

qwen3.5-0.8b\newline seed 42 &
+0.55 / 0.86 &
0.955 /\allowbreak{} 0.971 /\allowbreak{} 0.947 &
0.32 &
0.277 /\allowbreak{} 0.103 &
0.170 /\allowbreak{} 0.180 &
0.952 /\allowbreak{} 1.483 &
recorded training log \\

apertus-8b\newline seed 17 &
+0.55 / 1.07 &
0.934 /\allowbreak{} 0.956 /\allowbreak{} 0.920 &
0.39 &
0.268 /\allowbreak{} 0.053 &
0.131 /\allowbreak{} 0.214 &
0.448 /\allowbreak{} 0.678 &
recorded training log \\

apertus-8b\newline seed 42 &
+0.17 / 0.39 &
0.943 /\allowbreak{} 0.970 /\allowbreak{} 0.947 &
0.41 &
0.266 /\allowbreak{} 0.116 &
0.172 /\allowbreak{} 0.264 &
0.564 /\allowbreak{} 0.622 &
recorded training log \\

qwen3.5-9b\newline seed 17 &
+0.54 / 0.57 &
0.962 /\allowbreak{} 0.976 /\allowbreak{} 0.958 &
0.48 &
0.290 /\allowbreak{} 0.178 &
0.073 /\allowbreak{} 0.109 &
0.504 /\allowbreak{} 1.096 &
recorded training log \\

qwen3.5-9b\newline seed 42 &
+0.14 / 0.15 &
0.969 /\allowbreak{} 0.988 /\allowbreak{} 0.979 &
0.42 &
0.285 /\allowbreak{} 0.251 &
0.150 /\allowbreak{} 0.141 &
0.669 /\allowbreak{} 1.038 &
recorded training log \\

olmo3-7b\newline seed 17 &
+0.34 / 0.38 &
0.956 /\allowbreak{} 0.969 /\allowbreak{} 0.945 &
0.35 &
0.283 /\allowbreak{} 0.196 &
0.063 /\allowbreak{} 0.135 &
0.322 /\allowbreak{} 0.532 &
recorded training log \\

olmo3-7b\newline seed 42 &
+0.04 / 0.35 &
0.962 /\allowbreak{} 0.979 /\allowbreak{} 0.964 &
0.45 &
0.276 /\allowbreak{} 0.156 &
0.136 /\allowbreak{} 0.193 &
0.395 /\allowbreak{} 0.603 &
recorded training log \\

qwen3-4b\newline seed 17 &
+0.28 / 0.60 &
0.955 /\allowbreak{} 0.968 /\allowbreak{} 0.941 &
0.58 &
0.283 /\allowbreak{} 0.198 &
0.069 /\allowbreak{} 0.168 &
0.445 /\allowbreak{} 1.039 &
recorded training log \\

qwen3-4b\newline seed 42 &
-0.18 / 0.61 &
0.953 /\allowbreak{} 0.970 /\allowbreak{} 0.945 &
0.56 &
0.271 /\allowbreak{} 0.120 &
0.174 /\allowbreak{} 0.211 &
0.547 /\allowbreak{} 0.910 &
recorded training log \\

\bottomrule
\end{tabular}
\end{table*}
% }
% \end{landscape}
% \clearpage

% \clearpage
% {\fontsize{8}{10}\selectfont\setlength{\tabcolsep}{3pt}
% \renewcommand{\arraystretch}{1.18}
% \renewcommand{\theHtable}{supp.11.a}
\setcounter{table}{10}
\begin{table*}[tbp]
\centering
\small
\caption{Compute by experiment and stage, and frozen configurations.}
\label{tab:S11_compute_and_configuration}

\setlength{\tabcolsep}{3pt}
\renewcommand{\arraystretch}{1.12}

\begin{tabular}{
@{}
>{\raggedright\arraybackslash}p{5cm}
>{\raggedright\arraybackslash}p{3.6cm}
>{\centering\arraybackslash}p{2.2cm}
>{\centering\arraybackslash}p{3.2cm}
>{\centering\arraybackslash}p{1.4cm}
@{}
}
\toprule
\textbf{Experiment} &
\textbf{Stage} &
\textbf{Evaluations} &
\textbf{GPU hours (logged)} &
\textbf{Models} \\
\midrule

architecture\_\allowbreak{}replication &
replication & 8 & 40.7 & 4 \\

architecture\_\allowbreak{}screen &
screening & 19 & 97.8 & 19 \\

association\_\allowbreak{}methods &
confirmation\_\allowbreak{}sensitivity & 29 & 23.7 & 1 \\

clinc150\_\allowbreak{}frozen\_\allowbreak{}transfer &
transfer & 7 & 9.6 & 2 \\

clinc150\_\allowbreak{}tuned &
confirmation & 2 & 5.9 & 1 \\

clinc150\_\allowbreak{}tuned &
control & 3 & 4.5 & 1 \\

clinc150\_\allowbreak{}tuned &
replication & 5 & 25.1 & 1 \\

counterfactual\_\allowbreak{}reference &
reference & 194 & 545.7 & 20 \\

development\_\allowbreak{}selection &
development & 646 & 705.9 & 1 \\

development\_\allowbreak{}selection &
development\_\allowbreak{}reference & 8 & 17.6 & 1 \\

label\_\allowbreak{}semantics\_\allowbreak{}banking77 &
exploratory & 12 & 16.0 & 1 \\

label\_\allowbreak{}semantics\_\allowbreak{}clinc150 &
exploratory & 9 & 36.9 & 1 \\

matched\_\allowbreak{}suppression &
exploratory & 7 & 16.9 & 1 \\

method\_\allowbreak{}comparison &
confirmation & 6 & 6.0 & 1 \\

method\_\allowbreak{}comparison &
replication & 32 & 34.2 & 1 \\

pilot\_\allowbreak{}milestone &
pilot & 15 & 13.3 & 1 \\

seed42\_\allowbreak{}strength &
exploratory & 10 & 45.0 & 2 \\

test\_\allowbreak{}fixture &
excluded & 31 & 0.0 & 1 \\

\bottomrule
\end{tabular}

\par\smallskip
\noindent
\parbox{\textwidth}{\footnotesize
Logged evaluation costs are incomplete. In particular,
zero entries for full-model evaluations omit full training,
and recovery is not fully accounted for. These numbers
do not establish the total compute cost or a retraining speedup.
}
\end{table*}
% }
% \clearpage

% \clearpage
% {\fontsize{8}{10}\selectfont\setlength{\tabcolsep}{3pt}
% \renewcommand{\arraystretch}{1.18}
% \renewcommand{\theHtable}{supp.12.a}
\setcounter{table}{11}
\begin{table*}[tbp]
\centering
\footnotesize
\caption{Audit record: gates, missing or reserved results, reconciliation and repairs.}
\label{tab:S12_audit_and_limitations}

\setlength{\tabcolsep}{5pt}
\renewcommand{\arraystretch}{1.15}

\begin{tabular}{
@{}
>{\raggedright\arraybackslash}p{3cm}
>{\raggedright\arraybackslash}p{\dimexpr\textwidth-3cm-4\tabcolsep\relax}
@{}
}
\toprule
\textbf{Item} & \textbf{Upstream audit record} \\
\midrule

Acquisition &
Opaque full models: all seven checks pass (gate 0.70 /\allowbreak{} 0.50). Semantic seed-42 full model: forgotten class transfer\_\allowbreak{}timing (BANKING77 bundle 02) at 0.65 development accuracy, below the 0.70 class threshold; targets were gated at the primary seed only. \\

Privacy audit &
S2-Rank stopped at its preregistered acquisition gate at seed 17 (G1 0.550, G4 0.555); the inverse-rank attack was not tested. \\

Missing /\allowbreak{} completed coverage &
Uniformization at matched suppression (no feasible development checkpoint). CLINC150 bundle 06 under opaque labels was added on 2026-10-18 after the figure brief named it as a contrast bundle (acquisition passed); coverage of the planned designs is complete. \\

Reserved &
CLINC150 class\_\allowbreak{}05\_\allowbreak{}bundle\_\allowbreak{}08 held back unseen; no file of it is read by the build (the collector stops if one appears). \\

Excluded model &
checkpoint reload not reproducible (results/\allowbreak{}model\_\allowbreak{}checks/\allowbreak{}internlm3-8b\_\allowbreak{}reload\_\allowbreak{}diagnosis.json; docs/\allowbreak{}HANDOVER.md); jobs blocked by design \\

Reconciliation &
Row-level OD recomputed for 358 evaluations: max |recomputed - record.json| = 0.0e+00. Paired bootstrap: point estimates exact over 252 comparisons; intervals exact where n\_\allowbreak{}boot matches (5000); stored 2000-draw summaries differ by at most 0.0075 (Monte Carlo); no category changes. \\

Filename collisions &
Seed-42 and opaque outputs once overwrote seed-17 /\allowbreak{} semantic analysis files; repaired by the seed /\allowbreak{} opaque run tag and regeneration (PROTOCOL tooling notes). The build keys every result by dataset, model, label scheme, seed, forget set and run label and asserts uniqueness. \\

Reporting-code fix &
The paper collector initially mapped 'grad\_\allowbreak{}diff\_\allowbreak{}kl' labels to GradDiff by prefix; fixed (longest match) before any asset was finalised; no stored result was affected. \\

Recovery reruns &
8-9B recovery runs initially ran out of memory at the second checkpoint (model not released); fixed and rerun per checkpoint. Row-level reruns for Qwen3.5-0.8B and the Apertus-8B controls were added to provide uncertainty for the control contrasts. \\

Selection notes &
CLINC150 tuning ran without the intervention gate (no same-data replicas); the selected point lies on the grid edge (beta 10, 10 epochs), extending it was held. The best matched-strength epoch (seed-42 diagnostic) is post hoc. \\

Records &
CLINC150 bundles 04-07 results are recorded in the analysis files and this build; PROTOCOL.md pre-registered their analysis (2026-10-14) but has no results entry for them. \\

\bottomrule
\end{tabular}

\par\smallskip
\noindent
\parbox{\textwidth}{\footnotesize
The opaque-check count is corrected to seven using Table S5.
Other upstream integrity checks and protocol claims are reported
rather than independently reproduced from this export-only archive.
The unchanged source table and CSV preserve the original wording.
}
\end{table*}
% }
% \clearpage

%%%%%%%%%%%%%%%%%%%%%%%%%%%%%%%%%%%%%%%%%%%%%%%%%%%%%%%%%%%%%%%%%%%%%%%%%%%%%%%
%%%%%%%%%%%%%%%%%%%%%%%%%%%%%%%%%%%%%%%%%%%%%%%%%%%%%%%%%%%%%%%%%%%%%%%%%%%%%%%

\end{document}